\documentclass{aa}  
\usepackage{hyperref}
\usepackage{graphicx,graphics,rotating,amssymb,amsmath}
\usepackage{amsmath}
\usepackage{physics}
\usepackage{amssymb}
\usepackage{xcolor}
\usepackage[utf8]{inputenc}
\usepackage{graphicx}
\usepackage{lipsum} 
\usepackage{pdfpages}

\usepackage{txfonts}
\newcommand{\Cloudy}{\textsc{Cloudy}}

\newcommand{\Pynbody}{\textsc{Pynbody }}
\newcommand{\Starburst}{\textsc{Starburst 99 }}

\newcommand{\Gasoline}{\textsc{Gasoline }}
\newcommand{\Ponos}{\textsc{Ponos}}
\newcommand{\Sigame}{\textsc{Sigame }}

\def\kramsesrt{\textsc{Kramses-rt}}
\def\ramsesrt{\textsc{Ramses-rt}}
\def\romulus25{\textsc{Romulus25 }}

\def\krome{\textsc{Krome}}

\def\starburst99{\textsc{Starburst99 }}

\usepackage{lscape}
\usepackage{longtable} 
\usepackage{array}

\definecolor{orange}{HTML}{fc8e2d}

\newcommand{\revv}[1]{{#1}}
\newcommand{\rev}[1]{{#1}}
\begin{document} 

    \title{High resolution modelling of [CII], [CI], [OIII] and CO line emission from the ISM and CGM of a star forming galaxy at $z\sim6.5$}
  
   \author{A. Schimek
          \inst{1}
          \and D. Decataldo
          \inst{1}
          \and S. Shen
          \inst{1}
          \and C. Cicone
          \inst{1}
          \and B. Baumschlager 
          \inst{1}
          \and E. van Kampen
          \inst{2}
          \and P. Klaassen
          \inst{3}
          \and P. Madau
          \inst{4}
          \and L. Di Mascolo
          \inst{5,6,7}
          \and L. Mayer
          \inst{8}
          \and I. Montoya Arroyave
          \inst{1}
          \and T. Mroczkowski
          \inst{2}
          \and J. Warraich
          \inst{1}
            }

   \institute{Institute of Theoretical Astrophysics, University of Oslo, PO Box 1029, Blindern 0315, Oslo, Norway \\ \email{alice.schimek@astro.uio.no}
   \and European Southern Observatory, Karl-Schwarzschild-Straße 2, 85748 Garching bei München, Germany 
   \and UK Astronomy Technology Centre, Royal Observatory Edinburgh, Blackford Hill, Edinburgh United Kingdom
   \and Department of Astronomy \& Astrophysics, University of California, 1156 High Street, Santa Cruz, CA 95064, USA
   \and Astronomy Unit, Department of Physics, University of Trieste, via Tiepolo 11, 34131 Trieste, Italy
   \and  INAF - Osservatorio Astronomico di Trieste, via Tiepolo 11, 34131 Trieste, Italy
   \and IFPU - Institute for Fundamental Physics of the Universe, Via Beirut 2, 34014 Trieste, Italy
   \and Center for Theoretical Astrophysics and Cosmology, Institute for Computational Science, University of Zurich, Winterthurerstrasse 190, CH-8057 Zurich, Switzerland
    }

   \date{Received ; accepted }

  \abstract 
   {The circumgalactic medium (CGM) is a crucial component of galaxy evolution, but thus far its physical properties are highly unconstrained. As of yet, no cosmological simulation has reached convergence when it comes to constraining the cold and dense gas fraction of the CGM. Such components are also challenging to \rev{observe directly}, as they require sub-millimeter (sub-mm) instruments with a high sensitivity to extended \rev{and mostly diffuse emission.} 
   We present a state-of-the-art theoretical effort at modelling the [CII]~$158 \mu$m, [CI](1-0)~$609 \mu$m, [CI](2-1)~$370 \mu$m, CO(3-2)~$867 \mu$m, and [OIII]~$88 \mu$m line emissions that arise from the ISM and CGM of galaxies, \rev{with the goal of studying the contribution from different cold ($T<10^4$~K) components of galaxy halos.} We use the high-resolution cosmological zoom-in simulation \Ponos~(m$_{gas}$ = 883.4 M$_{\odot}$), which represents a typical star forming galaxy system at $z = 6.5$, composed of a main disc with stellar mass $M_*=2\times10^9~M_{\odot}$ that is undergoing a major merger. We adopt different modelling approaches based on the photoionisation code \Cloudy. Our fiducial model uses radiative transfer post-processing with \ramsesrt~and \krome~(\kramsesrt) to create more realistic FUV radiation fields, which we then compare to other sub-grid modelling approaches adopted in the literature.
    We find significant differences in the luminosity and in the contribution of different gas phases and galaxy components between the different modelling approaches. 
    [CII] is the least model-dependant gas tracer, while [CI](1-0) and CO(3-2) are very model-sensitive. In all models, we find a significant contribution to the emission of [CII] (up to $\sim$10\%) and [OIII] (up to $\sim$21\%) from the CGM. Our fiducial RT model produces a lower density, $T \sim 10^{4} $~K tail of [CII] emission that is not seen in the other more simplistic models and that resides entirely in the CGM, \rev{ionised by the FUV background and producing the extended halos observed in [CII] at high-z}. 
    Notably, [CII] and [OIII] trace different regions of the CGM: [CII] arises from an accreting filament and from the tidal tails connecting the main disc and its merging satellites, while [OIII] traces a puffy halo surrounding the main disc, probably linked to supernova feedback.
    \rev{We discuss our results in the context of sub-millimeter observations. Using simulated spectra and mock maps, we show that, despite the rather compact angular extent of \Ponos’s CGM, deep ALMA observations would not detect this component, even in [CII] which is the brightest available tracer. Instead, a next generation single-dish observatory such as the Atacama Large Aperture Submillimeter Telescope (AtLAST) could detect \Ponos' CGM in [CII] at high S/N, and possibly even in [OIII].} 
   } 
   

\titlerunning{High-res modelling of FIR and sub-mm line emission from the ISM and CGM of the \Ponos~simulation at z=6.5}
\authorrunning{Schimek et al.}

\maketitle

\section{Introduction}\label{Intro}

\subsection{Studying the circumgalactic medium of galaxies}

We loosely define the circumgalactic medium (CGM) of a galaxy as the gaseous halo surrounding the disc and extending up to the virial radius (R$_{\rm vir}$). It is often assumed that the CGM consists of mostly hot and warm gas \citep{Tumlinson17}. However, the discoveries of massive (\rev{up to a few $\sim 10^{9}~{\rm M}_{\odot}$}) and fast ($v>1000$~km~s$^{-1}$) outflows of cold and dense molecular gas extending by kpc in local starbursts and active galaxies \citep{Feruglio10, Fischer10, Sturm11, Cicone14, Veilleux20, Herrera21}, and of halos of cold atomic and molecular gas extending from tens to hundreds of kpc at high redshift \citep{Cicone15, Emonts16, Ginolfi17, Fujimoto20,Li21,Cicone21,Meyer22, DeBreuck22, Emont23, Scholtz23, Jones23}, strongly suggest that there could be a significant cold (T $< 10^{4}$~K) gas component in the CGM. In the current model of galaxy formation and evolution, the CGM consists of (i) gas that has been expelled from galaxies by outflows, (ii) inflows from the intergalactic medium (IGM), and (iii) gas stripped from the interstellar medium (ISM) of satellites and merging companions \citep{Dekel09,Tumlinson17,Cicone19}. Feedback processes inside galaxies, their merger histories, accretion, inflow and outflow events all leave their imprint in the CGM, which makes it a key component to study the baryonic cycle of galaxies and so decipher galaxy evolution.
In a parallel struggle to observational efforts, cosmological simulations have not been able to reproduce cold ($T<10^4~K$) gas beyond the inner ISM region of galaxies, because of limited resolution, and so have had little to no predicting power on the presence of a cold and dense CGM or IGM components. However, the situation is improving, thanks to increased resolution and improved modelling techniques \citep{Scannapieco15, Schneider17,Mandelker18, Mccourt18, Hummels19, Suresh19, VanDeVoort19,Sparre19,Nelson21, Rey23}.

Because of the diffuse and low surface brightness nature of the ionised medium that - up to now - is believed to account for most of the CGM mass, the CGM has been studied preferentially by analysing absorption lines in the spectra of background sources (e.g. \citealt{Prochaska14, Werk14, Lehner15, Bowen16, Cooper19, Dutta20}). However, any denser neutral and ionised CGM components could be observed in emission through atomic fine structure lines and molecular rotational transitions at far-infrared (FIR) and sub-mm wavelengths. In particular, many FIR lines such as [CII]~$\lambda 158 \mu$m that are not observable from ground at $z\sim0$ due to the Earth's atmospheric absorption, get red-shifted into the sub-millimeter atmospheric windows at higher-z, hence becoming observable with ground-based telescopes such as the Atacama Large Millimeter/submillimeter Array (ALMA), the Atacama Pathfinder Experiment (APEX), the Northern Extended Millimeter Array (NOEMA) and, in the future, the Atacama Large Aperture Submillimetre Telescope (AtLAST)\footnote{https://www.atlast.uio.no} \citep{Carilli13, AtLAST2020,Cicone19}. 


\subsection{The AtLAST project and the CGM science case}

\rev{Our study is part of a larger effort dedicated to deriving theoretical predictions for future AtLAST observations.}
AtLAST is a concept for a 50-m-class, single dish, (sub)mm telescope to be built in the 2030s in the Atacama desert, powered by renewable energy \citep{AtLAST2020}. Interferometers such as ALMA are ideal to study at sub-arcsec resolution targets that have at most a projected size of a few arcsec on the sky, but the small field of view (FoV) and poor sensitivity to low surface brightness emission make even a powerful telescope such as ALMA inadequate to capture diffuse emission on scales of $\gtrsim10$~arcsec at sub-millimeter wavelengths or to perform a swift and extensive mapping \citep{Carniani20, Cicone19}\footnote{See also Table~7.1 of the \href{https://almascience.nrao.edu/documents-and-tools/cycle9/alma-technical-handbook}{ALMA Cycle 9 Technical Handbook}}. AtLAST will fill this technological gap and enable us to map  portions of the sky that are tens of degrees in size at an angular resolution better than $<5$~arcsec at sub-mm wavelengths, by capturing also the diffuse large-scale structures \citep{AtLAST2020, AtLAST22}. One of the scientific drivers of AtLAST is the study of the CGM of galaxies, \rev{which is the focus of this paper}. 

Because AtLAST will be able to \rev{directly image extended reservoirs - including the cold and dense gas - from ISM to IGM scales}, providing robust constraints on future AtLAST line observations requires an investigation that \rev{embraces {\it both} the} galaxies and their surrounding halos, as well as their cosmological context, \rev{with the highest possible resolution}. \rev{Indeed, for this particular science goal, we cannot give up neither the resolution, because we need a realistic treatment of cold gas, nor the cosmological context, since explaining the origin of cold gas in the CGM requires tracking mergers, outflows, accreting filaments, and the interaction of these processes across cosmic times.}


\subsection{\revv{Submm/FIR line modelling efforts}}

In this work we trace cold and warm gas components by modelling the following emission lines: [CII]~$\lambda 158 \mu$m, [CI](1-0)~$\lambda 609 \mu$m, [CI](2-1)~$\lambda370 \mu$m, CO(3-2)~$\lambda 867 \mu$m, and [OIII]~$\lambda 88 \mu$m. 

\rev{Modelling FIR/sub-mm lines presents many challenges, as they arise from a multi-phase medium (HII, HI, H$_2$), and depend on a wide range of physical processes acting from sub-pc to Mpc scales.}
\rev{In recent years there have been several efforts focusing on the exploration and analysis of cold gas tracers in galaxies, although they focused on reproducing the emission of the ISM, not the extended CGM.}
\rev{These studies were done by using semi-analytical models \citep[e.g.][]{Lagache18,Popping19, Pizzati20,Pizzati23}, applying complex sub-grid modelling on lower resolution (m$_{\rm gas}~\sim$ 10$^{5}$ - 10$^{9}$ M$_{\odot}$) simulations to model the dense gas and to reflect a Milky Way (MW)-like population of giant molecular clouds (GMC) \citep[e.g.][]{Olsen15, Olsen17, Vallini18, Leung20}, using radiative transfer post-processing \citep[e.g][]{Vallini15, Arata20}, applying state of the art on-the-fly radiative transfer on high resolution (spatial resolution $\sim$~10~pc) simulations  \citep[e.g.][]{Pallottini19, Katz19, Lupi2020, Pallottini22, Katz22}, and trying to refine the sub-grid cold gas cloud modelling \citep[e.g.][]{Vallini18,Pallottini19,Pallottini22}.}
\rev{Even for high resolution, state-of-the-art simulations, sub-grid modelling is necessary when trying to model the cold gas that resides in temperature and density regimes not probed by the simulation. Such sub-grid models can suffer from confirmation biases. It is worth stressing that the choice of sub-grid models can affect strongly the results, as shown by e.g. \cite{Popping19}, who compared the emission of giant molecular clouds (GMCs) by assuming different internal density profiles. }
\rev{In the current generation of high resolution simulations, it became evident that previous sub-grid recipes could not be applied in the same way and that new modelling approaches are needed (e.g more refined sub-grid recipes and on-the-fly radiative transfer), which come with their own complications and challenges \citep{Vallini17, Bisbas23}.}


\subsection{This paper}
We use the \Ponos~simulation \citep{Fiacconi17} at $z=6.5$, which represents the high-z progenitor of a local massive galaxy. \rev{\Ponos~has a resolution high enough ($<4$~pc) to begin to resolve GMCs, which is crucial for modelling cold gas, although the simulation is still far from reaching typical GMC densities, or from resolving the complex sub-parsec filaments within GMCs (e.g. \citealt{Arzoumanian+11}). }
\rev{In this work, we apply different models for unresolved GMC structures via sub-grid density profiles, and compare their results with more recent approaches of radiative transfer calculations, by studying the effects on the derived synthetic emission lines within the same simulation. The results are then used to produce theoretical predictions that can inform future sub-mm observations. In contrast to previous modelling efforts, we especially focus on the emission originating from the CGM component of the galaxy.} 

This paper is structured as follows. We introduce the \Ponos~simulation in Section~\ref{Sim}, and then detail the modelling approaches of the emission lines in Section~\ref{Model}. In Section~\ref{Results} we discuss the differences on the resulting emission entailed by different models, and set a range of emission line luminosities that can be expected from a similar source. In Section~\ref{sec:comp_obs} we compare our results with currently available observational and theoretical constraints from the literature. In Section~\ref{sec:comp_sim} we discuss our results in the framework of other simulation studies. Finally, in Section~\ref{Dis} we discuss future prospects in this field, specifically in the context of the AtLAST project. Our Conclusions are summarised in Section~\ref{sec:conclusions}.

\begin{figure}[tb]
\centering
    \includegraphics[clip=true,trim=0.25cm 1.4cm 0.5cm 1.4cm,scale=.314]{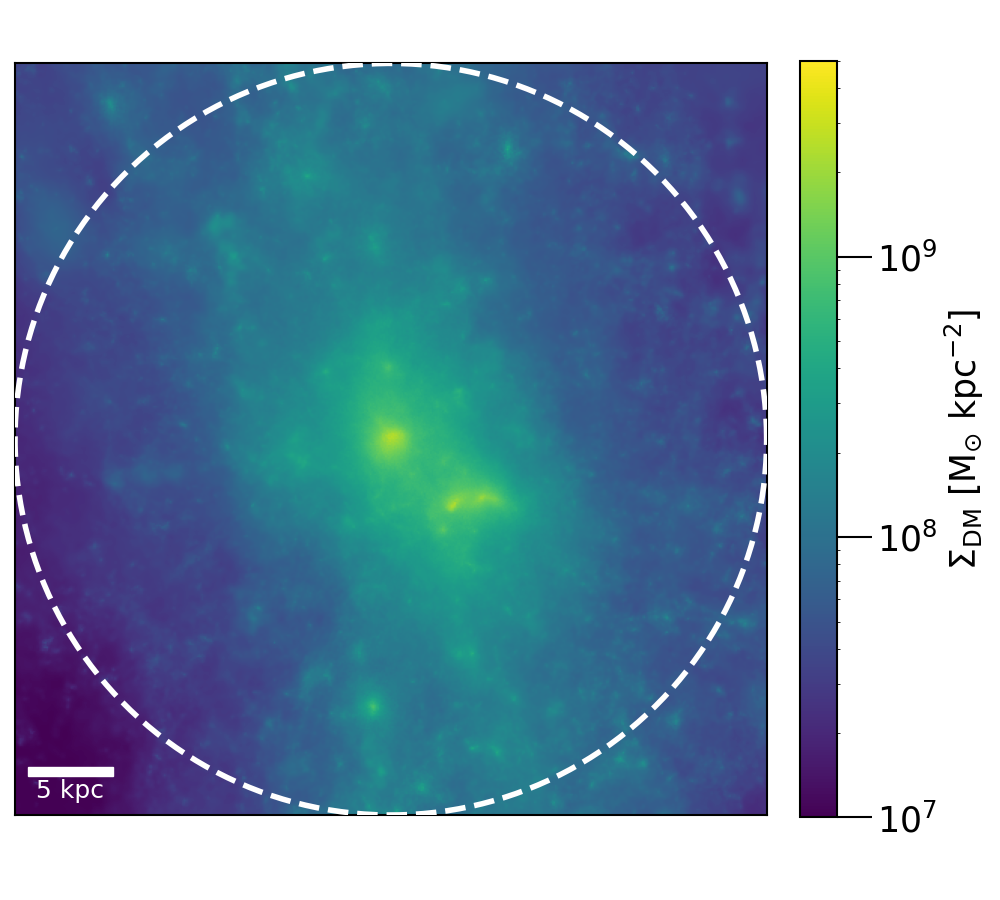}
    \quad
	\includegraphics[clip=true,trim=0.0cm 1.4cm 0.5cm 1.4cm,scale=.32]{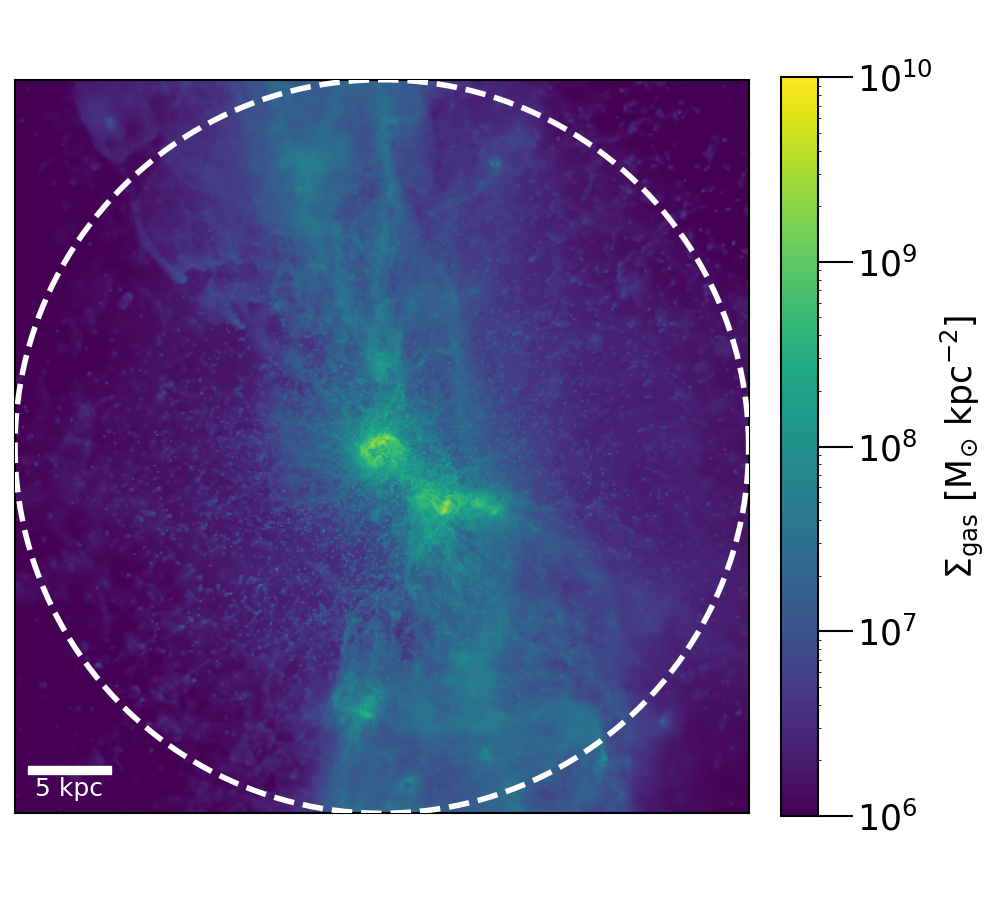}
	\quad
	\includegraphics[clip=true,trim=0.0cm 1.4cm 0.5cm 1.4cm,scale=.32]{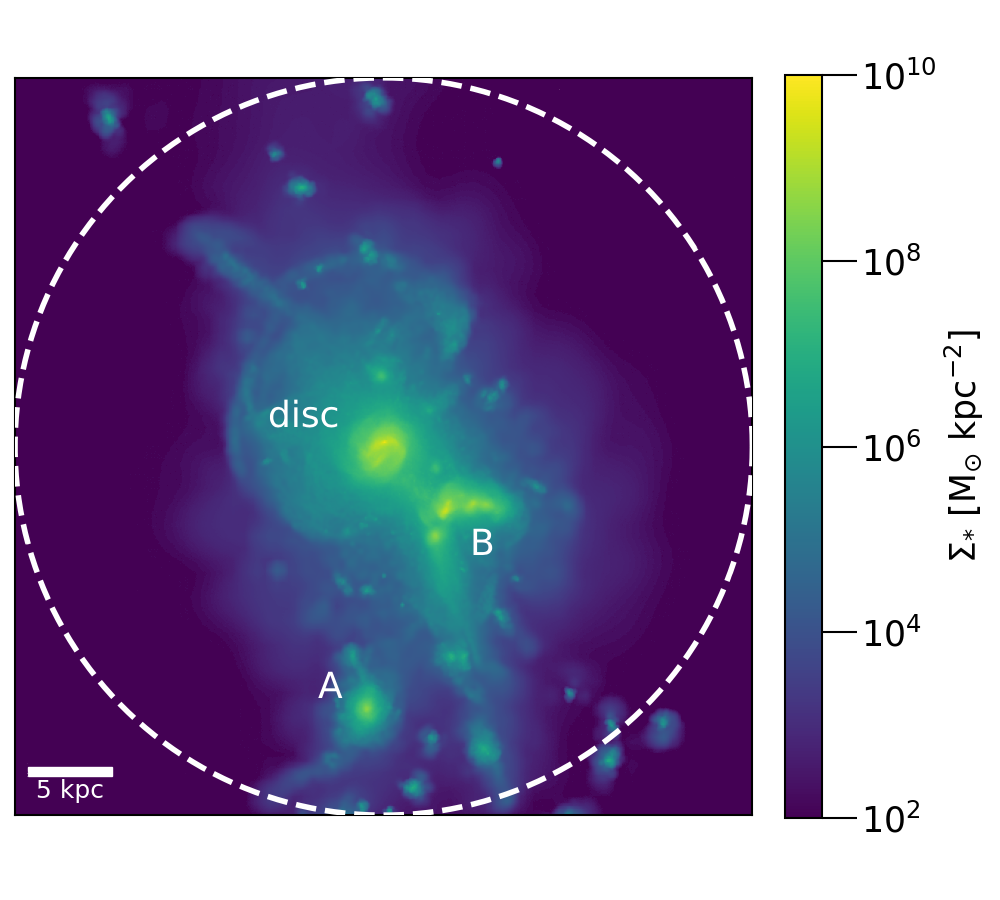}\\
      \caption{Maps of the \Ponos~simulation at redshift z = 6.5, showing the dark matter surface density ({\it top panel}), the gas surface density ({\it middle panel}), and the stellar surface density ({\it bottom panel}). The white circle marks R$_{\rm vir}$. The highest density structures trace the central disc and the three major satellites, two of which are close to the disc and merging with each other (merger B). Another merger can be seen in the south (A). Extended structures embedding both gas and stars lie between the main disc and the mergers.}
    \label{fig:ponos}
\end{figure}
\section{Simulation}
\label{Sim}

In this study we analyse the \Ponos~simulation by \cite{Fiacconi17}, which is a high-resolution cosmological zoom-in simulation run with the \Gasoline smoothed particle hydrodynamics (SPH) code \citep{Wadsley04} down to z = 6.5. The simulation has a gas mass resolution of m$_{\rm gas} = 883.4 $~M$_{\odot}$, a stellar particle mass of m$_{*} = 0.4 ~\cdot$~m$_{\rm gas} = 353.4 $~M$_{\odot}$, and uses WMAP~7/9 cosmology ($\Omega_{m,0} = 0.272$, $\Omega_{\Lambda,0} = 0.728$, $\Omega_{b,0} = 0.0455$, $\sigma_{8} = 0.807$, n$_{s} = 0.961$, and H$_{0} = 70.2$~km~s$^{-1}$~Mpc$^{-1}$) \citep{Komatsu11, Hinshaw13}. This leads to a minimum smoothing length of around 3.6~pc in the disc of the galaxy. The simulation includes a non-equilibrium chemical network for HI, HII, HeI, HeII and HeIII, and a uniform redshift-dependent UV radiation background due to stellar and quasar reionisation, according to \cite{HM12}. Cooling due to metals under the UVB is calculated and tabulated using the photoionisation code \Cloudy~\citep[last described in][]{Cloudy} assuming photoionisation equilibrium \citep{Shen10}. A turbulent diffusion model is included for thermal energy and metals \citep{Shen10}, which reduces artificial surface tension near strong density gradients and allows instabilities to develop \citep{Agertz07, Wadsley17}. Star formation occurs when the gas density exceeds a threshold $n_{\rm SF}=10~{\rm cm}^{-3}$ and the gas flow is converging and Jeans unstable. Star formation is modelled with a stochastic approach, where the SFR follows the local Schmidt-Kennicutt relation $\dot{\rho}_{\star} = \epsilon_{\rm SF} \rho_{\rm gas}/t_{\rm dyn} \propto \rho_{\rm gas}^{1.5} $, with a star formation efficiency parameter $\epsilon_{\rm SF} = 0.05$. Each stellar particle represents a stellar population with a \cite{Kroupa01} initial mass function. Feedback due to Type Ia and Type II Supernovae is modeled following \cite{Stinson06}. The simulation tracks metal production from stellar winds, SN Ia and SN II with yields detailed in \cite{RVN96}. We track the $\alpha$-elements and iron-peak elements separately, an approach that has been adopted with previous simulations (\citealt{Guedes11, Shen14, Kim14}, AGORA comparison project). However, abundances between different $\alpha$-elements are assumed to follow the solar abundance ratio \citep{Asplund09}. 

\Ponos~is the progenitor of a massive galaxy with a mass of M$_{\rm vir}(z=0)=1.2 \cdot 10^{13}$~M$_{\odot}$ at $z=0$ (M$_{\rm vir}(z=6.5)=1.22~\cdot~10^{11}$~M$_{\odot}$ at $z=6.5$). For the analysis presented in this paper, we chose the snapshot at $z = 6.5$, where the galaxy is undergoing a merger (stellar mass merger ratio 1~:~2.7). In this snapshot, the central simulated galaxy has a stellar mass of M$_{*}= 2 \cdot 10^{9} $~M$_{\odot}$, a virial radius of R$_{\rm vir} = 21.18$~kpc, and a star formation rate SFR $ = 20$~M$_{*}$~yr$^{-1}$. These properties make \Ponos~a typical star forming galaxy at $z=6.5$.

The simulation data were analysed using  \Pynbody \citep{Pynbody}. 
In Fig.~\ref{fig:ponos} we show maps of the dark matter surface density ($\Sigma_{\rm DM}$), gas surface density ($\Sigma_{\rm gas}$), and stellar surface density ($\Sigma_{\rm *}$) in the \Ponos~simulation snapshot studied in this work. The maps capture the chaotic state of the galaxy, due to the ongoing major merger. 

\subsection{Component separation}\label{compsep}

\begin{figure}
\centering
	\includegraphics[clip=true,scale=.3]{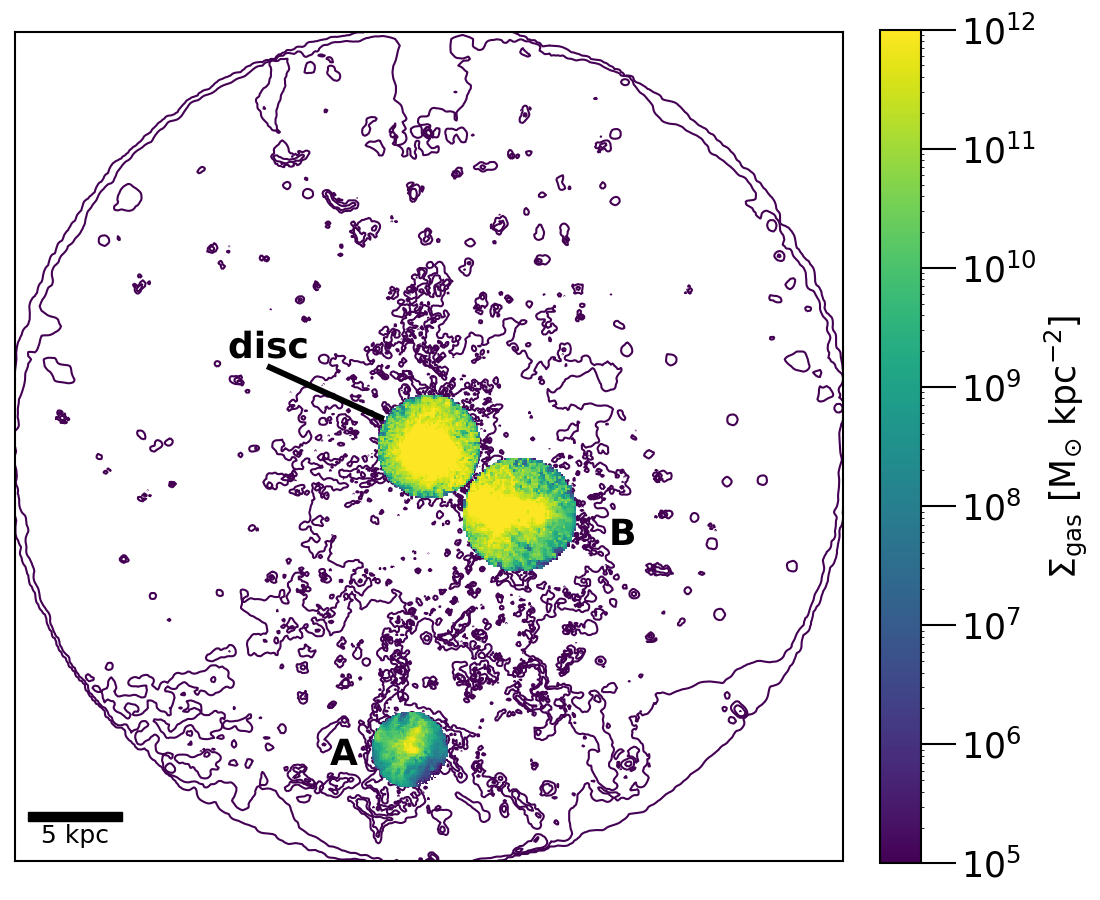}\quad
    \includegraphics[clip=true,scale=.3]{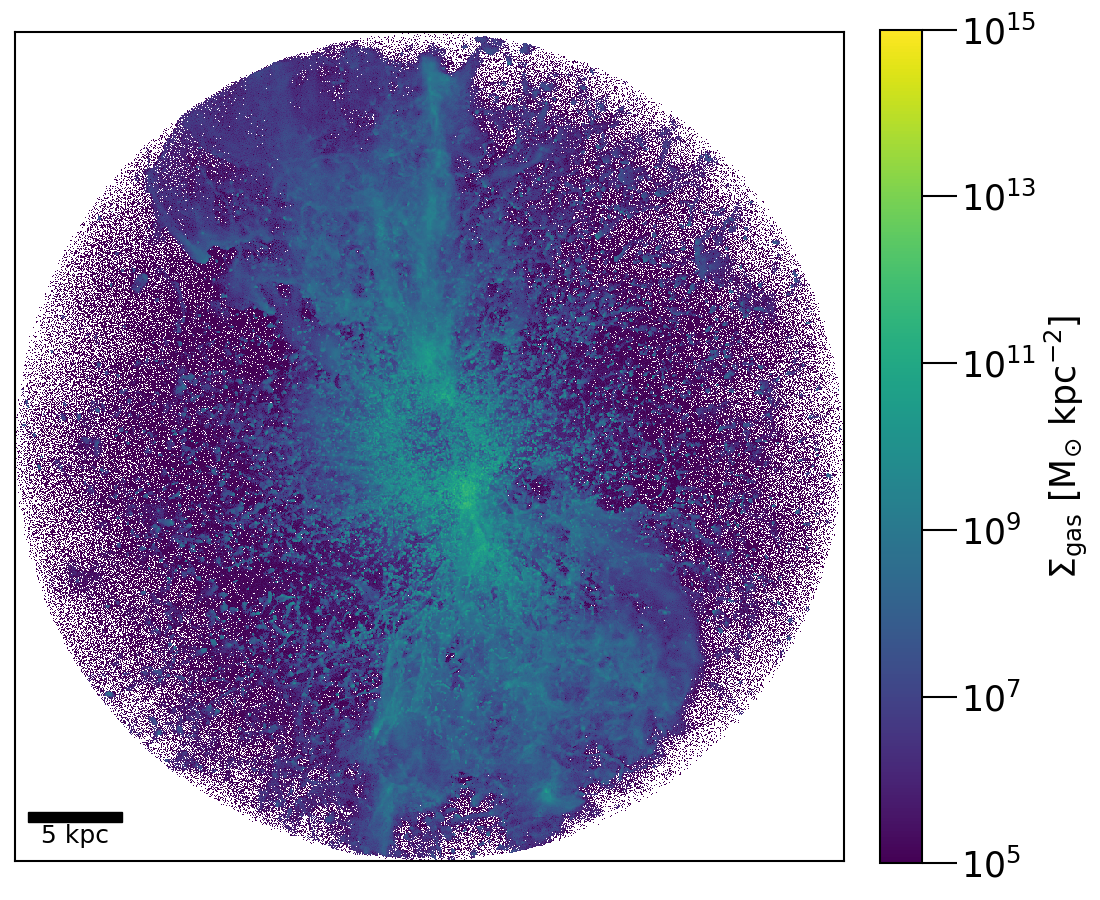}\\
      \caption{Gas density maps, showing the identification and de-blending of ISM and CGM gas components in the \Ponos~simulation snapshot. The top panel shows the ISM components, which include the main galaxy disc (including all gas particles within twice the half light radius) and the discs of the merging satellites. The gas density of the `non-ISM' particles is shown by the purple contours. The bottom panel shows the CGM component, made up by all gas within the R$_{\rm vir}$ after masking out the galaxy discs.}
         \label{fig:ponos_div}
\end{figure}

In order to analyse the ISM in the galaxy and its satellites separately from the diffuse CGM, we categorize the gas in the simulation into different components: main galaxy disc, mergers, CGM.
We ascribe all the particles contained within twice the half-light radius to the main galaxy disc. The mergers are masked out using a sphere with a 2 kpc radius for the southern merging companion (merger A), centred on the nucleus of the satellite, and a sphere with a 3 kpc radius for the two merging satellites closer to the disc (merger B), which are also in the process of merging with each other, and thus can not be deblended. 
\rev{These spherical regions are chosen to include 98\% of the stars of merger A, and 91\% of the stars of merger B.}
The diffuse CGM is defined as all the remaining gas particles contained within the R$_{\rm vir}$, excluding the already classified particles belonging to the main disc and to the ISM of the merging companions. Therefore, in our analysis, the definition of CGM does not include the discs of the main halo and its satellites, although observationally small merging companions would probably be unresolved and so ascribed to the CGM.
Fig.~\ref{fig:ponos_div} illustrates the results of this classification in the \Ponos~snapshot, where the top panel shows the gas density as a contour, overlaid with the masked areas defining the ISM components due to the main disc and to the discs of the merging satellites, while the bottom panel shows the gas density map of the CGM components.

Throughout this paper we divide gas particles into two phases, based on their temperature and density. Particles with temperatures T~$ \lesssim 10^{4} $~K and densities n$_{H} \gtrsim 10 $~cm$^{-3}$ are classified to be dense, and trace gas that would fall into the star formation criteria in the simulation. The diffuse phase traces gas particles that would not form stars. We will use the terms of dense gas phase and star forming gas interchangeably, and use the term diffuse gas for all gas that is not classified as dense and cold, even though it might be dense and hot. This definition is used for all models, although these two gas phases are only treated separately in the "multi-phase model" approach in Section \ref{sec:MPM}. Additionally, we define the diffuse gas to be ionised, when the gas particle has an electron fraction $\geq 0.5$. The dense, star forming gas makes up around 20\% of the total baryon mass of the simulation.

\section{modelling}
\label{Model}

\begin{figure}
\centering
   \includegraphics[width=\columnwidth]{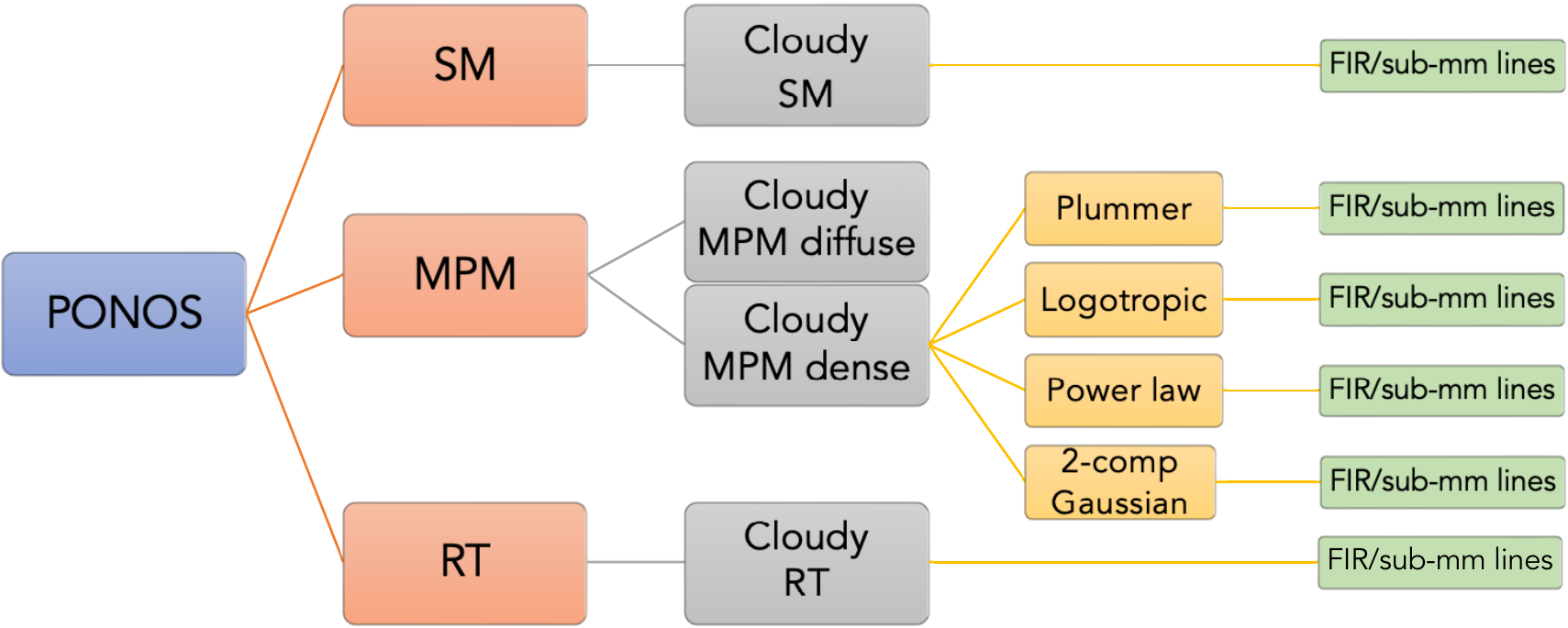}
      \caption{Flowchart illustrating our three different modelling approaches: simple model (SM), Multi-phase model (MPM), and Radiative transfer (RT) approach, which are described in detail in Sections~\ref{sec:SM}-\ref{sec:RT}. The RT is our fiducial model. The corresponding Cloudy parameters are reported in Table~\ref{table:1}.}
    \label{fig:flowchart}
\end{figure}

In this study, we are comparing three different approaches for modelling the unresolved gas phases in the \Ponos~simulation, with varying levels of complexity, which in total deliver six different predictions for the explored emission lines. For all models, the \Cloudy~photoionisation code  was used to derive synthetic line emission based on gas properties. The following sections will describe each model in detail. Fig.~\ref{fig:flowchart} illustrates the steps of the different approaches.

\begin{table}
\caption{\Cloudy~grid values for all different models. $T$ is the gas temperature, $n_{H}$ is the gas density, $Z$ is the gas metallicity, and $G_{0,bg}/G_{0,MW}$ is the background radiation field (see Eq.~\ref{eq:G0}. For the MPM dense gas $M$ is the mass of the cloud, and $P$ is the internal pressure of the cloud. $G_0$ in the RT model represents the UV radiation field the gas experiences (see Section~\ref{sec:RT})}             
\label{table:1}      
\centering                          
\begin{tabular}{l r r r}        
\hline\hline                 
SM \& MPM Diffuse  & min & max & steps \\    
 \hline
   log(T) [K] & 2.0 & 6.0 & 0.4 \\ 
   log(n$_{H}$) [cm$^{-3}$] & -3.5 & 4.0    & 0.2 \\ 
   log(Z) [Z$_{\odot}$] & -3.0 & 1.0     & 0.4 \\
   $G_{0,bg}/G_{0,MW}$ = 71.96 &  &    &  \\
\hline 
    MPM Dense  \\ \hline 
    log(T) [K] & 2.0 & 4.5 & 0.3 \\ 
   log(Z) [Z$_{\odot}$] & -3.0 & 1.0     & 0.3 \\ 
   log(M) [M$_{\odot}$] & 2.5 & 4.9    & 0.5 \\ 
   log(P) [K cm$^{-3}$] & 1.1 & 9.5    & 1.0 \\ 
   $G_{0,bg}/G_{0,MW}$= 71.96 &  &    &  \\ 
\hline  
RT \\ \hline 
   log(T) [K] & 2.0 & 6.0 & 0.4 \\ 
   log(n$_{H}$) [cm$^{-3}$] & -3.5 & 4.0    & 0.2 \\ 
   log(Z) [Z$_{\odot}$] & -3.0 & 1.0     & 0.6 \\ 
   log(G$_0$) [$1.6*10^{-3}$ erg cm$^{-2}$ s$^{-1}$] & -3.7 &  4.2  & 0.6 \\
   \hline 
\end{tabular}

\end{table}

\subsection{Simple Model}\label{sec:SM}

In the Simple Model (SM), we take the gas properties (density, temperature, metallicity) directly from the simulation, and use those as an input for \Cloudy. This will mainly serve as a simple test for the comparison with more advanced models. Instead of carrying out a \Cloudy~run for every single gas particle in the simulation, multi-dimensional grids were created, spanning over parameter ranges taken from the simulation (see Table~\ref{table:1}), and then interpolated over the values of SPH particles. \rev{Such grid approach is commonly used in studies to create synthetic emission of galaxies \citep{Olsen15,Olsen18,Pallottini19,Katz19,Lupi2020, Pallottini22,Katz22}}.

\rev{We use a slab geometry, so that each particle is considered as a gas slab using the given density, temperature and metallicity from the grid.}
\Cloudy~divides the computational domain (i.e. one cloud) into \rev{thin layers (concentric shells for spherical geometry, thin layers for slab geometry)}, called 'zones'. For the SM, we \rev{use the thickness of the cloud as a stop criterion, setting it equal to the maximal smoothing length, and we iterate until convergence}. The depletion of heavier elements onto dust is self-consistently included in the \Cloudy~model, and is set to scale linearly with the metallicity of the particles, \rev{which is used as an input parameter in our grid.} The size distribution and abundances of the dust are appropriate for a MW-like galaxy, and reproduce observed extinction properties \Cloudy, with the assumption of a MW-like gas to dust ratio and solar abundance ratios. At the redshift of the \Ponos~snapshot, the CMB temperature is around 20~K, which is quite high in relation to the temperature of the gas emitting the FIR and sub-mm lines targeted in this study. Therefore CMB radiation is included in the models, and assumed to be isotropic. 

The radiation field $G_{\rm 0,bg}$, \rev{which is defined to be the ionising FUV radiation field between 6 and 13.6 eV in units of Habing ($1.6 \cdot 10^{-3}$ erg cm$^{-2}$ s$^{-1}$) \citep{Habing68},} is assumed to be uniform and it is scaled to the average star formation rate density of the galaxy $\Sigma_{\rm SFR}$ following:
\begin{equation}\label{eq:G0}
    G_{\rm 0,bg} = G_{\rm 0,MW} \dfrac{\Sigma_{\rm SFR}}{\Sigma_{\rm SFR,MW}}, 
\end{equation} 
where $G_{\rm 0,MW} = 9.6\cdot10^{-4}$ erg cm$^{-2}$ s$^{-1}$ ( = 0.6 Habing) and $\Sigma_{\rm SFR,MW}= 0.0024$~M$_{*}$ yr$^{-1}$ kpc$^{-3}$ are the average FUV flux and the average SFR density in the MW \citep{Seon11}. \rev{Including FUV radiation in the modelling of emission lines originating from PDRs is necessary, as photons in this energy range are responsible for the photo-dissociation and ionisation of molecules and atoms. Additionally, FUV radiation is important for regulating the temperature in PDRs \citep{HT99,Wolfire22}. The FUV field, determined with the \Cloudy~{\it ISM} table, has the spectral shape of the solar neighbourhood, and is considered to be the the interstellar radiation field (ISRF) of the galaxy. }
\rev{We note that this is different from the \Sigame~approach by \cite{Olsen17}, where the FUV radiation field for the dense phase also includes fluxes from individual star particles within the smoothing length of each gas particle, without absorption. While this approach is likely accurate for lower resolution simultions where individual gas particles have a similar or larger mass than a GMC, it cannot be directly applied to \Ponos, in which each gas particle is a part of a larger GMC complex structure, and absorption along the line of sight would be important. Indeed, simply counting radiation from within the smoothing length results in a significant overestimation of the local radiation field. Wihin the framewok of post-processing, the most accurate way to compute the FUV field (both intensity and spectra shape) is the global radiative transfer modelling, and we explore this in our fiducial model (Section~\ref{sec:RT}). Thus, for simplicity, our simple and multi-phase model only includes the ISRF. We stress that these two models only serve as a first (and rather crude) exploration of the challenges of modelling FIR/Submm line emission without a global RT calculation for simulations that start to resolve the GMC structure. They are not intended to replicate \cite{Olsen17,Olsen18}.}

Cosmic rays are included, with a rate $\zeta_{\rm CR}$ scaled to $G_{\rm 0,bg}$: 
\begin{equation}\label{eq:CR}
    \zeta_{\rm CR} = \zeta_{\rm CR,MW} \dfrac{G_{\rm 0,bg}}{G_{\rm 0,MW}}, 
\end{equation}  
where $\zeta_{\rm CR,MW} = 3 \cdot 10^{-17}$~s$^{-1}$ is the average CR rate in the MW (\citealt{Webber98}).

The parameters used for the \Cloudy~runs for the SM are summarized in Table \ref{table:1}.
\subsection{Multi-Phase Model}\label{sec:MPM}

\begin{figure} 
   \includegraphics[width =\columnwidth]{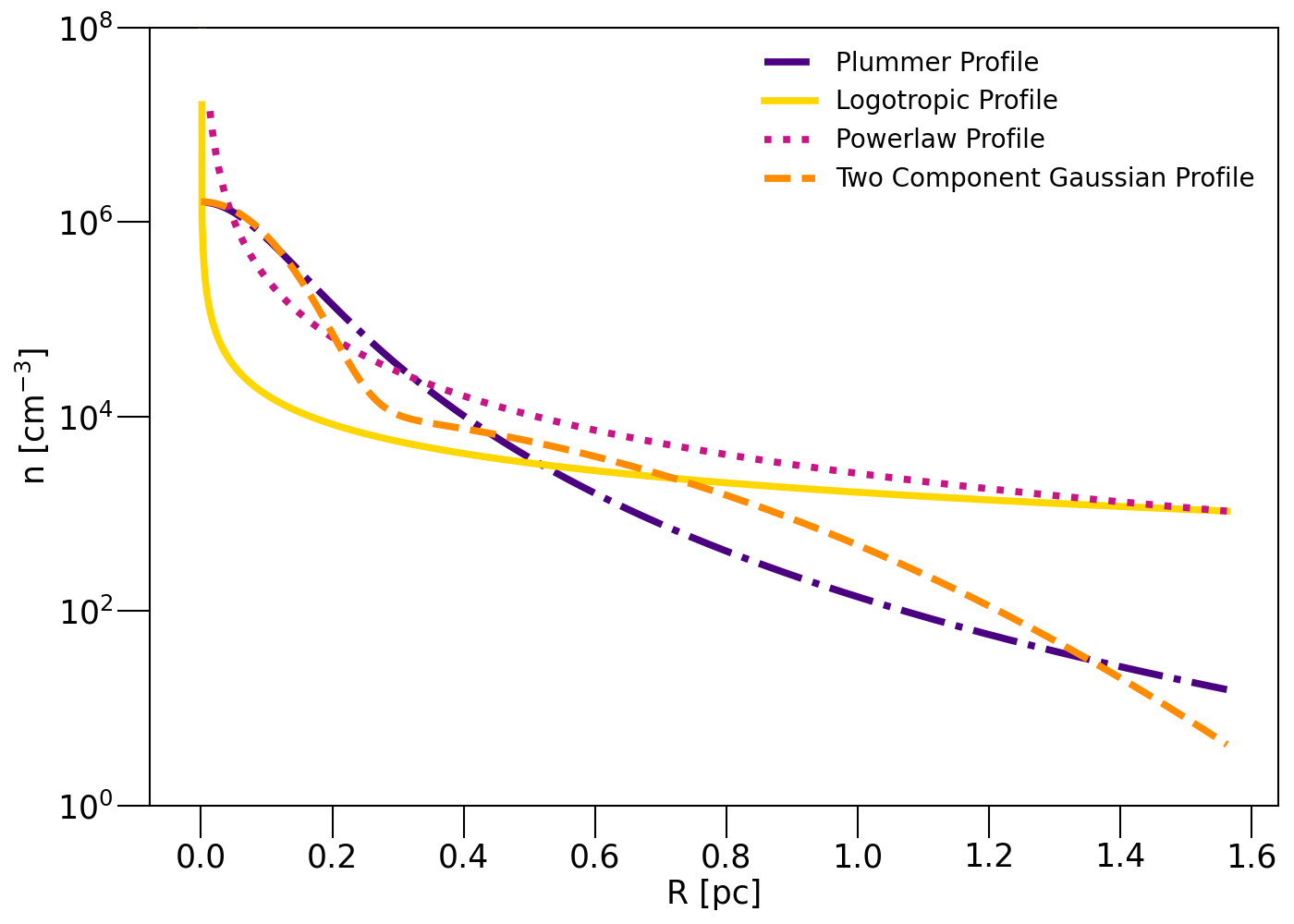} 
      \caption{A qualitative representation of the different sub-grid cloud density profiles applied in the MPM.}
         \label{fig:profiles}
\end{figure}

The Multi-Phase Model (MPM), which is loosely based on the approach by \cite{Olsen17}, differs from the SM in the sense that the particles are already divided into a dense and a diffuse phase before the \Cloudy~grid is applied, using the same dense gas criterion that was introduced in Section~\ref{compsep}. The two gas phases are treated differently, and separate \Cloudy~grids are applied. For the diffuse phase, we use the same approach as in the SM, interpolating over \Cloudy~runs across the same parameter grid created previously for the SM. 

For the dense phase, we adopt a different approach. Every gas particle that has been classified as being dense and cold is interpreted as an individual GMC.
This approach is commonly adopted in lower resolution simulations, as typical local GMC masses range from $10^{4} - 10^{6}$~M$_{\odot}$, \rev{whereas the mass resolution of \Ponos~} is m$_{gas} = 883.4$ M$_{\odot}$. \rev{As mentioned in the previous section, precisely due to this resolution challenge, there exist substantial differences in the calculation of FUV field between our MPM model and \Sigame. Nevertheless, we still} find it instructive to assess the performance of such MPM model applied to \Ponos, and compare the results obtained with different sub-grid GMC density models. 
We consider four possible sub-grid density profiles\rev{\footnote{We note that, in contrast to \cite{Pallottini19} and \cite{Vallini18} we do not include density distribution profiles in our analysis, as this is a different modelling approach from the sub-grid density profiles that we apply in this model. }} for the GMCs, hence each giving rise to a different MPM `sub-model': (1) a Plummer sphere profile, (2) a logotropic profile, (3) a power-law profile and (4) a two-component Gaussian. \rev{Profiles 1, 2 and 3 have already been tested by \cite{Popping19} in their semi-analytical models (SAMs). Since these authors found that even small changes in the sub-grid prescription can lead to significant differences in the FIR line emission, we want to test here how the use of different density profiles affect the emission in the context of a galaxy simulation. Each of the profiles is described in more detail below.} 
\rev{We also include turbulence within the \Cloudy~model, adopting the internal velocity dispersion $\sigma_v$ derived by \cite{Olsen15}: }

\begin{equation}\label{eq: velocity dispersion}
    \sigma_v = 1.2 \text{km s$^{-1}$} \left( \frac{P/k_B}{10^4 \text{cm$^{-3}$ K}} \right)^{1/4} \left( \frac{R_{\rm GMC}}{\text{pc}} \right)^{1/2}, 
\end{equation}
\rev{where P is the pressure of the SPH particle and R$_{\rm GMC}$ is the cloud radius, derived from the SPH particle mass and the pressure P, following the pressure-normalized scaling relation for virialised GMCs (e.g. \citealt{Field11}): }

\begin{equation}\label{eq: R_GMC}
    R_{\rm GMC} = \left( \frac{P/k_{\rm B}}{10^4\, \text{cm$^{-3}$ K}} \right)^{-1/4} \left( \frac{M_{\rm GMC}}{290  \ \text{M$_{\odot}$}} \right)^{1/2}.
\end{equation}

Fig.~\ref{fig:profiles} shows a comparison of the different radial density profiles. We implement these profiles in \Cloudy~using 30 radial bins, \rev{and using the "dlaw" \Cloudy~ parameter, which allows to input custom density profiles \citep[see \Cloudy~documentation]{Cloudy}. 
In contrast to the diffuse phase, for the GMCs we assumed a spherical geometry, and derived the emission in each shell as a function of the radius, which was summed up to give the total emission of the cloud.}

\subsubsection{Plummer profile}
The Plummer profile \citep{Plummer11}, first introduced to fit observations of globular clusters, was suggested by \cite{Whitworth01} to also fit GMCs \citep{Zucker21}, pre-stellar core and class 0 proto-star density profiles \citep{Popping19}. The density $n$ as a function of radius $R$ is given by:
\begin{equation}
    n(R) =  \dfrac{3 M_{\rm GMC}}{4 \pi R_{\rm p}^{2}} \left(1 + \dfrac{R_{\rm GMC}^{2}}{R_{\rm p}^{2}}\right)^{-5/2},
\end{equation}
where $R_{\rm p} = 0.1$ $R_{\rm GMC}$, M$_{\rm GMC}$ is the cloud mass (i.e. the SPH particle mass from the simulation).

\subsubsection{Logotropic profile}
A logotrope is a limiting form of a polytrope, where the polytropic index $\gamma = 0$ \citep{Logotrope}. A logotropic profile was used by \cite{Olsen15} and (\citeyear{Olsen17}) in their modelling, using the \Sigame approach. \rev{The larger particle mass allowed the authors to divide their gas particles/fluid elements into GMCs according to the GMC mass function. The higher resolution of  \Ponos~does not allow us to apply this method, and instead we ascribe one individual cloud to each gas particle}. The radial density profile is given by:
\begin{equation}
    n(R) =  n_{\rm ext} \dfrac{R_{\rm GMC}}{R},
\end{equation}
where: 
\begin{equation}
n_{\rm ext} = \frac{2}{3}~\dfrac{M_{\rm GMC}}{4/3~\pi~m_{\rm H}~R_{\rm GMC}^3}
\end{equation}
is the external density. We note that in this recipe, the density is infinite at the center of the cloud (see also Fig.~\ref{fig:profiles}). 

\subsubsection{Power law profile}
The third density profile that we explore is a power law profile:
\begin{equation}
    n(R) =  n_{0} \left(\dfrac{R}{R_{\rm GMC}}\right)^{-\alpha}.
\end{equation}

using an with exponent of $\alpha = 2$. 
According to the findings by \cite{Walker90}, this type of profile fits the density profiles of molecular cloud cores, which was also tested by \cite{Popping19}. Similar to the logotropic profile, the power law profile has an infinite central density, which needs to be handled properly when using these profiles in \Cloudy, by choosing some `cut off' density value. We found that the results are very sensitive to the choice of such parameter, and a slight difference can change the resulting integrated line luminosity even by three orders of magnitude. We calculated the value of the cut-off density by choosing a radius close to the centre of the cloud, where the density values are within realistic central densities of 10$^{5}$ - 10$^{6}$ cm$^{-3}$ (e.g. see findings by \cite{Bergin97} and recent observations by \cite{Zhang18}), and allowing densities up to 10$^{9}$~cm$^{-3}$. This was done by estimating if the most inner radial bin still had realistic density values. If the density was too high, a radius half way between the last and second to last radial bin value was chosen, to calculate the density, which was used as the cut-off density.

\subsubsection{Two-component Gaussian profile}
Finally, we explored also a two-component Gaussian (2CG) profile, motivated by \cite{Zucker21}, who found that this provides a good fit to local GMCs measured with the Gaia satellite \citep{Gaia, Leike20}. This profile is described by:
\begin{equation}
    n(R) =  a_{1} \exp{-\frac{R^{2}}{2\sigma_{1}^{2}}} + a_{2} \exp{-\frac{R^{2}}{2\sigma_{2}^{2}}}.
\end{equation} 

The four free parameters (a$_1$, a$_2$, $\sigma_1, \sigma_2$) were made dependent on cloud properties. In particular, $\sigma_{1}$ and  $\sigma_{2}$ depend on the cloud radius ($\sigma_{1} = 0.05 R$ and  $\sigma_{2} = 0.2 R$), while $a_{1}$ and $a_{2}$ was set to depend on the central density of the Plummer profile ($a_{1}$ being equal to the central density in the Plummer profile, and $a_{2} = \sqrt{a_{1}}$). Both the Plummer profile and the 2CG profile have finite central densities and, based on \cite{Zucker21}, we set them to reach the same central densities, as shown in Fig.~\ref{fig:profiles}. 
Physically, the transition between the inner and outer Gaussians in the 2CG profile represent a shift in temperature, density, or chemical composition of the gas in the cloud \citep{Zucker21}. The shift in density could represent the transition between the molecular and atomic gas phases, and the one in temperature could be a transition between an unstable warm neutral medium and a cold neutral medium.

\subsection{\rev{Global} radiative transfer model}\label{sec:RT}

The \rev{global} radiative transfer (RT) approach provides our fiducial modelling of the FIR and sub-mm line emissions from \Ponos. \rev{Although the post-processing with \Cloudy~is also a RT technique, we now apply a global RT calculation over the whole simulation, taking into account the propagation of photons from all emitting sources, so that we can have a more accurate estimate of the UV flux in each point of the simulation box. The \Ponos~simulation was} post-processed with \kramsesrt~\citep{Pallottini19, Decataldo19, Decataldo20}, a customised version of \ramsesrt~\citep{Ramses, RamsesRT} where a non-equilibrium chemical network generated via the package \krome~\citep{KROME} has been implemented. 
In this way, the radiation field in each grid cell is computed with an accurate scheme of radiative transport accounting for gas self-shielding. This is different from using a uniform background radiation scaled with $\Sigma_{\rm SFR}$, which was done in the simpler SM and MPM approaches. The resulting radiation flux is then directly used as input in the \Cloudy~grid, after being sampled using ten flux bins.
In the RT approach, a sub-grid modelling of the dense gas phase (GMCs) is not necessary, since the RT naturally creates a multi-phase medium, and so the temperature, metallicity and density values needed to compute the resulting line emissions are taken directly from the post-processed simulation data. 
However, due to the finite resolution, our maximal reachable density is $n\sim10^3$~cm$^{-3}$, hence, despite having mitigated this issue compared to lower-resolution cosmological simulation, the resulting molecular gas fractions should be regarded as lower limits.

\subsubsection{RT post-processing} 

\begin{figure} 
   \includegraphics[width =\columnwidth]{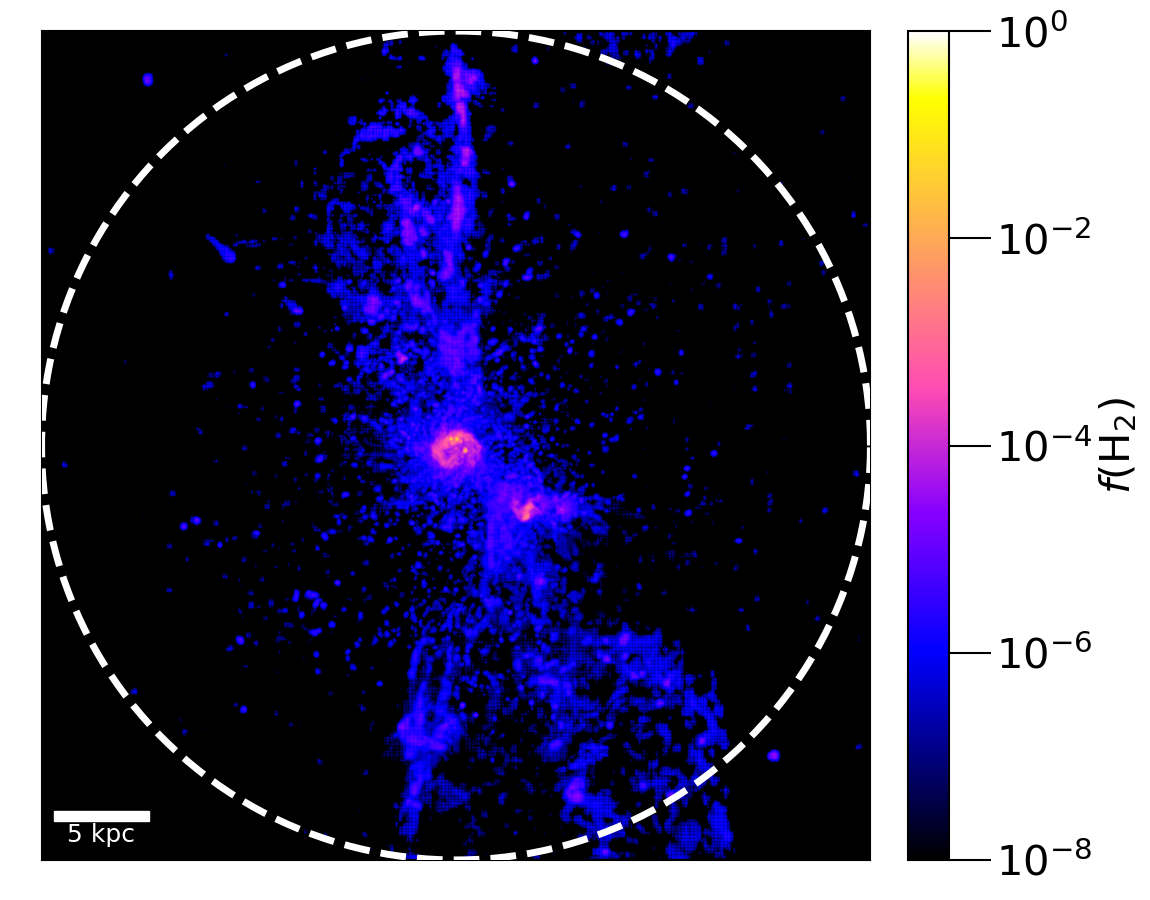}     
      \caption{H$_2$ gas fraction in \Ponos~obtained from the RT post-processing with \textsc{Kramses-rt}. Most of the H$_2$ resides in the main disc and in the merging component B, although clouds of molecular gas are also formed in the CGM, especially in the northern accreting filament.}
    \label{fig:H2}
\end{figure}

\begin{figure}
   \includegraphics[width =\columnwidth]{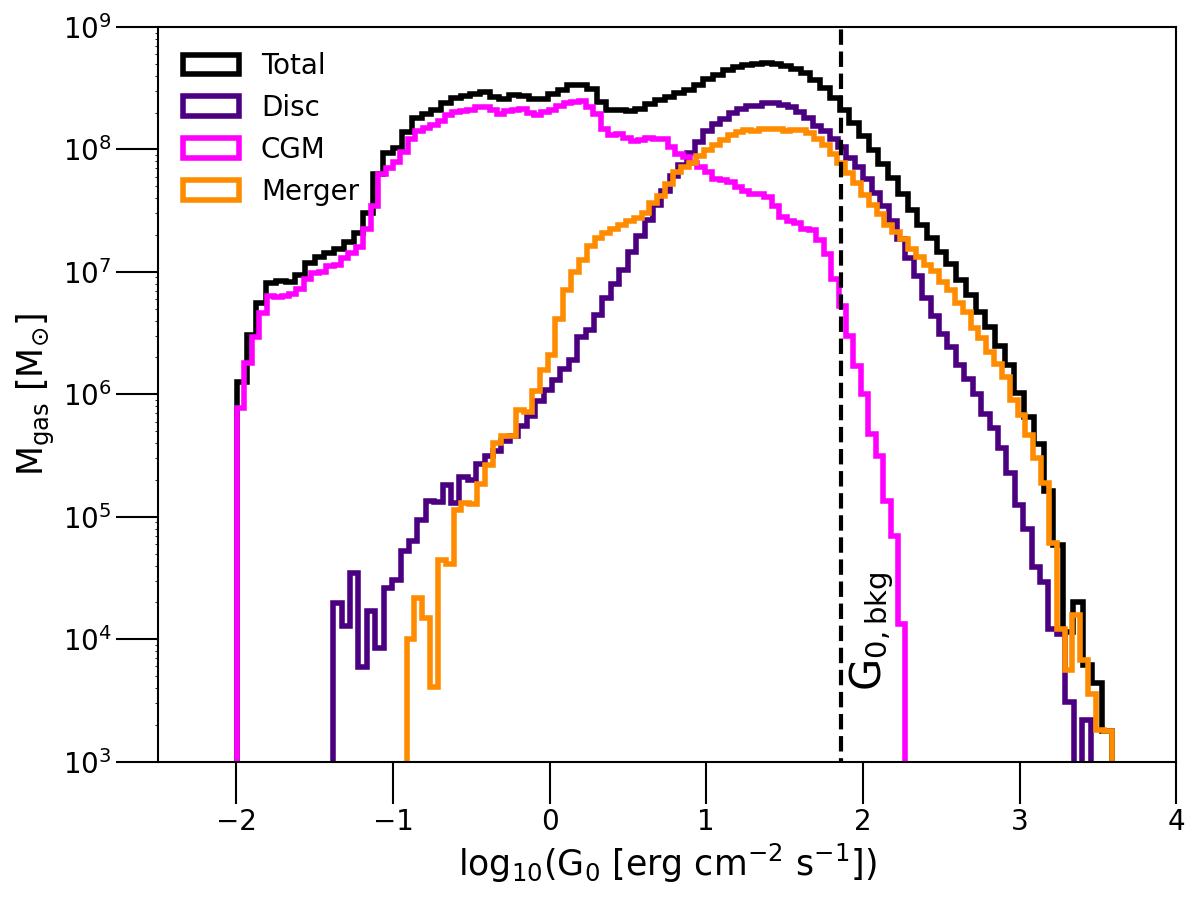}
      \caption{The distribution of the log(G$_0$) values of the \Ponos~simulation after RT post-processing. In black the distribution for the whole halo is shown, while in purple the disc, in orange the merger component, and in pink the CGM is represented. The high log(G$_0$) values are part of the disc and mergers, where the gas experiences UV radiation from young stars. \rev{The black dashed line represents the background radiation used in the SM and MPM.}}
         \label{fig:G0}
\end{figure}

To post-process the \Ponos~simulation with \kramsesrt, we first convert the SPH simulation into a grid. To do this, we select particles with coordinates within a box of size 50 kpc, which is slightly larger than twice the $R_{\rm vir}$ and thus encompasses the whole region defined as the CGM. The simulation box is initially tiled with $32^3$ cells (coarse resolution). We assign values of density, thermal pressure, metallicity and different elements abundances (HI, HII, He, HeI, HeII, electrons) from the SPH particles to the cells using a cloud-in-cell (CIC) scheme. The resolution is increased whenever the gas mass in a cell is larger than 32 times the gas mass resolution in the original SPH simulation, by using up to ten additional levels of refinements and repeating recursively the CIC interpolation to assign values to the refined cells. This is designed to always resolve the smoothing length of each particle in the SPH simulation. The final AMR grid has a resolution ranging from 195.3~pc in the outer low density regions, to a maximum of 1.52~pc in the dense disc gas. 

Since the original \Ponos~simulation does not include H$^-$, H$_2$ and H$_2^+$, which are instead accounted for in \krome, we assume initial zero abundance for those species, and then run the chemical network over the whole box for a time $t=20$~Myr in isothermal conditions and with a uniform UV background (UVB) as the only radiation source (from the \citealt{HM12} tables at $z = 6.49$) attenuated according to the cell's optical depth. The chemical evolution is accomplished in small time-steps of $dt=0.1$~Myr, and the flux in each cell is attenuated at every time-step according to the opacity of the cell. We verified in single-cell tests with \krome~that 20~Myr are enough to reach the equilibrium abundances for all chemical species, for a range of initial temperature and gas densities that represent the values found in \Ponos. 

After these preparation steps, we let the simulation evolve with radiative transfer and chemical evolution, but no dynamical evolution (RT post-process step). Stellar spectra are computed with \Starburst (SB99) \citep{SB99}, according to the age and the metallicity of the stellar particle, and then rescaled to its mass. To reduce the computational cost of loading photons for each stellar particle, particles located in the same cell are merged together, reducing the number of particles by 65\%, without changing the results as the total amount of photons propagated into the surrounding cells are unchanged. 
Stellar radiation propagates into the grid according to \ramsesrt's moment-based scheme, with full speed of light (we do not use a reduced speed of light).

Additionally to stellar radiation, a uniform UVB (\citealt{HM12} tables at $z = 6.49$) is included and attenuated in each cell according to the column density of each chemical species. Since the density of cells at the center of the box (i.e. in the galaxy disc) is high enough to completely absorb the UVB, this approach is meant to approximate a more realistic model where the UVB radiation is propagating inwards from the outside of the galaxy and absorbed on its way to the center. We notice that including an UVB is important especially for low density CGM gas, yielding lower H$_2$ abundances in the outskirts of the galaxy.
We allow the temperature to evolve together with the abundances during the post-process, but we do not allow gas hotter than $T=10^6$~K to cool down, since we assume that such gas was heated because of shocks due to stellar mechanical feedback (which was included in the original SPH simulation, while there is no dynamical evolution in the RT post-process) and we do not want to loose its imprint on gas temperature and chemical composition.
We evolve in this way until the variation of chemical abundances is less than 0.1~\% in two consecutive timesteps, which was around $t \sim 1$ Myr, more than five times the light-crossing time of the entire box.

The results of the post-processing are shown in Fig.~\ref{fig:H2}, reporting the surface density map of the H$_2$ gas fraction, and in Fig.~\ref{fig:G0}, showing the gas mass weighted distribution of G$_0$, divided into the different galaxy components. Most of the H$_2$ resides inside the main disc and in the close merging companion (component B in Fig.~\ref{fig:ponos}), with some filaments and clouds in the CGM also displaying a non-negligible H$_2$ fraction. Due to the radiation of young stars, the highest G$_0$ values are found in the disc and in the merger components, while the radiation field in the CGM is weaker.

\subsubsection{\Cloudy~post-processing}

It has to be noted that there are some inconsistencies when combining the post-processing with \kramsesrt~and \Cloudy. More specifically, \ramsesrt~in combination with \krome~solves non-equilibrium photo-chemistry, while \Cloudy~(which we use for further analysis) assumes a photoionisation equilibrium. 
The parameters of the \Cloudy~grid used in the RT model are reported in Table~\ref{table:1}. With respect to previous models, the FUV flux $G_0$ is not assumed proportional to the $\Sigma_{\rm SFR}$, but it can be inferred from the RT simulation. 
\rev{Since adding another parameter for each radiation band is computationally expensive, we compute the FUV flux into the range [6, 13.6] eV and add the corresponding $G_0$ as parameter to the \Cloudy~grid models. In this way, the total amount of \Cloudy~models is 42560. Nevertheless, we retain the information about the shape of the spectra, by computing the median spectra among all fluid elements residing within a $G_0$ bin and importing it in \Cloudy~via the "table SED" command (which allows to import a custom spectrum).}

\rev{To account for the AMR structure of the grid, we integrate the emission originating from \Cloudy~to a depth corresponding to the cell size, multiplying it by the surface area of the cell:}

\begin{equation}
    J = \int_{0}^{l_i} I(l)~dl \cdot A \,,
\end{equation} 
\rev{where $J$ is the total integrated emission, $I(l)$ is the emission of a thin slab at distance $l$, $l_i$ is the size of a cell at the refinement level $i$, and $A = l_{i}^{2}$ is the surface of a grid cell.} 
Temperatures, gas densities, and metallicities are taken from the \kramsesrt~post-processed simulation. Emission line luminosities are derived from the simulation by interpolating over the \Cloudy~grid, \rev{accounting for the cell size}. 

\section{Results}
\label{Results}

\subsection{Comparison of the six different model results} \label{sec:modelcomp}

\begin{figure} 
\centering
   \includegraphics[width =\columnwidth]{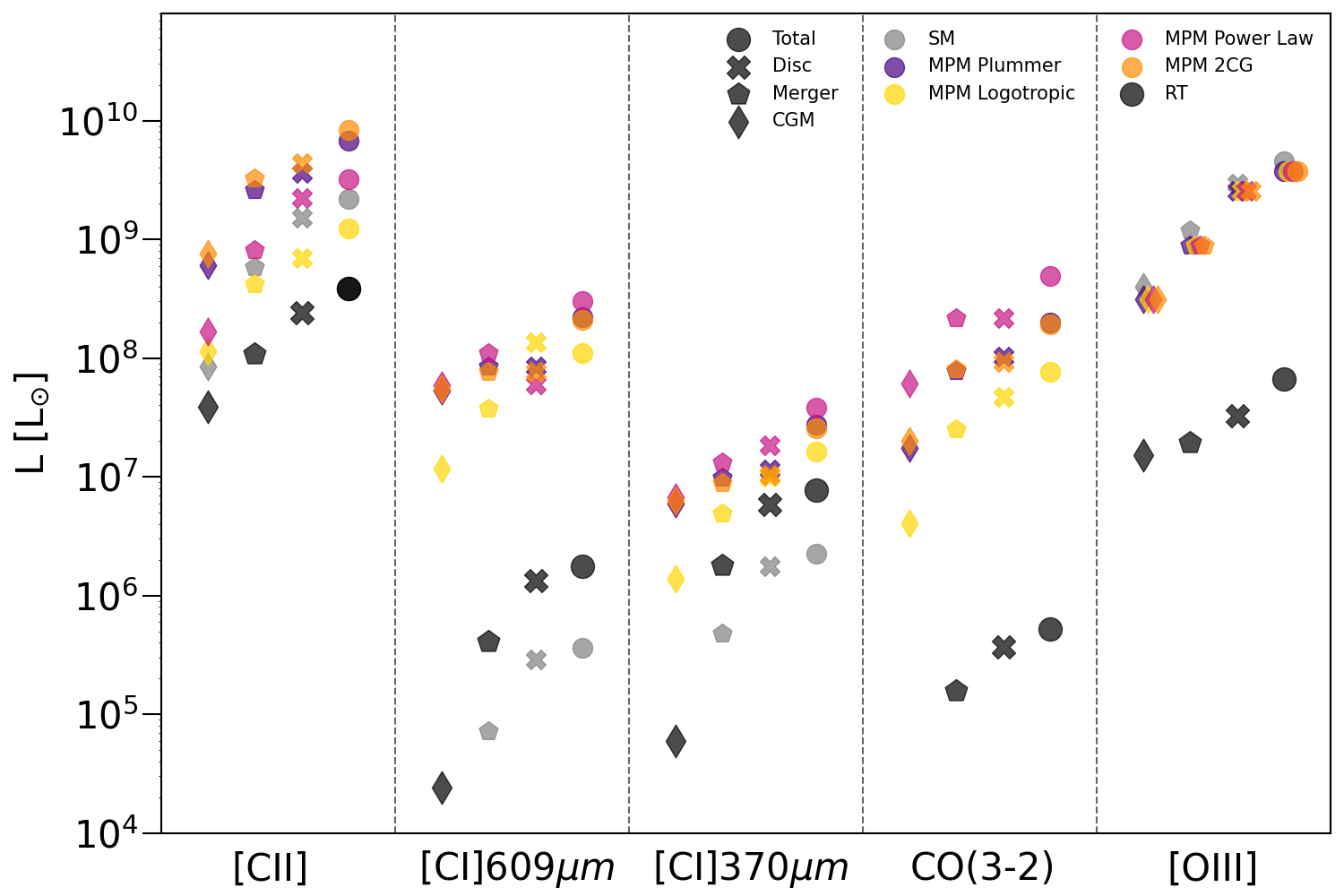}
      \caption{Comparison of line luminosities (in solar luminosities) obtained with our six modelling strategies (see summary in Fig.~\ref{fig:flowchart}), computed for different halo components. Each model is represented by a different colour: SM (grey), MPM+Plummer (purple), MPM+Logotropic (yellow), MPM+Power Law (magenta), MPM+2-component Gaussian (orange), and RT (black). Different halo components are indicated using symbols: circle markers for the whole \Ponos~halo, crosses for the main disc, pentagons for the mergers, and diamonds for the CGM (see Section~\ref{compsep} or the definition of the components). For the SM, dense gas tracers such as CO are unrealistically low and outside the ranges shown in this plot.}
         \label{fig:lums}
\end{figure}

\begin{figure*}
\centering
   \includegraphics[clip=true,trim=1cm 2cm 0.5cm 0.1cm,scale=.33]{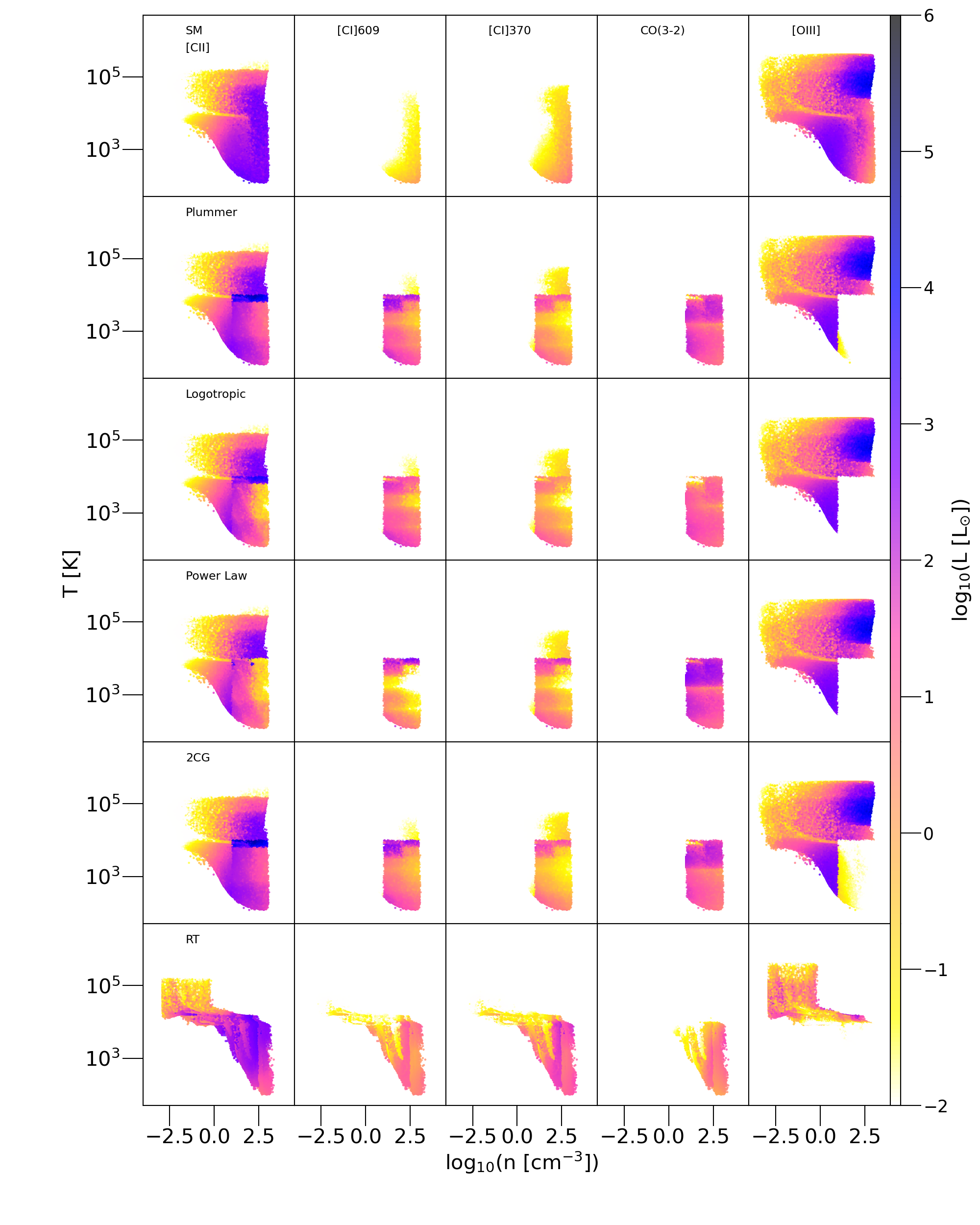}
      \caption{Density-Temperature ($n$-$T$) phase diagram of all six models used in this analysis. The rows show the different models, while the columns represent the different synthetic emission lines. The colours represent the emission in a particular line. \rev{For the MPM the densities do not directly correspond to those taken directly from the simulations, as they are given sub-grid density profiles, as described in Section~\ref{sec:MPM}.}The bottom row corresponds to the fiducial RT model, and it shows how the RT post-processing produces densities and temperatures that differ from the original simulation.}
    \label{fig:phase}
\end{figure*}

Fig.~\ref{fig:lums} shows the total luminosities in each line ([CII], [CI]609$\mu$m, [CI]370$\mu$m, CO(3-2) and [OIII]) obtained with the six different models. The different symbols represent the different halo components, with circles corresponding to the total line luminosity from the whole computational box. The corresponding luminosity values for all models are listed in Table~\ref{table:linelums} in Appendix~\ref{app:table}. 

\revv{The left-most panel of Fig.~\ref{fig:lums} shows the [CII] luminosity values obtained with the different models. The [CII] emission line is one of the brightest gas tracers in the FIR, accounting for 0.1 - 1\% of the total FIR emission of a star-forming (SF) galaxy \citep{Stacey91}. Single ionised carbon has a lower ionisation potential (11.6 eV) than hydrogen (13.6 eV), and is sensitive to the ultraviolet (UV) radiation emitted by young, massive stars. [CII] is one of the main coolants in the ISM \citep{Tielens85, Wolfire22}, the main cooling transition for the cold atomic medium \citep{Veilleux20}, and traces both the neutral and ionised gas. C$^{+}$ is collisionally excited by molecules, atoms or electrons, depending on its environment, as it can reside co-spatially with molecular, neutral atomic and ionised hydrogen, due to its low ionisation potential \citep{Olsen15,Olsen17, Leung20, Lupi2020}. In the figure, it can be seen that the [CII] luminosity values obtained with the different models are not too dissimilar from each other, differing at most by one and a half orders of magnitude. The RT model (plotted in black) produces the lowest [CII] emission, with over 60\% originating from the disc, and 56\% originating from the gas classified as dense. The second lowest [CII] emission results from the MPM that employs a} logotropic density profile for the dense gas in GMCs (plotted in yellow in Fig.~\ref{fig:lums}).
\rev{In this model,} most of [CII] originates from the disc, and over two thirds of the emission \rev{arises} from gas classified as dense. \rev{We thus obtain the opposite trend with respect to \cite{Popping19}, who found the logotropic density profile to produce the highest luminosities, followed by the power law profile and finally the Plummer profile. Instead, in our study, the 2CG and Plummer profile result in the highest [CII] luminosity values. We ascribe such effect to (i) the increased PDR covering fraction, which is seen in all MPM models, and is due to the assumption that every SPH particle represents a GMC; and (ii) to the shape of the density profile itself. Indeed, both the Plummer and the 2CG profile have large wings spanning densities of n$_H~\sim$ 10$^{0}$ - 10$^{4}$ cm$^{-3}$, which are typical densities for [CII] emission for the temperatures where the sub-grid density profiles are applied \citep{Goldsmith12}.}

\revv{The atomic carbon ([CI]) lines, shown in the second and third columns of Fig.~\ref{fig:lums}, display} large differences between the tested models, especially in the lowest frequency transition ([CI](1-0)~609$\mu$m). 
\revv{Atomic carbon is thought to reside mostly in molecular gas, as it needs some degree of shielding due to the low ionisation potential of C$^{+}$ of $h\nu = 11.6$~eV. [CI] lines are optically thin and usually do not trace a large reservoir of carbon, as most is either ionised to C$^{+}$, or bounded into a CO molecule \citep{Popping19,Valentino20,Dunne22}. \cite{Papadopoulos22} found that the [CI](2-1) and (1-0) lines are sub-thermally excited, and their ratio thus does not reflect the molecular gas temperatures of galaxies. Although classically believed to arise from a narrow layer in photon dominated regions (PDRs), it has been shown that [CI] emission is well mixed and co-spatial with CO, thus confirming that [CI] represents a valid alternative H$_2$ tracer \citep{Papadopoulos22, Thi09, Hartigan22,Montoya-Arroyave23}.} 
\revv{In our modelling, the} SM produces the lowest [CI] emission in all components. In all four MPM models, [CI]~609$\mu$m is brighter than [CI]~370~$\mu$m, which is an  opposite trend to the one produced by the SM and RT models. \rev{This is due to the fact that [CI](1-0) probes higher densities than [CI](2-1), which are reached in the MPM but not in the SM and RT.} The MPM models do not show significant differences from one another in [CI] line emission. \rev{[CI]~370~$\mu$m appears to be less model sensitive than [CI]~609$\mu$m.}

\revv{The fourth column of Fig.~\ref{fig:lums} shows the synthetic emission of CO(3-2), which also displays a strong model dependency. Carbon monoxide (CO) rotational lines are the best proxies for tracing the cold H$_2$ gas in external galaxies, as they are bright, and can be studied over a wide range of redshifts. 
When available in combination with [CII] data, CO(3-2) line observations can help us determine what fraction of the [CII] emission originates from molecular gas.
We choose the CO(3-2) line as it is the lowest-J CO transition that is observable with sub-mm telescopes at $z>6$.
CO(3-2) traces the surface of warmer and denser molecular gas, compared to lower J transitions \citep{Wilson08, Carilli13, Leroy22}. Despite its higher excitation requirements compared to lower-J CO lines, in starburst galaxies this line can trace the bulk of the H$_2$ gas reservoir \citep{Bothwell13, Montoya-Arroyave23}, and it has the advantage of a higher contrast against the Cosmic Microwave Background (CMB)  \citep{dacunha13,Zhang16}.}
 
The large variance of [CI] and CO values delivered by the different models is expected, since these low energy transitions are produced in the densest gas (concomitant with H$_2$), which is very sensitive to post-processing recipes in all cosmological simulations as they cannot resolve the scales relevant for H$_2$ formation ($<0.1$~pc). As a result, the MPM models, which include ad-hoc sub-resolution treatments of dense gas, estimate higher [CI] and CO values than the other models.  
\revv{The SM values of CO(3-2)} are so low that they are outside the lower range included in the plot ($L_{\rm CO(3-2)}=4.3 \cdot 10^{-4}$ L$_{\odot}$ for the SM). The MPM approach produces reasonable values of CO luminosity, as in the MPM models every dense cloud (particle) is assigned a sub-grid density profile that reaches densities high enough for sizable molecular gas emission. Instead, the RT model relies on the post-processed density values of shielded clouds, which do not reach the typical densities expected in GMCs (n~$\sim~10^{4}~-~10^{5}$~cm$^{-3}$) because of the finite resolution limit of \Ponos, \rev{which results in low CO emission. Thus, the resulting CO values in the fiducial model should be interpreted as lower limits.} 

\revv{The right-most panel of Fig.~\ref{fig:lums} shows the results of our modelling of the [OIII]88$\mu$m line, the only tracer of warm ISM included in our analysis (T$\sim10^{4}-5\cdot10^{5}$~K).
[OIII] is a common tracer of HII regions, as O$^{++}$ has a high ionisation potential of 35 eV and so it needs hot, massive stars to be ionised. Thus, this line 
can be used to map recent star formation, and it is an extinction-free probe for the conditions of the gas \citep{Ferkinhoff10}.}
\revv{[OIII]88$\mu$m is one of the brightest FIR fine structure lines and it can be
observed with ease from ground from $z\geq8$, hence it is, together with [CII], a major player in follow-ups of very high-z candidates \citep{DL14,Inoue16, Carniani18,Hashimoto18,Hashimoto19,Carniani20, Harikane20, Fujimoto22,Witstok22, Algera23, Popping22,Ren23}.} 
In contrast to the often used optical [OIII] emission lines (e.g. 5007$\AA{}$), the FIR [OIII] lines have a different dependence on the physical conditions of the gas, being less sensitive to the gas electron temperature and most sensitive to the electron density, as their critical density for collisional de-excitation to dominate is quite low \citep{Dinerstein85}. 
Since [OIII] probes warm gas in HII regions, this tracer is not affected by the implementation of sub-grid density profiles \revv{in our study}, and as a consequence all four MPM models deliver the same [OIII] luminosity values (as shown by the overlapping MPM symbols in Fig.~\ref{fig:lums}). 
The [OIII] values produced by the RT model are lower than in the SM and MPM models because [OIII] emission is weakened by \rev{the heating and cooling} processes in place during in the post-processing of the RT run, which is investigated further below.


Fig.~\ref{fig:phase} shows the density-temperature ($n$-$T$) phase diagrams resulting from the different models, colour-coded according by line luminosities. Each column represents a different emission line, and the rows show the different models. The phase diagram allows us to to understand which portion of the gas (in terms of densities and temperatures) is responsible for the emission of each line. For the MPM models, it is important to note that the densities shown in Fig.~\ref{fig:phase} are those taken directly from the simulations, so they do not correspond to the values passed to \Cloudy~in order to model unresolved GMCs, \rev{which have sub-grid density profiles}. 
Because of this, the diagrams look quite similar for all models except for the RT one, due to the temperature evolution of the gas during the post-processing where heating and cooling processes are at work. In particular, the hot and dense gas phase (top-right corner of each sub-plot) \rev{has disappeared after the RT post-processing due to cooling, while some gas at lower densities is heated up by stellar radiation. The hot and dense phase is a result of the stellar feedback in the simulation, where SNe shocks heat up the gas in the ISM. Such gas has a very short cooling time, and thus would not be long lasting. In the \Ponos~simulation a blast wave SNe feedback is employed, where the cooling is delayed \citep{Stinson06}, leading to this gas phase being present in the snapshot. During the RT post-processing this hot dense gas is allowed to cool. Additionally, the cold and diffuse gas phase is heated up by stellar radiation.}

The first column of Fig.~\ref{fig:phase} confirms the multi-phase origin of the [CII] emission, which traces gas spanning at least three orders of magnitude in $T$ and $n$, with a major contribution from denser gas.
It can also be seen that while the emission varies between the models, the differences are, as was already seen in Fig.~\ref{fig:lums}, not as strong as for other lines, and [CII] retains its multi-phase nature in all modelling approaches. Interestingly, the RT run produces a lower density, $T\sim10^4~K$ tail of [CII] emission that is not seen in the other models, and which resides entirely in the CGM component. \rev{The latter corresponds to cold streams accreting onto the main disc, and to tidal tails connecting the main disc and the mergers (which can also be seen in the H$_2$ structure in Fig~\ref{fig:H2}). This gas is mostly ionised by the UVB, and not stellar radiation.}

The [CI] lines (second and third column in Fig.~\ref{fig:phase}) are weaker than [CII] and less-multi-phase, but still probe a wide range of densities and temperature especially in the RT model, where however most of the [CI] emission arises from high density gas. In the MPM models, [CI] comes only from the particles classified as dense ($n_H>10~\rm cm^{-3}$) which undergo our sub-grid density profile modelling, hence the sharp transition seen in the phase diagrams. 

A similar effect can be seen for the CO(3-2) emission, which is actually only sizeable in the MPM models, as already discussed in relation to Fig.~\ref{fig:lums}. Compared to [CI], CO emission is spread more uniformly across the particles classified as dense ($\log_{10}(n_H)>1$).
In our fiducial RT model, CO emission is weak, and originates \rev{mostly} from the densest and coldest gas in the simulation, where we also find a low radiation field.

As already pointed out, [OIII] emission is almost model-independent in our study, as most of it originates from gas phases that are not affected by the additional sub-grid models.  
In the SM and MPM models, [OIII] is dominated by hot and dense gas, in addition to a more diffuse and colder gas phase, with a star forming gas contribution of less than 0.01\%, except for the SM where it is around 17\%. The RT model on the other hand differs from the previous models, as after the post-processing the hot and dense gas disappears, due to the cooling of gas previously heated up by supernova feedback. Such hot and dense phase erased by the RT model had an artificially delayed cooling according to the sub-grid blast wave feedback model of \Ponos, \rev{as described above}. At the same time, the RT run removes also the [OIII] contribution from colder gas at lower densities thanks to heating by stellar radiation. 
We consider the [OIII] values delivered by the RT model to be more realistic than the enhanced [OIII] values produced by the other models. O$^{++}$ needs a certain excitation energy to be ionised, and thus re-emitted as [OIII], hence a lower emission from low temperature gas is expected. Since gas temperature and density in the RT model are also linked to the G$_0$ value, higher radiation leads to higher temperatures, explaining the lower [OIII] emission rate generated by the RT approach. The FIR [OIII] emission line is sensitive to the critical density of the gas (i.e. the density where collisional de-excitation becomes the dominant de-excitation method) \citep{Dinerstein85}, so together with the needed high excitation energy for [OIII] we expect lower emission for gas at high densities, and at low temperatures. Most of the [OIII] emission originates from gas at $T\sim1.5 \cdot 10^{4}$~K and $n_{H}\sim 10$~cm$^{-3}$, which is in the range of typical values for HII regions. 

\subsection{Results from the fiducial RT model}
\label{sec:RTModel}

\begin{table}
\caption{Line luminosities (in units of L$_{\odot}$) obtained from the fiducial RT model, divided by halo component. The corresponding values obtained from all the other models are reported in Table~\ref{table:linelums} in Appendix~\ref{app:table}.}             
\label{table:2}      
\centering  
\small
\begin{tabular}{lcccc}       
\hline\hline                 
Line                & Total                 & Disc                  & Merger                & CGM \\    
\hline                        
$\rm [CII]$         & $3.91\cdot 10^8$      & $2.41\cdot10^{8}$     & $1.08\cdot10^{8}$     & $3.87\cdot10^{7}$ \\ 
$\rm [CI]609\mu$m   & $1.76\cdot10^{6}$     & $1.33 \cdot 10^{6}$   & $4.07 \cdot 10^{5}$   & $2.33 \cdot 10^{4}$ \\ 
$\rm [CI]370\mu$m   & $7.70 \cdot 10^{6}$   & $5.84 \cdot 10^{6}$   & $1.80\cdot10^{6}$     & $5.83 \cdot 10^{4}$ \\ 
$\rm CO(3-2)$       & $5.25 \cdot 10^{5}$   & $3.66 \cdot 10^{5}$   & $1.58 \cdot 10^{5}$   & $4.16 \cdot 10^{1}$  \\
$\rm [OIII]$        & $7.09 \cdot 10^{7}$   & $3.25 \cdot 10^{7}$   & $1.95 \cdot 10^{7}$   & $1.52 \cdot 10^{7}$  \\ 
\hline                                   
\end{tabular}
\end{table}
\begin{figure}
\centering
   \includegraphics[width =\columnwidth]{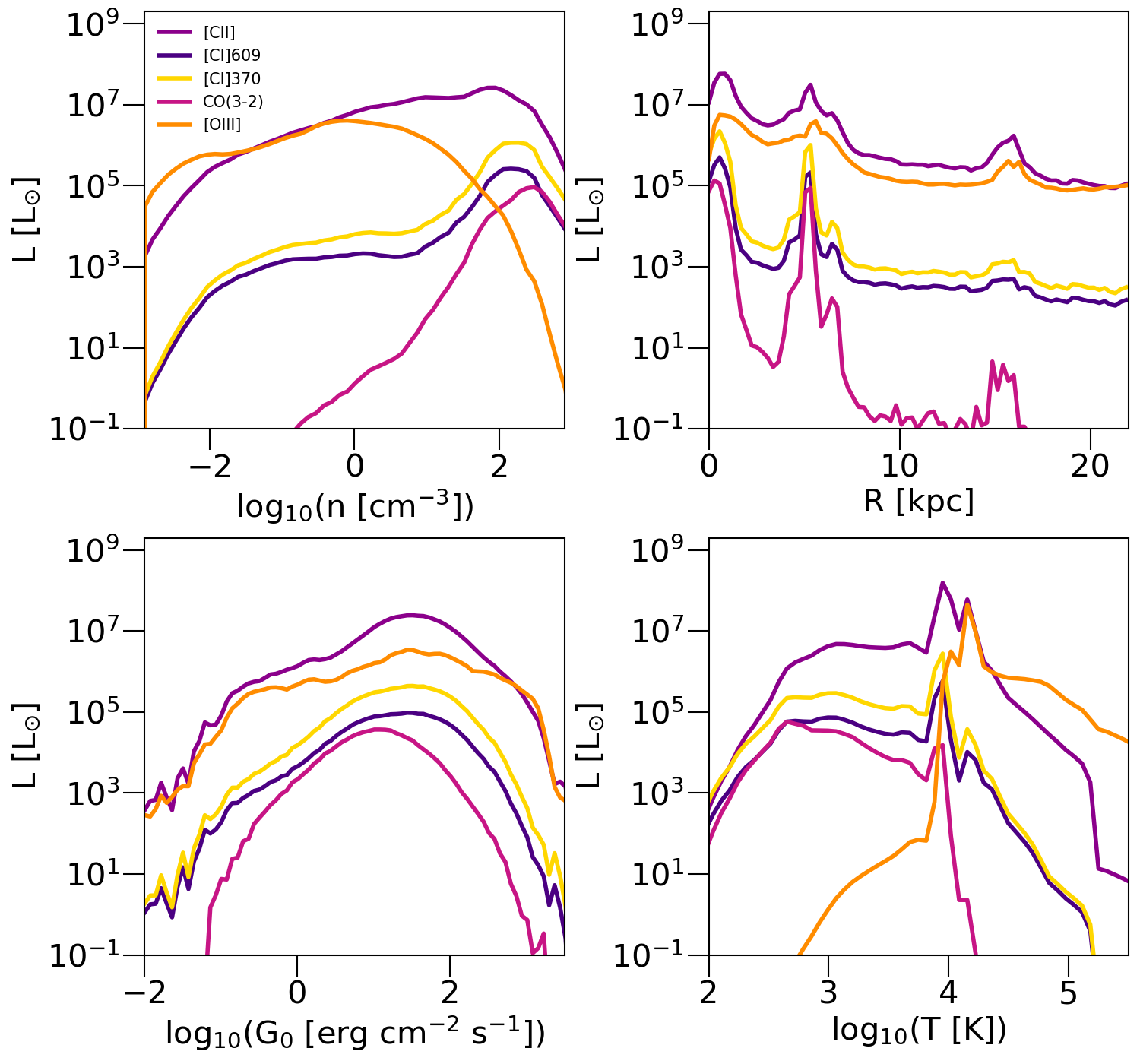}\\
      \caption{[CII], [CI], CO, and [OIII] line luminosities of the RT model as a function of gas number density ($n_{H}$, top-left panel), distance from the center of the simulation box (top-right panel), radiation field (bottom-left panel), and gas temperature (bottom-right panel).}
      \label{fig:Hist}
\end{figure}

In this section we analyse in more detail the results of our fiducial RT model. Fig.~\ref{fig:Hist} shows the different line luminosities plotted as a function of gas density, radial distance (\rev{where $R$ is computed} from the centre of the \Ponos~simulation), Habing radiation field, and gas temperature. The contributions from the main disc and from the minor mergers (B at $r\sim6$~kpc and A at $r\sim16$~kpc) are clearly seen as peaks in [CII], [OIII], \rev{[CI] and CO in the radial profiles (top-right panel of Fig.~\ref{fig:Hist}).} 

Focusing on [CII], we observe stronger emission from denser gas, though the dependence on density is not as steep as for other lines, which again hints at the multi-phase nature of [CII]. The [CII] emission peaks at log(G$_0$) = 1.5. \rev{Only over half of [CII] (around 56\%) originates from the dense, star-forming phase, while such dense and cold gas makes up 20\% of the total baryon mass. The [CII] emission peaks at a temperature of $T\sim10^{4}$~K.} 

[OIII] also seems insensitive to gas density \rev{at lower densities}, but drops noticeably towards high densities, where [CII], \rev{[CI] and CO} peak. \rev{Compared to the other lines}, [OIII] emits strongly from high temperature gas that experiences a high radiation field. \rev{99\%} of [OIII] emission originates from the diffuse gas phase ($n_H<10~$cm$^{-3}$). 

The two [CI] transitions behave similarly in the RT model, with their luminosities deviating in median by 2.5 and by a factor of 9 at most, especially from the high-$n$, lower-$T$ gas, where the [CI]~370~$\mu$m shows slightly stronger emission especially towards low temperatures, high radiation field strength and high densities. Over 90\% of the [CI] emission in both transitions originates from dense gas ($n_H>10~$cm$^{-3}$). \rev{Compared to [CII], the [CI] transitions clearly rise in emission towards higher densities.}

\rev{Even though the CO emission is weak, it can be clearly seen that it peaks towards high densities and low temperatures (over 99\% of the emission originates from gas classified as cold and dense). Compared to the other lines, there is little emission of CO in the CGM and between the galaxies (around 0.01\%), and the emission clearly peaks in the main disc and the mergers. The resolution within the CGM and the tidal tails is lower than in the main disc, suggesting that CO emission inside the CGM could be underestimated.}

\begin{figure*}
\centering
   \includegraphics[width = 0.75\textwidth]{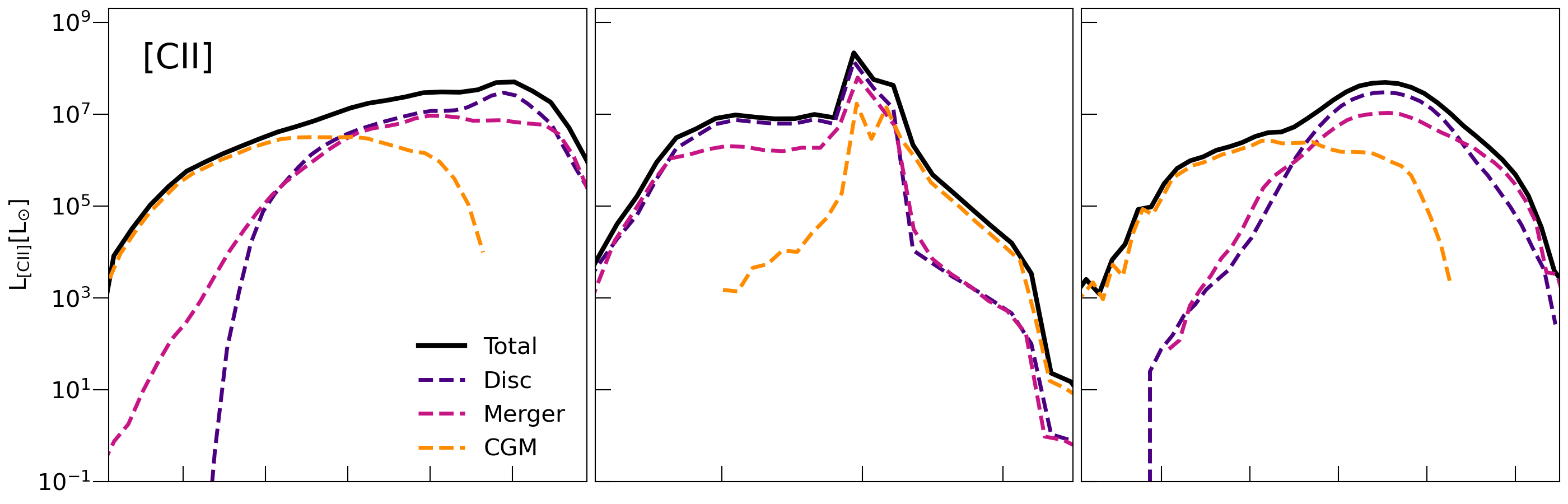}
   \includegraphics[width = 0.75\textwidth]{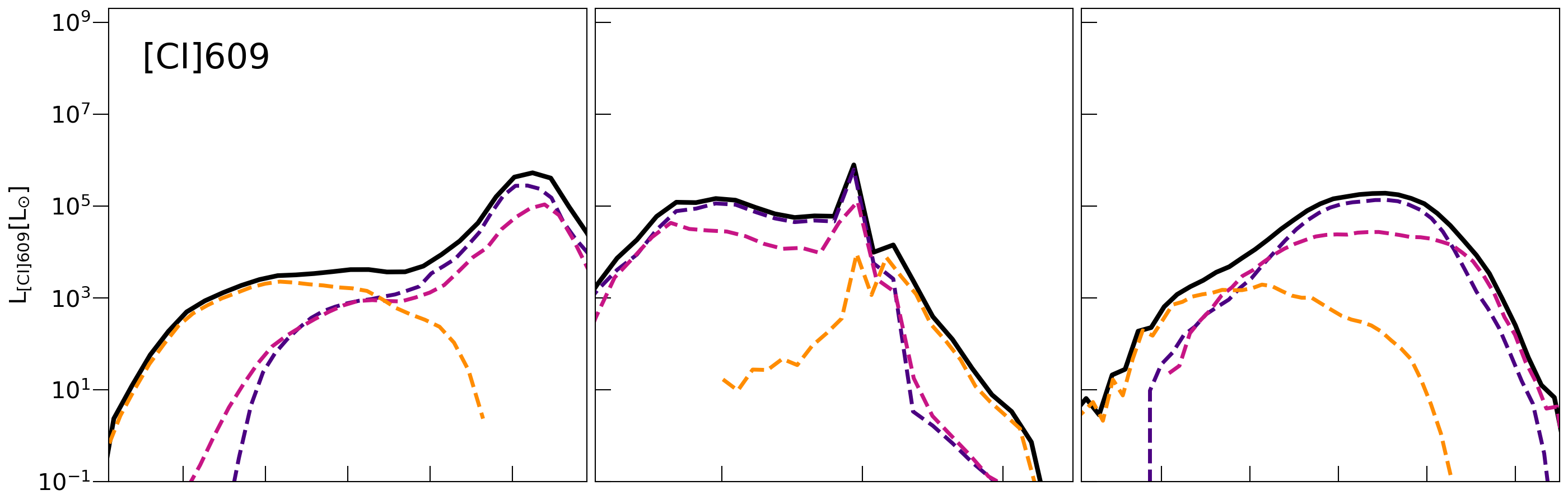} 
   \includegraphics[width = 0.75\textwidth]{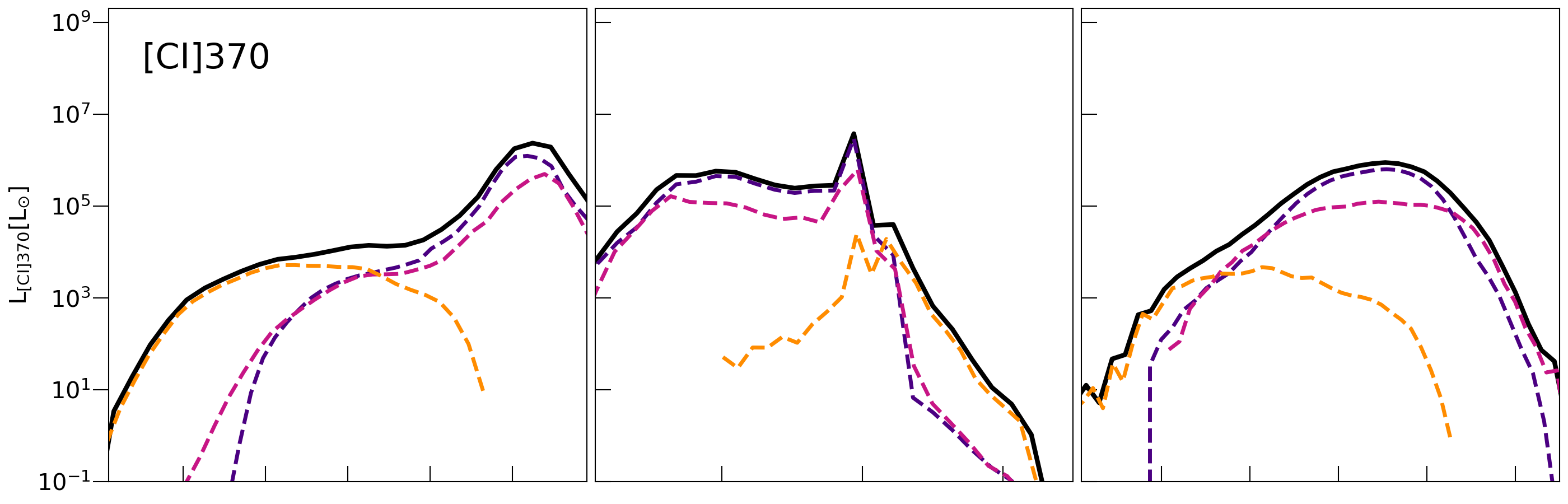}
   \includegraphics[width = 0.75\textwidth]{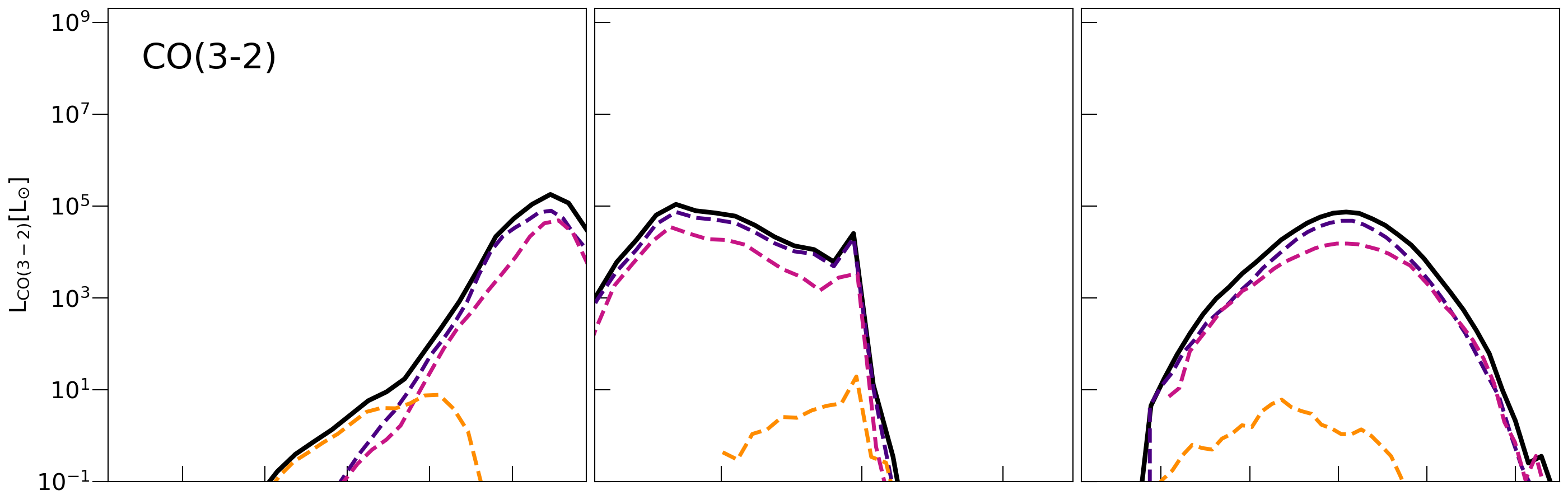}
   \includegraphics[width = 0.75\textwidth]{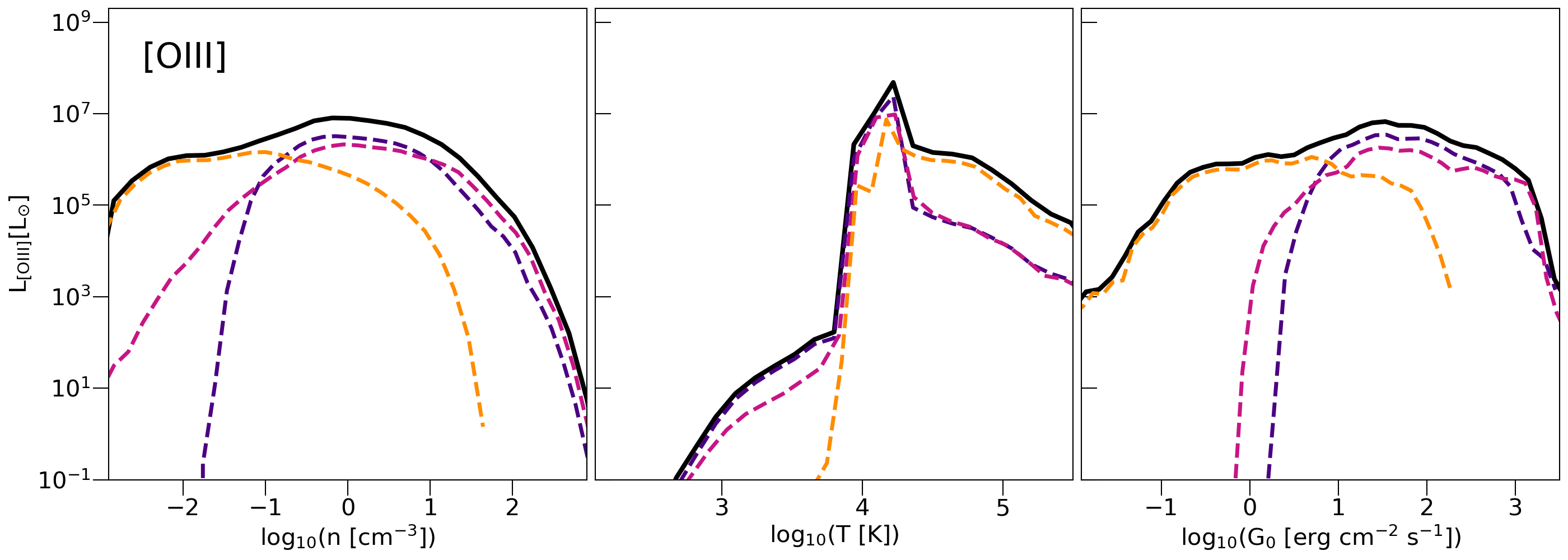}\\
      \caption{Line luminosities plotted as a function of gas number density (left column), gas temperature (middle column) and incident radiation field (right column). The lines are also divided into the contribution from different halo components, where the black solid line shows the total emission from the whole simulation box, the dashed purple line shows the main disc emission, the two mergers are plotted using a magenta dashed line, and the CGM is plotted using a dashed orange line.}
      \label{fig:Hist_mor}
\end{figure*}

Fig.~\ref{fig:Hist_mor} shows the different line luminosity profiles as a function of gas number density, temperature, and incident radiation field, plotted separately for each morphological component. The main disc, the mergers, and the CGM are plotted using dashed lines, while the total emission from the \Ponos~box is reported as a solid black line. In all these sub-mm/FIR tracers, the CGM component is by far the dominant emitter at low densities and low radiation fields. The density value below which the CGM becomes the dominant emitting component is \rev{equivalent} for CO and [CI] (n~$\sim 3$~cm$^{-3}$), higher for [CI] than for [CII] (n~$\sim 1$~cm$^{-3}$), and higher for [CII] than [OIII] (n~$\sim 0.2$~cm$^{-3}$). 
\rev{The CGM contribution to the total [CII] emission is around 10\%, and it dominates at low densities and low radiation fields.}
\rev{For the [CI] lines}, the contribution from lower density gas located in the CGM is \rev{$\sim1.3$\%} for the [CI] 609 $\mu$m transition and \rev{$\sim0.8$\%} for the [CI] 370 $\mu$m one. [CI] peaks at \rev{similar radiation field strengths as} [CII].  

As expected from the resolution limitations discussed earlier (\rev{which, unfortunately,} not even the RT approach can mitigate for the higher density molecular gas with $n>10^3~cm^{-3}$), the CO(3-2) emission is weak and peaks at the highest densities, lowest temperatures, and \rev{lower} radiation field values. As a consequence, CO(3-2) exhibits a very small contribution from the CGM, equal to \rev{0.01\%}. Most of the CO(3-2) emission originates from the main disc, making up 70\% of the total CO line luminosity.

[OIII] is less sensitive to the density, compared to other lines, but it shows a clear uptrend towards higher temperatures and higher radiation fields, which is due to the already discussed higher excitation energy needed for doubly ionised oxygen. The highest temperature, lowest density and lowest radiation field emission is dominated by the CGM, making up \rev{21\%} of the total [OIII] emission. 

\rev{For all analysed lines we find stronger emission for higher metallicity gas. At higher metallicities there are more metals in the gas that can emit in the analysed emission lines \citep{Vallini15}.}


\begin{figure*}
\centering
   \includegraphics[width = 0.85\columnwidth]{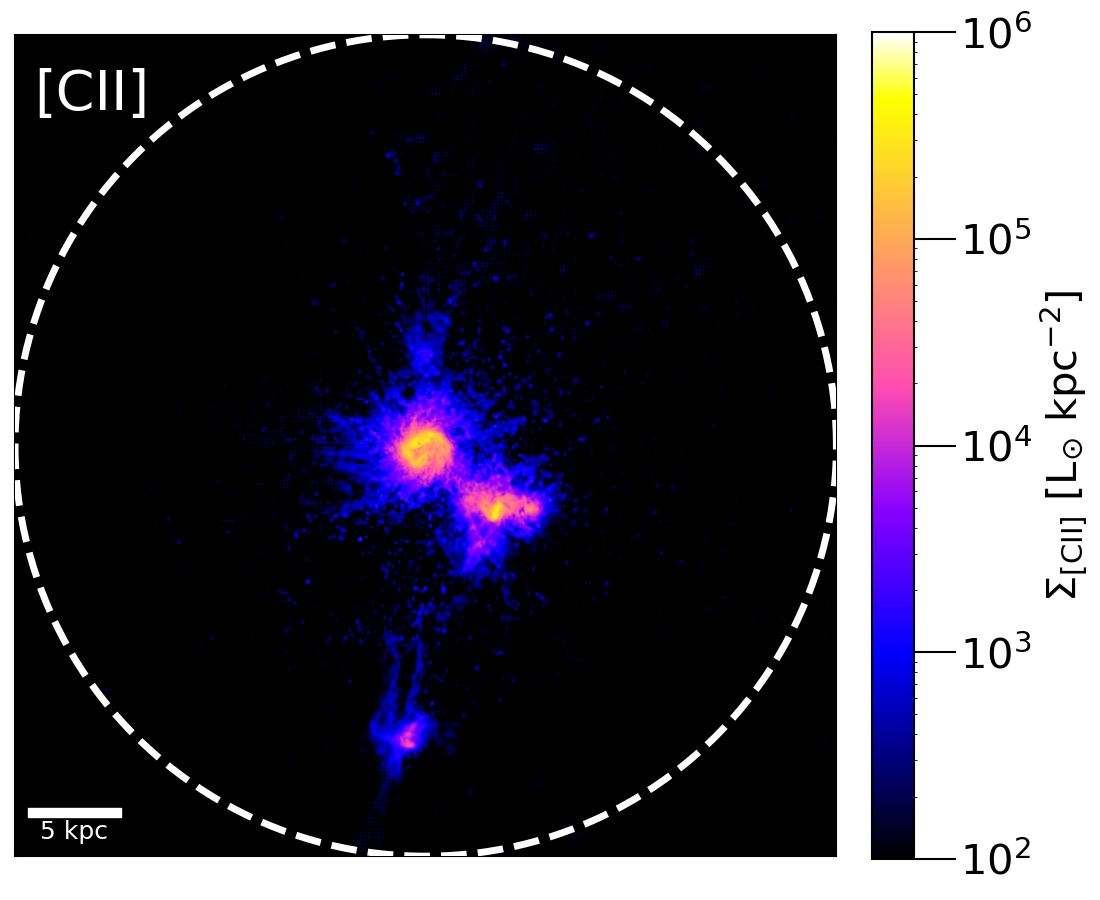}
   \includegraphics[width = 0.85\columnwidth]{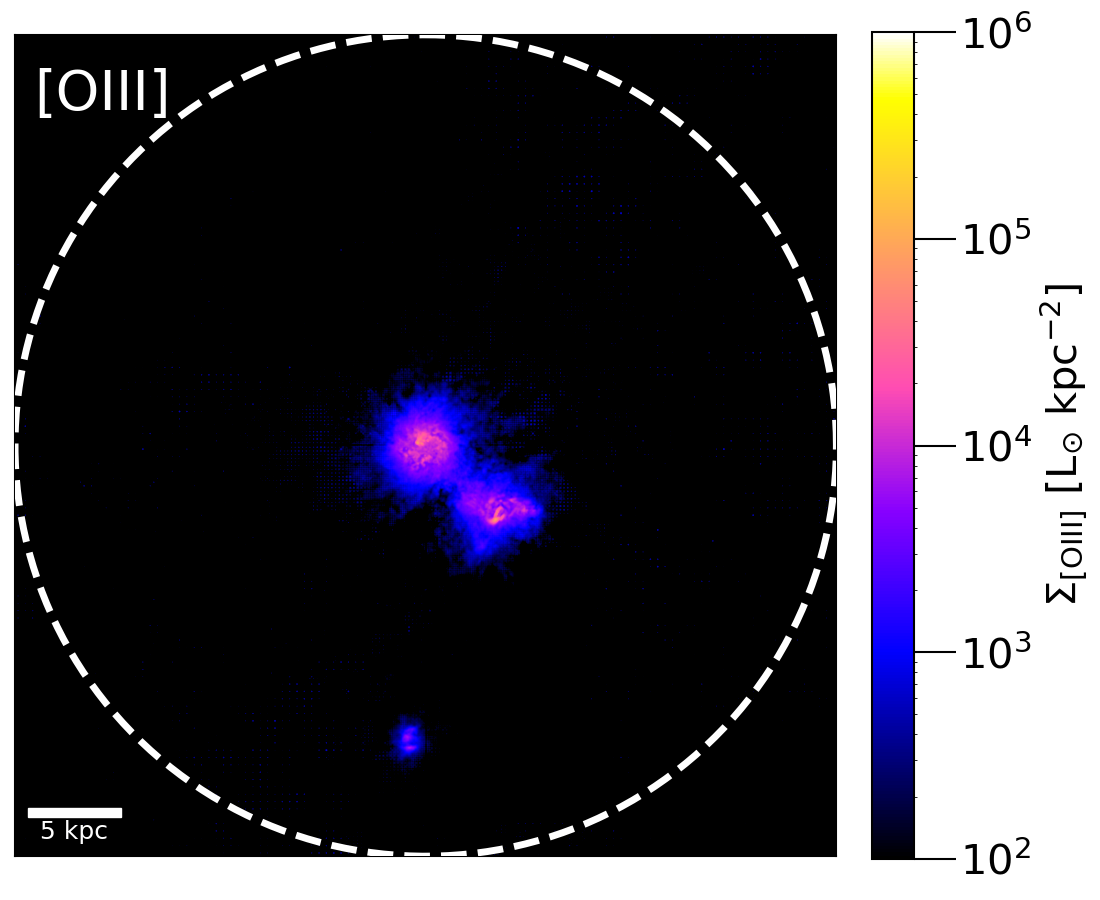}
   \includegraphics[width = 0.85\columnwidth]{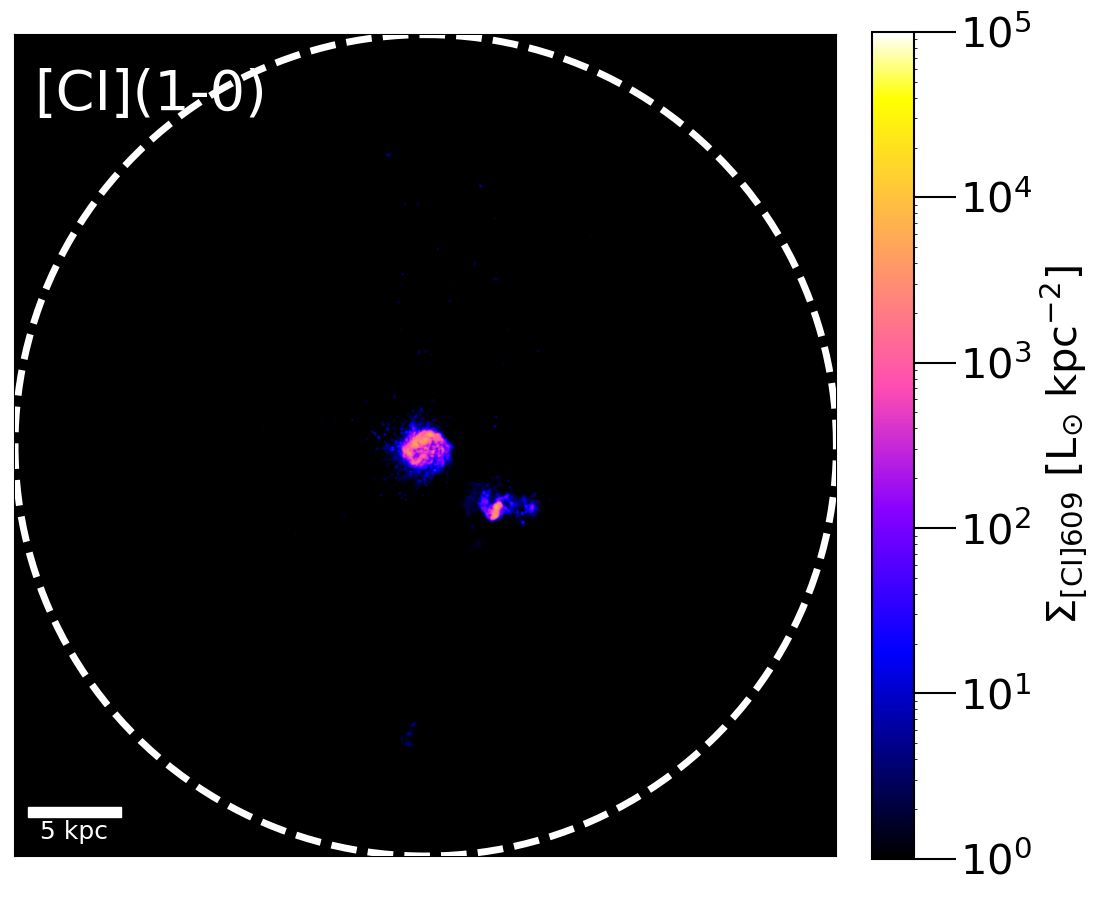}
   \includegraphics[width = 0.85\columnwidth]{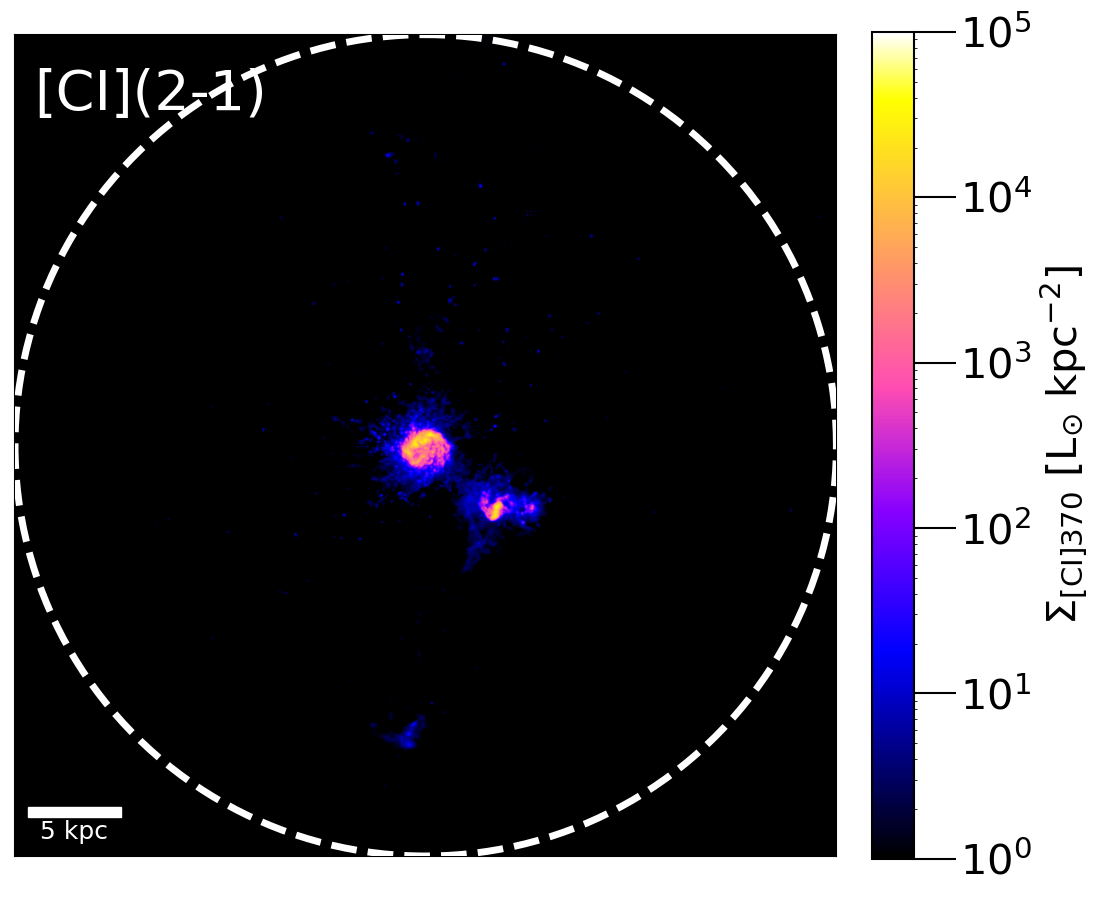}
   \includegraphics[width = 0.856\columnwidth]{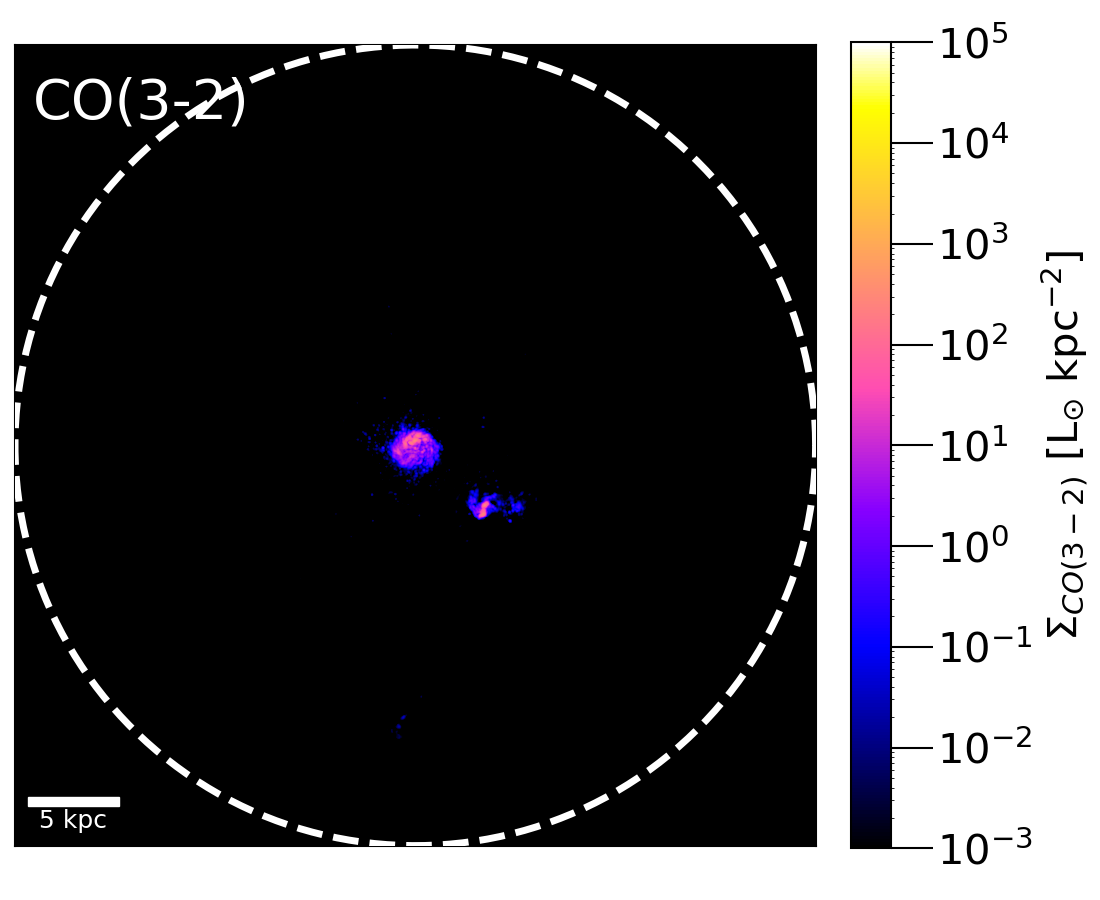}
   \includegraphics[width = 0.856\columnwidth]{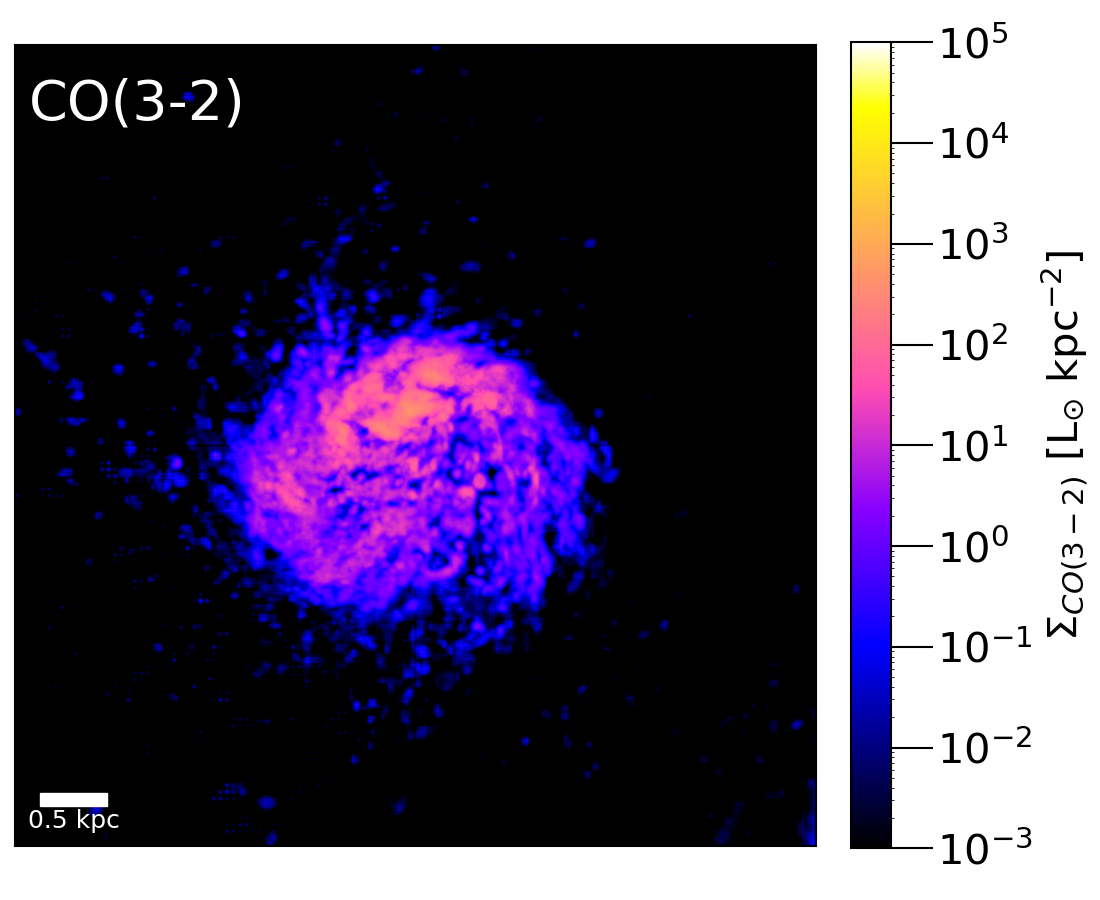}\\
      \caption{Line emission maps obtained with the fiducial RT model. The circles in each plot mark R$_{\rm vir}$. The top-left panel shows [CII], which is clearly extended and clumpy in appearance, with most of its emission originating from the main disc and merger components, and bridges of gas in between. The top-right panel shows the [OIII] map, which is fainter than in the other models tested in this study (see e.g. Fig.~\ref{fig:EMmapsMPM} in Appendix showing the same maps obtained for the MPM 2-component Gaussian model). [OIII] is extended, although less than [CII], and traces warm-ionised gas rather than the cold neutral gas traced by [CII]. The central maps show the [CI]$609\mu$m and [CI]$370\mu$m transitions, which are also extended, although fainter than [CII] (note the different surface luminosity scaling). The bottom panels show the CO(3-2) maps. The lowest plot on the right shows the CO(3-2) emission zoomed in on the main disc. As discussed in the main text, CO is faint in the RT model, and concentrated in the main disc and merger components. There are faint clouds of CO(3-2) emission in the CGM, overlapping with regions with relatively high [CI] emission.}
         \label{fig:EMmaps}
\end{figure*}

\begin{figure*}
\centering
   \includegraphics[width = 0.85\columnwidth]{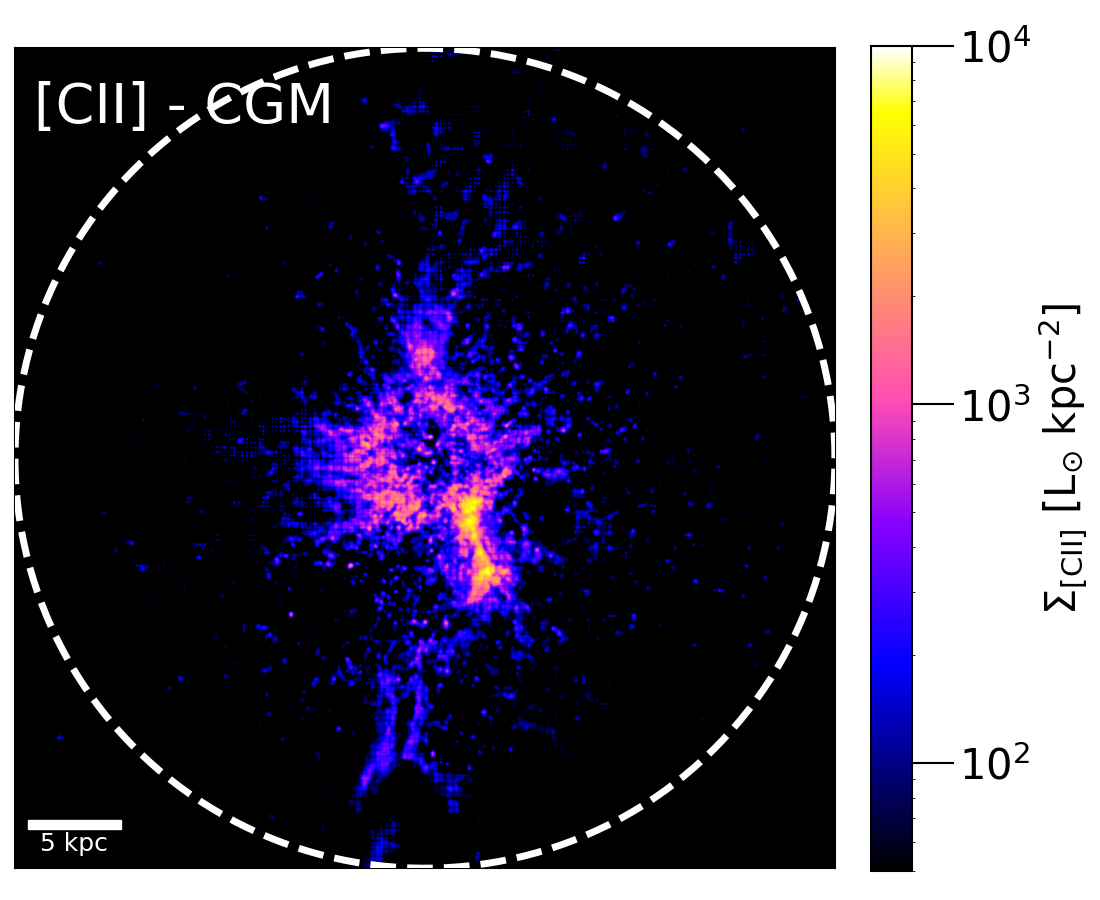}
   \includegraphics[width = 0.85\columnwidth]{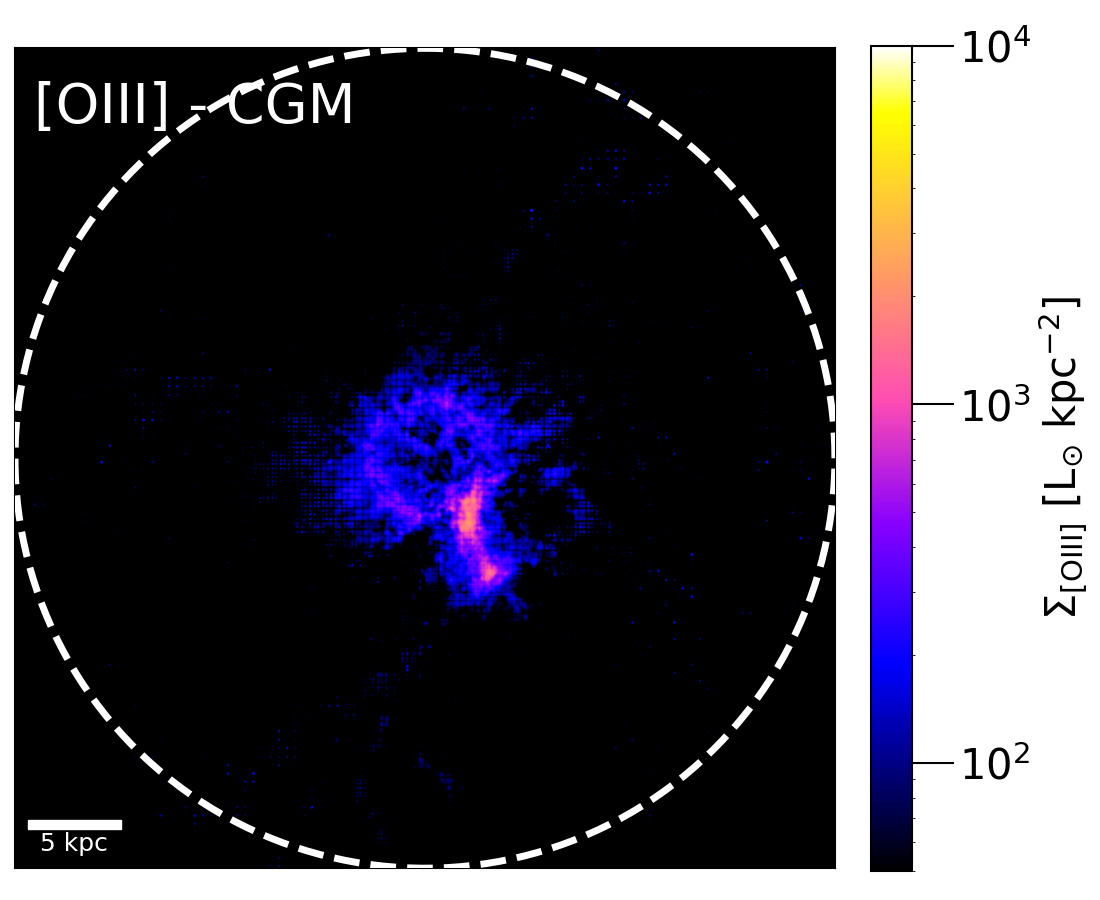}
      \caption{\rev{Line emission maps obtained with the fiducial RT model, only showing the CGM emission of Ponos for [CII] and [OIII].}}
         \label{fig:EMmaps_CGM}
\end{figure*}

The emission line maps obtained using the RT approach are shown in Fig.~\ref{fig:EMmaps}. The corresponding maps obtained from one of the alternative models, the MPM 2CG model, are reported in the Appendix in Fig.~\ref{fig:EMmapsMPM} for comparison. Because of the compactness of the CO emission, for this tracer we also plotted a version of the map zoomed-in on the central galaxy disc of \Ponos~galaxy \rev{(bottom left panel of Fig.~\ref{fig:EMmaps})}. The white dashed circle in each plot \rev{of Fig.~\ref{fig:EMmaps}} shows the virial radius (R$_{\rm vir}$). 
\rev{Fig.~\ref{fig:EMmaps_CGM} shows the [CII] and [OIII] maps obtained by selecting only the CGM components of \Ponos.}

\rev{For [CII] (top-left panels of Fig.~\ref{fig:EMmaps} and Fig.~\ref{fig:EMmaps_CGM}), the maps clearly show extended emission in the CGM, both in a diffuse halo around the central disc and in the gas bridges and tidal tails connecting the merging galaxies. The accreting filament (which hosts H$_2$ gas, see Fig.~\ref{fig:H2}) is responsible for the northern asymmetric extension of [CII], which is more visible in the CGM-only map.
These results are nicely consistent with recent observations of [CII] emission extending up to 10-15~kpc around high-z galaxies, reported by several authors  \citep{Fujimoto19,Fujimoto20,Ginolfi20a,Ginolfi20b,Fujimoto22,Fudamoto22}. Additionally there is a bright CGM contribution south of the main disc, which is gas being accreted from merger B onto the central disc.}

\rev{The [OIII] maps (top-right panels of Fig.~\ref{fig:EMmaps} and Fig.~\ref{fig:EMmaps_CGM})} show extended emission, although less extended than [CII]. Notably, the CGM emission in [OIII] \rev{arises} from different places than the CGM emission seen in [CII]. According to our computation and component separation, the CGM contribution to [OIII] is \rev{21\%}, while this number is lower \rev{(10\%)} for [CII]. \rev{However, Fig.~\ref{fig:EMmaps_CGM} shows that the northern accreting filament, and the bridges and tails connecting the merging galaxies are much brighter in [CII] than in [OIII], and [CII] extends as far as R$_{\rm vir}$}. Instead, [OIII] appears more puffy and concentrated in a halo surrounding the galaxy discs, possibly also due to SN feedback processes. [OIII] needs the presence of young, massive stars to be excited, so this is expected. 
\rev{Previous studies, both observational \citep{Carniani20, Fujimoto20,Ginolfi20a, Ginolfi20b,Fujimoto21, Fudamoto22} and theoretical \citep{Arata20,Pallottini19, Katz19, Pizzati20, Pizzati23}, have indeed found [CII] to be more extended than rest-frame UV discs, sub-mm continuum emission from dust, and than [OIII] emission. We find the same result in our simulation, both for the main disc and for mergers A and B.}
\rev{\cite{Fujimoto19} investigated the origin of the extended [CII] emission, dismissing the idea of satellite galaxies as a cause, and focusing of circum-galactic HII and PDR regions, as well as cold streams and outflows. In our study we find that the extended emission originates from cold gas falling into the galaxy from the filament and from tidal tails between the main disc and the satellites, similar to \cite{Ginolfi20b}, who estimated that 50\% of the [CII] emission of a high-z merger originated from the regions between the nuclei. \cite{Ginolfi20a} detected broad [CII] wings ascribed to outflows in normal high-z galaxies from the ALPINE sample, especially for SFR$>25~M_{\odot}~yr^{-1}$, and more extreme [CII] outflows have been observed in the luminous quasar population at these redshifts \citep{Maiolino12, Cicone15, Bischetti+19, Izumi21b, Meyer22}.
While \Ponos~has SF driven outflows, is not affected by extreme AGN driven outflows, which could lead to a higher fraction of extended emission. Moreover, how to address the formation and survival of molecular gas within the multi-phase outflow environments is still an open question for cosmological simulations, as this task requires extremely high resolutions that can only be achieved in idealised simulations \citep{Decataldo19, Nelson21}. }

\rev{Both the [CI] and CO maps in Fig.~\ref{fig:EMmaps} show mainly compact emission, tracing the dense and cold gas within the disc and merger components, with only minor contribution from the gas that we ascribe to the CGM. Therefore, the CGM of \Ponos~does not appear to be detectable in molecular gas tracers. However, there are two caveats to consider: (i) the lower numerical resolution on CGM scales which, as already mentioned, can lead to underestimating its H$_2$ content; and (ii) components that we ascribe to the ISM of the mergers (e.g. tidal tails, bridges, and the minor merging satellites themselves) and so we do not count as `CGM', may appear blended or unresolved in the observational data, and so they can contribute to what observations detect as a `diffuse' CGM emission. A more in depth discussion of observations is addressed in Sections~\ref{sec:comp_obs} and \ref{Dis}.}

\subsubsection{\rev{\textcolor{black}{Kinematics of the CGM traced by FIR/Submm lines}}}

\begin{figure*}
\centering
   \includegraphics[width = 0.75\textwidth]{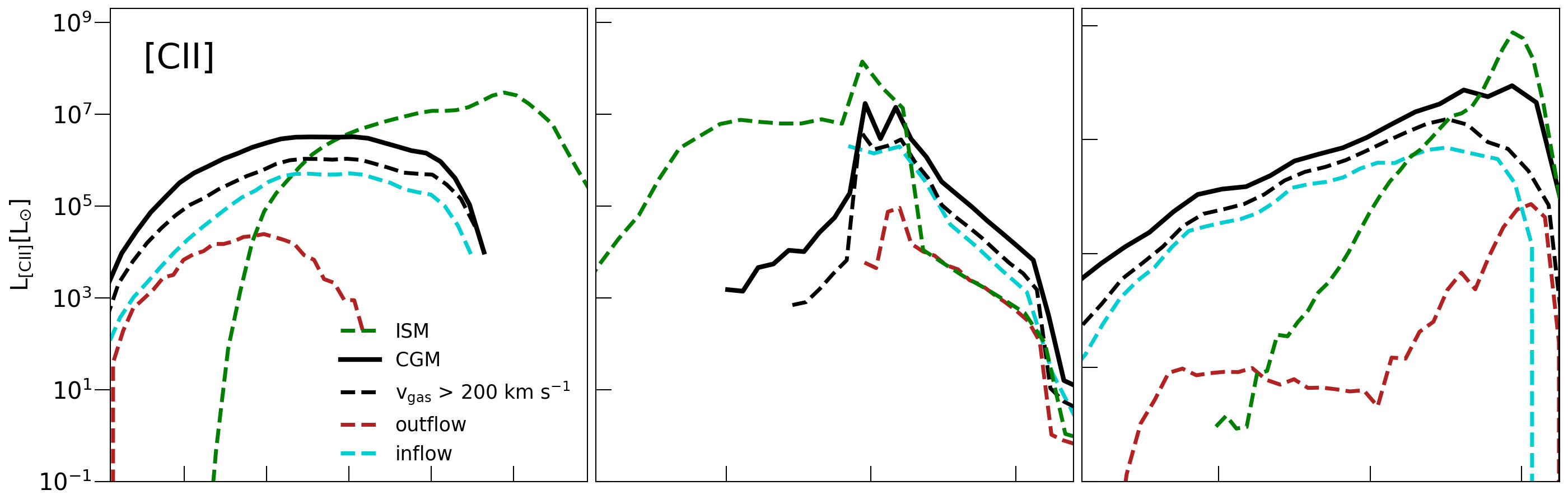}
   \includegraphics[width = 0.75\textwidth]{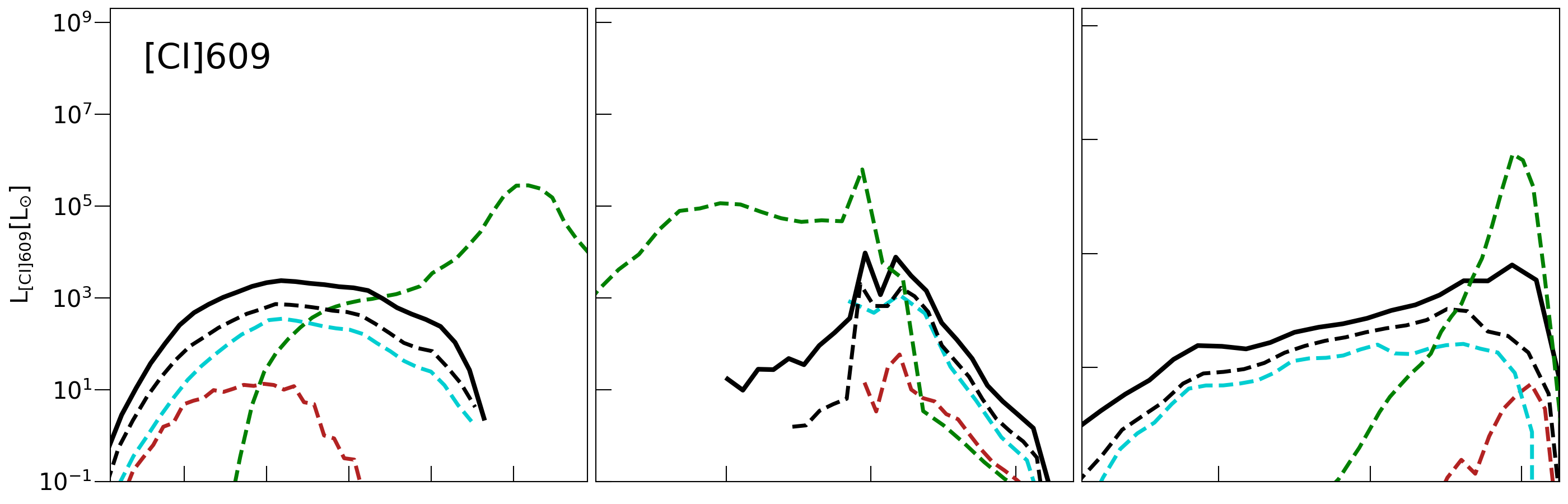} 
   \includegraphics[width = 0.75\textwidth]{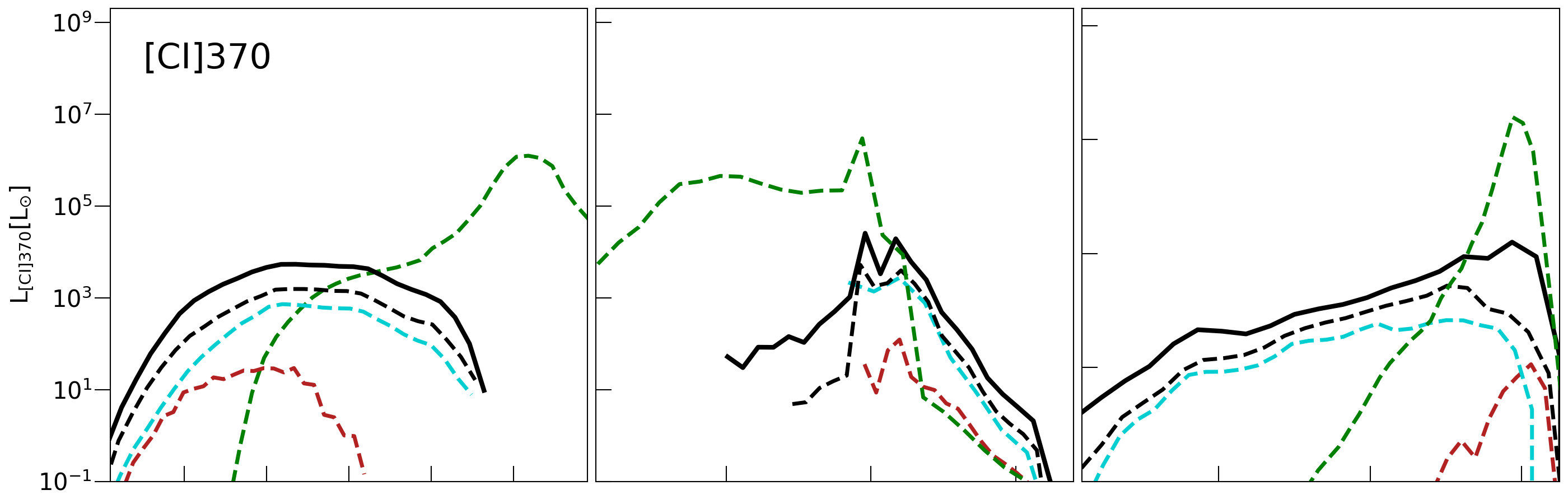}
   \includegraphics[width = 0.75\textwidth]{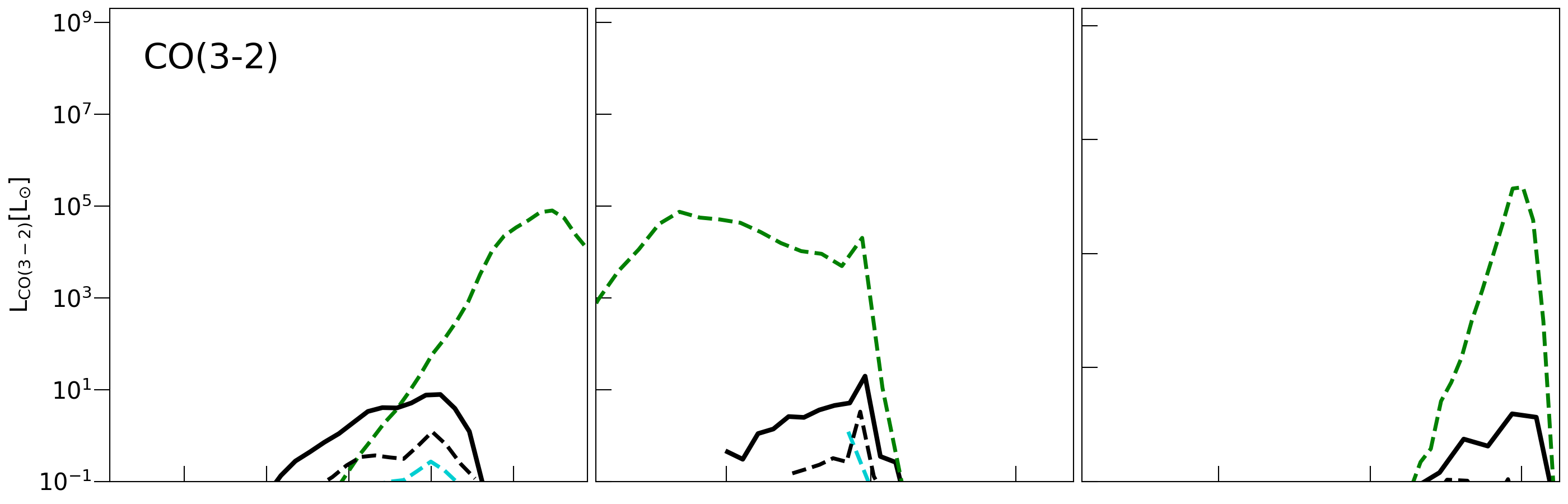}
   \includegraphics[width = 0.75\textwidth]{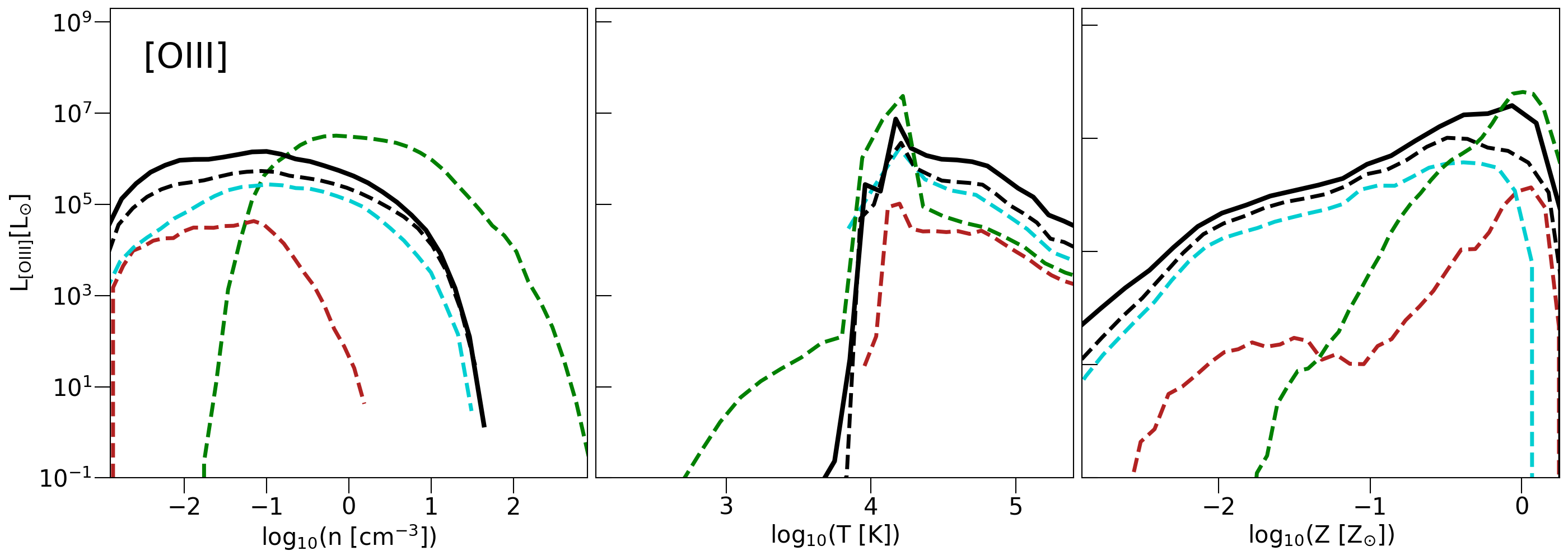}\\
      \caption{\rev{\textcolor{black}{Line luminosities of the CGM plotted as a function of gas number density (left column), gas temperature (middle column) and gas metallicity (right column). The lines are also divided into the contribution from different CGM components as well as the main disc, where the black solid line shows the total CGM emission, the dashed black line shows the emission of high velocity CGM gas (v $> 200$ ~km s$^{-1}$), outflowing high velocity gas in the dashed red line, inflowing high velocity gas in the blue dashed line and as a reference the disc/ISM component in dashed green.}}}
      \label{fig:CGM_hist}
\end{figure*}

\rev{\textcolor{black}{While the \Ponos~simulation does not have extreme, AGN driven outflows, there are outflows caused by stellar feedback processes, pushing gas from the ISM into the CGM. To study the CGM emission in the different lines in detail, we divided the gas into 'inflows' and 'outflows'. This was done by selecting gas at velocities of v $> 200$~km s$^{-1}$ (which is higher than the rotational velocity of the main disc, and is also in the range of velocities found by \cite{Pizzati23} for outflowing [CII] bright gas in their semi-analytical modelling) which we defined as being high velocity gas, and then dividing this high velocity gas into gas that is flowing away from the main disc (outflowing), and gas that is flowing towards the main disc (inflowing).}}
\rev{\textcolor{black}{As shown in Fig.~\ref{fig:CGM_hist} we plot the emission of the CGM (solid black), the high velocity CGM (dashed black), the inflows (dashed blue) and the outflow (dashed red) together with the disc/ISM emission (dashed green) as functions of the gas density, the gas temperature and the gas metallicity for the individual emission lines.}}

\rev{\textcolor{black}{In all lines (except CO, where we do not see significant emission from the outflowing component) the outflowing gas has low densities, high temperatures and high metallicity, which is consistent with gas expelled by stellar feedback. The outflowing gas is only dominant in emission in the CGM for the highest metallicities. The outflows within the \Ponos~are dominated by hot gas, but cold clouds within the outflows could be unresolved, which would lead to a stronger [CII] emission from the outflowing gas. It has to be noted, that outflowing gas very close to the disc, would not fall into the definition of the CGM in our study, as it would reside within twice the effective radius of the main disc.}}
\rev{\textcolor{black}{The inflowing gas has a wider range of densities, temperatures and metallicites, as it represents both pristine gas flowing in from the IGM, as well as gas accreted to the main disc from merger components in the form of gas bridges and tidal tails. The inflowing gas is the dominating component emitting in the CGM, which can also be seen in Fig.\ref{fig:EMmaps} and Fig.\ref{fig:EMmaps_CGM}, where the a accreting stream in the north and the gas bridges between the merging galaxies are bright in [CII]. At high redshift the cold gas is dominated by accreting gas \citep[e.g.][]{Decataldo23}, so it is not surprising that the inflows dominate in luminosity. For [OIII] the inflowing component is dominating as well, but this inflowing gas is mostly found within the bright tidal bridge between the main disc and merger B (see Fig.~\ref{fig:EMmaps_CGM}), where there is ongoing star formation and SN events that are traced by [OIII]. The rest of the emission originates from the outflowing gas.}}

\rev{Without additional information about the gas velocity and the chemistry, the interpretation of the CGM component is not straight forward. In our analysis we find that no gas tracer can be said to definitely trace inflows and outflows. A more statistical analysis of the CGM emission of similar systems could help to disentangle the different line tracers better.}

\subsubsection{The [OIII]/[CII] ratio map}

\begin{figure*}[tbh]
\centering
   \includegraphics[clip=true,trim=0.0cm 0cm 0.0cm 0.0cm,scale=.35]{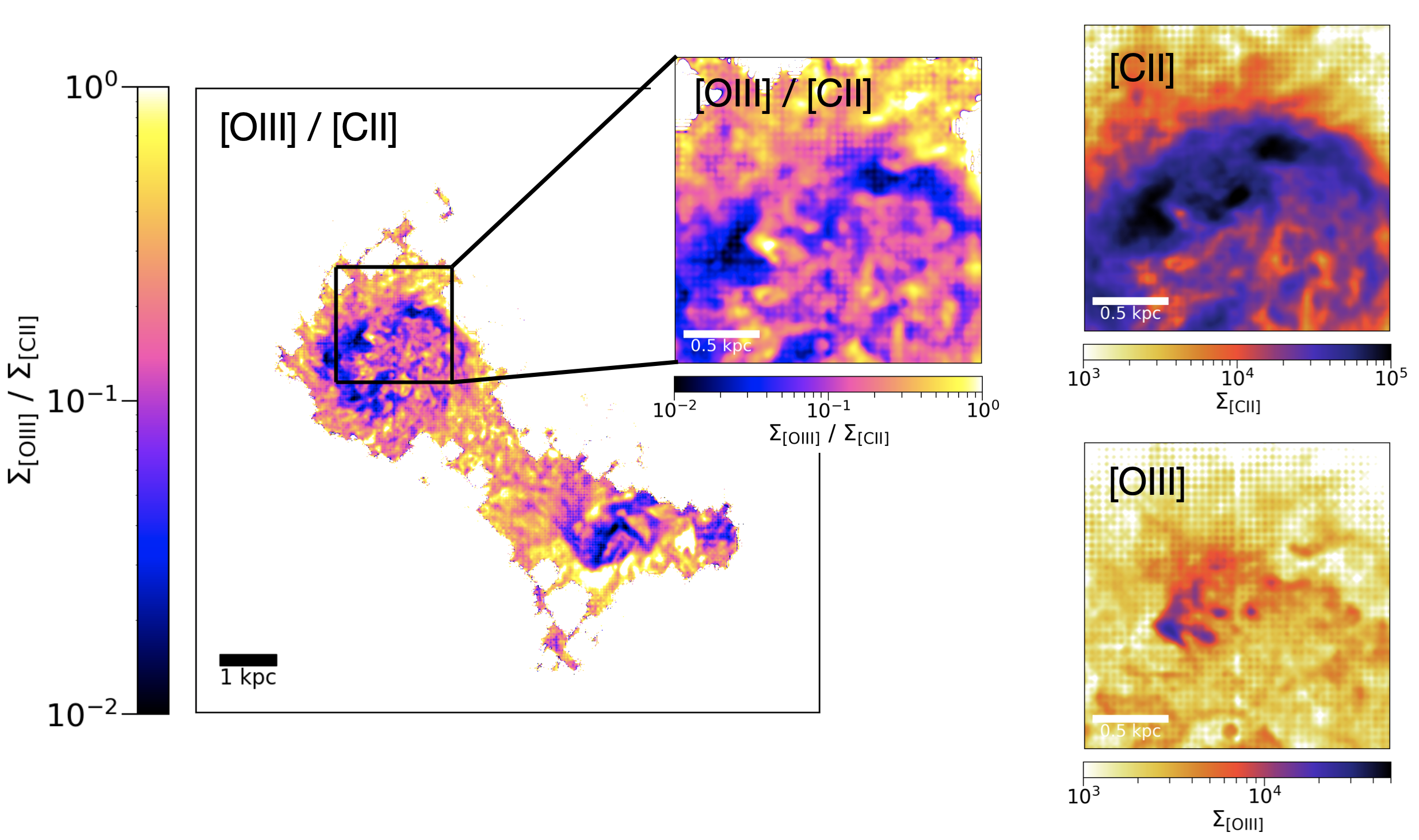}
      \caption{[OIII]/[CII] \rev{surface brightness} ratio map obtained from the RT run. The right panels show the [CII], [OIII], and [OIII]/[CII] plots in a zoomed-in portion of the main disc, to highlight the spatial offset between [OIII] and [CII]. \rev{The units of $\Sigma_{\rm line}$ are $[\rm L_{\odot}~kpc^{-2}]$.}} 
      \label{fig:O3C2}
\end{figure*}

\rev{Emission line ratios are a common tool to explore galaxy properties \citep{Rubin94,Nagao11, Nagao13,Pereira17,Santos17,Rigo18, Killi23}. The L$_{\rm [OIII]}$/L$_{\rm [CII]}$ ratio is especially interesting, as it involves two prominent FIR emission lines that arise from gas with different properties, and thus can be used to constrain the physical conditions of the high-z ISM and, potentially, CGM \citep{Carniani18,Hashimoto19, Harikane20,Arata20, Vallini21, Fujimoto22, Katz19,Katz22}.}
\rev{At $z\geq 6$  the measured $\rm L_{[OIII]}/L_{[CII]}$ ratios are typically high (between 1 - 20), due to a deficit in [CII] emission compared to local galaxies \citep{Inoue16,Hashimoto19,Carniani20,Harikane20,Vallini21, Witstok22, Kumari23}. The latter is generally ascribed to extreme physical conditions of the ISM in early galaxies, such as very low PDR covering fractions due to compact morphologies and high densities, strong radiation fields and high turbulence, as well as different abundance ratios and low metallicities \citep{Vallini15,Olsen17,Pallottini17,Lagache18,Pallottini19, Lupi2020, Arata20,Vallini21, Katz22}. Due to their different wavelengths,  angular resolution and sensitivity effects can bias the ratios measured by observations. In particular, the more extended nature of [CII] compared to [OIII], makes it more prone to its flux being underestimated by interferometric observations performed with high angular resolution (the surface brightness dimming discussed by \cite{Carniani20})}.

\rev{To calculate the L$_{\rm [OIII]}$/L$_{\rm [CII]}$ ratio for \Ponos, we first mask out gas with weak emission ($\Sigma_{\rm line} < 1 L_{\odot}$~kpc$^{-2}$) in either of the lines.
Such cut has a negligible effect on the total ratio, but it affects spatially-resolved ratio maps. We note that this effect applies to observational data as well, as they all have a sensitivity threshold. Because of the relative compactness of [OIII] and of the above cut, our line ratio map, reported in Fig.~\ref{fig:O3C2} (computed from the fiducial RT run), shows only the main disc and merger B, which are the regions from which most of the [OIII] emission originates (see Fig.~\ref{fig:EMmaps}).} 
\rev{The total L$_{\rm [OIII]}$/L$_{\rm [CII]}$ ratio measured in \Ponos~is around 0.17, which is lower than reported by observations at similar redshifts, and more in line to the $\leq 1$ values typically measured in local starburst galaxies \citep{Brauher08,DL14,Cormier15}.}
\rev{\cite{Katz22} found that using different abundance ratios in the \Cloudy~modelling, which would be more representative of the early Universe, can lead to lower [CII] and higher [OIII] emission, thus delivering higher L$_{\rm [OIII]}$/L$_{\rm [CII]}$ ratios more similar to the values observed at high redshift. }

\rev{In addition, as also pointed out by \cite{Pallottini19}, the line ratios can vary within the galaxy itself, depending on the radiation field experienced by the gas and other gas properties.} 
\rev{The [OIII]/[CII] ratio map in Fig.~\ref{fig:O3C2}} shows a clear offset between these two tracers in the dense star forming regions in the \Ponos~main disc.
In the densest \rev{portions of the disc}, [CII] appears more widespread, while [OIII] is bright in \rev{compact regions that corresponds to cavities of the [CII] map, possibly created by SN feedback} (see comparison between the two top-right zoomed-in panels showing the [CII] and [OIII] maps). 
\rev{There is a clear morphological difference between the two tracers, with [OIII] appearing `puffier' and more concentrated around the main galaxy and the ISM material of the mergers, and [CII] being more extended and also tracing filamentary structures between the mergers \citep[as also found by e.g.][]{Ginolfi20b}.} 
\rev{In summary, higher L$_{\rm [OIII]}$/L$_{\rm [CII]}$ are found in regions with higher G$_0$ values within the main disc and merger B, coinciding with areas of high star formation. Our results support the use of spatially resolved observational studies of the L$_{\rm [OIII]}$/L$_{\rm [CII]}$ ratio to study the physical properties of the ISM in high-z galaxies.}

\rev{Line ratios, including different tracers, will be further explored in a follow-up publication.}

\section{Discussing our results in the context of observational works}\label{sec:comp_obs}

\begin{figure*}[tbh] 
   \includegraphics[clip=true,trim=0.5cm 1.5cm 0.1cm 0.1cm,scale=.32]{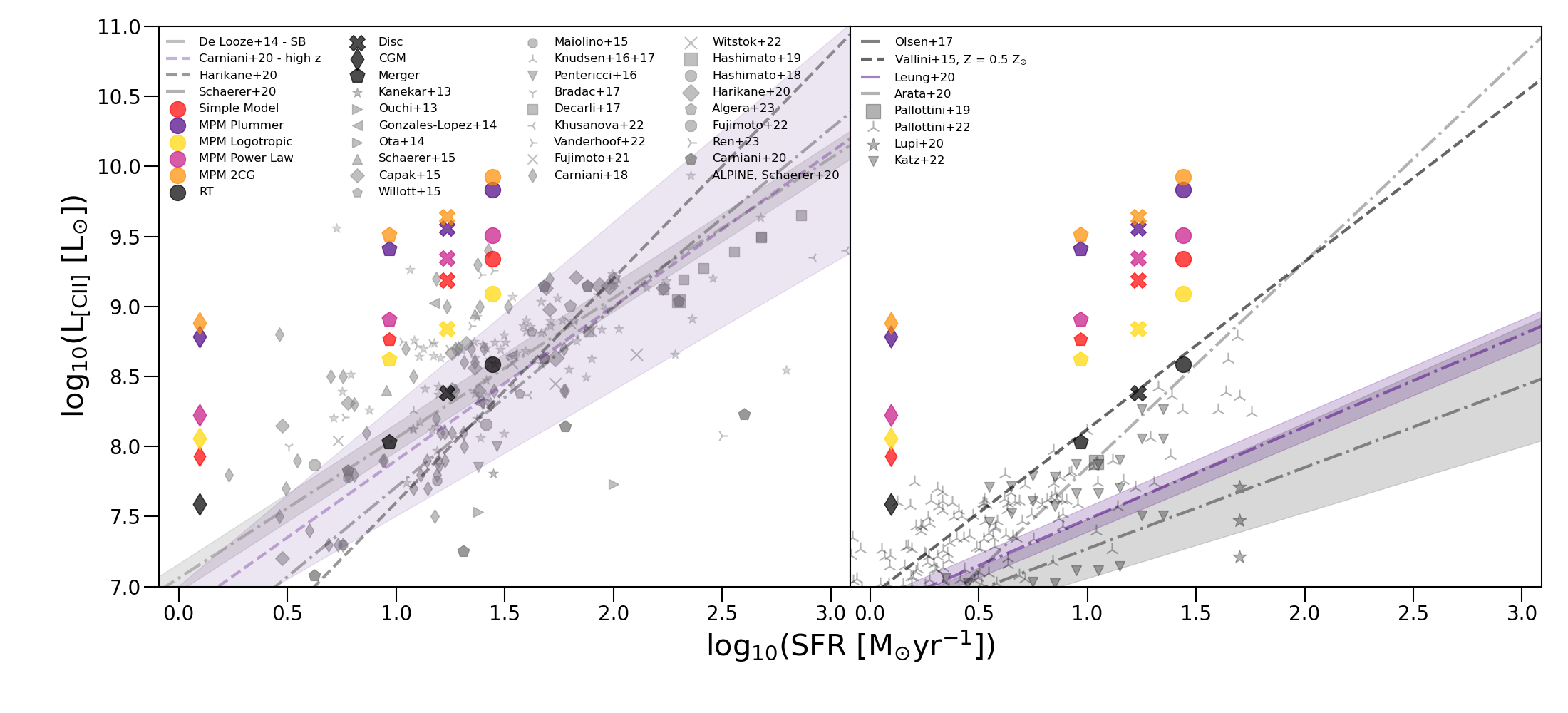}
    \caption{[CII] line luminosity as a function of the SFR. In the left panel, our models are compared with observational constraints available from the literature at a similar redshift, while the right panel shows the comparison with other theoretical works. The circles represent the whole halo, crosses represent the main disc, and the pentagon the merging companions A and B. The CGM is shown using diamond symbols. The observational data are taken from: \cite{Kanekar13} at z = 6.5, \cite{Ouchi13} at z = 6.5, \cite{GL14} at z = 6.5 - 11, \cite{Ota14} at z = 6.9, \cite{Schaerer15} z = 6.8 - 7.5, \cite{Capak15} for z = 5.1 - 5.7, \cite{Willott15} for z = 6, \cite{Maiolino15} for z = 6.8 - 7.1, \cite{Knudsen16} and \cite{Knudsen17} for z = 6 - 7.6, \cite{Pentericci16} for z = 6.6 - 7.1, \cite{Bradac17} for z = 6.7, \cite{Decarli17} for z = 6 - 6.6, \cite{Khusanova22} for z = 6, \cite{Vanderhoof22} for z = 4.5, \cite{Fujimoto21} for z = 6, \cite{Witstok22} for z = 6 - 7, \cite{Hashimoto19} for z \textasciitilde 7, \cite{Harikane20} for z \textasciitilde 6, the ALPINE data for z = 4 - 8 \citep{Schaerer20}, \cite{Algera23} for z = 7.3, \cite{Fujimoto22} for z = 8.4,\cite{Ren23} for z = 7.2, and \cite{Carniani18} for z = 5 - 7. The relation shown in grey is by \cite{DL14} for SB galaxies, and the relation shown in purple by \cite{Carniani20} for high redshift galaxies. In addition, on the right panel, we compare our results to simulation data by \cite{Pallottini19, Pallottini22, Katz22} and \cite{Lupi2020}. The relation by \cite{Vallini15} based on physical motivated models for a metallicity of Z = 0.5 Z$_{\odot}$ can be seen in grey dashed, \rev{the relation by \cite{Arata20} in dot-dashed}, the relation found by \cite{Leung20} in purple, and the \cite{Olsen18} data in grey dot-dashed.}
         \label{fig:CIIvsSFR}
\end{figure*}

\begin{figure*}[tbh] 
   \includegraphics[clip=true,trim=0.5cm 1.5cm 0.1cm 0.1cm,scale=.32]{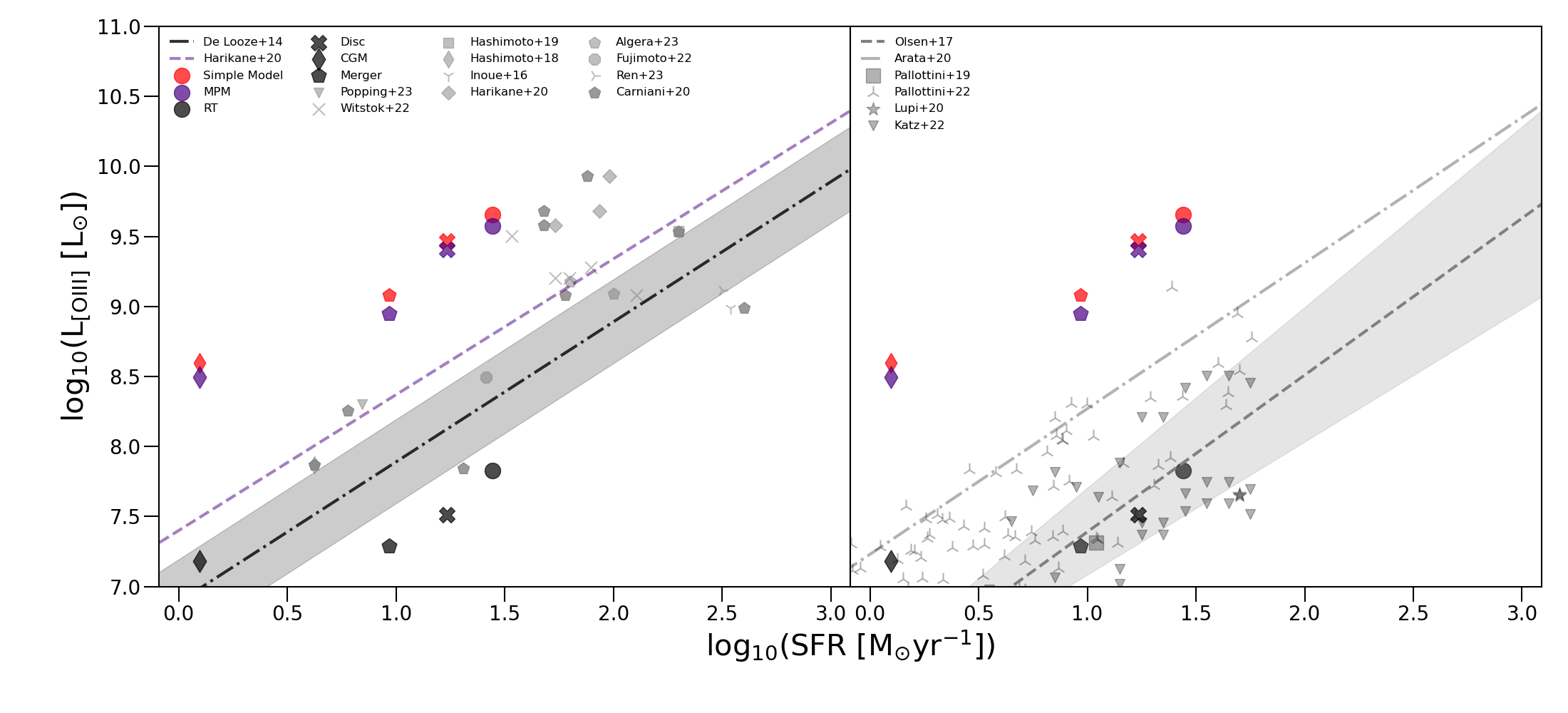}
      \caption{[OIII]88$\mu$m line luminosity as a function of SFR. Our modelling results based on the \Ponos~simulation are compared with observations in the left panel, and with other theoretical works in the right panel. Symbols and colour-coding are the same as in Fig.~\ref{fig:CIIvsSFR}. As was also seen in Fig.~\ref{fig:lums}, the [OIII] luminosities of the MPMs does not change much for different sub-grid density profiles for the dense gas, and thus we show the MPM emission in one data point. The MPMs are represented in indigo. The observational data in the left panel are taken from: \cite{Popping22} (upper limit of a z = 13 galaxy), \cite{Witstok22} (z = 6 - 7), \cite{Hashimoto19} (z \textasciitilde 7), \cite{Hashimoto18} (z = 9), \cite{Inoue16} (z \textasciitilde 7), \cite{Algera23} (z = 7.3), \cite{Fujimoto22} (z = 8.4), \cite{Ren23} (z = 7.2), and \cite{Harikane20} (z\textasciitilde 6). The relation reported in the left panel are the one derived by \cite{DL14} \rev{and by \cite{Harikane20}}. On the right panel, we compare our results to the simulation results by \cite{Pallottini19,Pallottini22, Katz22}, and \cite{Lupi2020}, \rev{the relation by \cite{Arata20} in dot-dashed}, and the relation reported is the one fitted by \cite{Olsen17} for their z = 6 simulated galaxies.}
         \label{fig:OIIIvsSFR}
\end{figure*}

\begin{figure}[tbh] 
   \includegraphics[clip=true,trim=0.0cm 0.0cm 0.cm 0.cm,scale=.3]{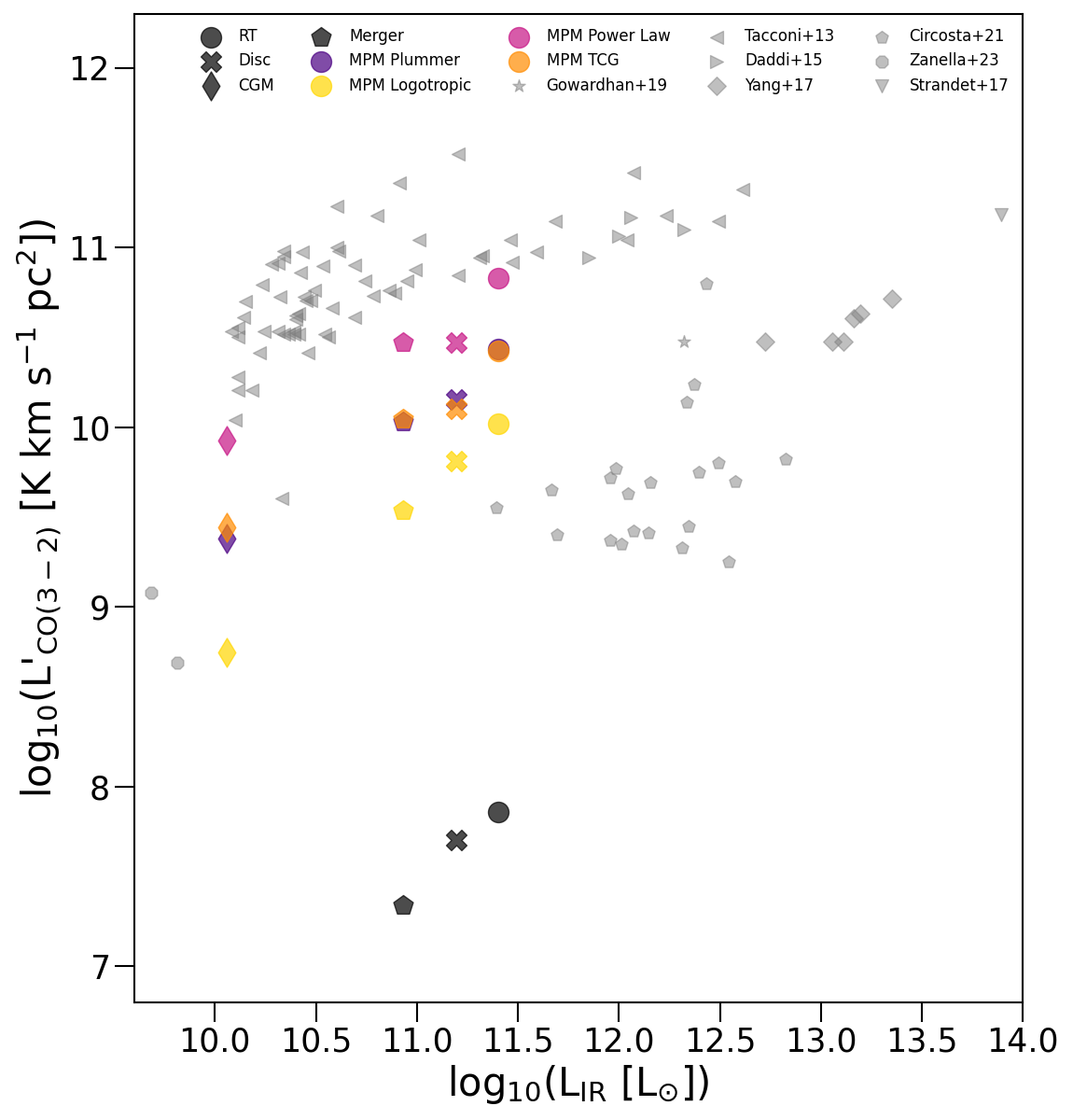}
      \caption{\rev{CO(3-2) line luminosity as a function of total infrared luminosity of \Ponos, obtained with the different models, compared with observational data. Symbols and colour-coding are as in Fig.~\ref{fig:CIIvsSFR}. The SM was not included because of its very low CO luminosity. Due to the scarcity of observations CO(3-2) data at $z\gtrsim6$, most data shown in this figure (except for \cite{Strandet17} at $z = 6.9$) represent galaxies at $z = 1 - 4$: \cite{Gowardhan19} at z$\sim$3.2, \cite{Tacconi13} at z = 1 - 3, \cite{Daddi15} at z = 1.5, \cite{Yang17} at z = 2 - 4, \cite{Circosta21} at z = 2 - 2.5 and \cite{Zanella23} at z$\sim$1.}}
         \label{fig:COvsIR}
\end{figure}

When comparing theoretical predictions with observational results, it is important to keep in mind that the observational setup plays a major role in the outcome of the observation. As discussed and demonstrated in Section~\ref{Dis_mocks} through ALMA mock observations of \Ponos, the choice of a single dish telescope vs an interferometer, the array configuration (which determines the angular resolution and the maximum recoverable scale of the source), the integration time, and the observed frequency, all have a major impact on the recovery of the physical properties of the target.
For the comparison between our results and observational constraints available in the literature, it is also important to note that, while our \Ponos~modelling delivers several data points (for the disc, merging companions, CGM, and total halo of the simulation), high-redshift observations usually only provide one data point per target, with only one associated estimate of SFR (or M$_*$, etc). There could be several minor mergers and CGM components contributing to those single data points, which went undetected by the observation.

\subsection{[CII] vs SFR}

The empirical correlation between [CII] luminosity and SFR of galaxies has been extensively investigated in the literature \citep{DL14, Herrera15,Capak15,Schaerer20,Carniani20}, and widely used to validate theoretical predictions against observational constraints. This relation is explored in Fig.~\ref{fig:CIIvsSFR} for our six different models, where the left panel focuses on the comparison with available high-redshift observational constraints, and the right panel on the comparison with simulation data and other theoretical models. The large coloured markers in both plots represent the data from our study, using the intrinsic SFR of \Ponos~and the derived line luminosities. As it was done in Fig.~\ref{fig:lums}, different components are represented with different symbols (the circles represent the total luminosity). The observational data and the simulation data of other studies are plotted in grey, while relations derived by other studies are plotted with grey and purple lines, the shaded areas representing their corresponding scatter. All observational and simulation data points are obtained for high-z objects, with the exact references and redshifts reported in the caption of Fig.~\ref{fig:CIIvsSFR}. The relations on the left panel are taken from \cite{DL14} for starburst galaxies, which \cite{Schaerer20} found to hold up for high-redshift galaxies, and from \cite{Carniani20} for high redshift galaxies. The relations on the right panel are by \cite{Olsen17}, \rev{\cite{Arata20}}, and \cite{Leung20}, based on their simulations, and by \cite{Vallini15} based on their theoretical modelling. 

\rev{By focusing on the comparison with observational data (left panel of Fig.~\ref{fig:CIIvsSFR}), we note that our SM, MPM power law, and MPM logotropic models lie above the \cite{Schaerer20} relation obtained for ALPINE data, but are still consistent with the upper envelope of the distributions of observational data points. In contrast, the MPM Plummer and MPM 2CG models strongly overestimate the total [CII] emission compared to available observational constraints. Our fiducial RT model is the most consistent with observations, and it even broadly follows the \cite{Schaerer20} [CII]-SFR best-fit relation, except for the CGM component, which lies above the relation. This could indicate that star formation (SFR$\sim\,1\,{\rm M}_\odot\,{\rm yr}^{-1}$ in the CGM) is not the only source of ionisation of the extended gas component, but that the UVB plays a vital role, as also suggested by other studies \citep{Fujimoto19,Pizzati20}.} 

\rev{In the right panel of Fig.~\ref{fig:CIIvsSFR}, we compare our results to other theoretical works. The RT model, represented by the same black symbols as in the left panel of Fig.~\ref{fig:CIIvsSFR}, is consistent with the relation by \cite{Vallini15}, except for its CGM component which lies above this relation. Our other models lie all above the results from other theoretical works, as explained in Section~\ref{sec:comp_sim}. } 

\subsection{[OIII] vs SFR}

Fig.~\ref{fig:OIIIvsSFR} shows the [OIII]~$88\mu$m luminosity as a function of SFR, compared with observational data on the left panel, and with theoretical predictions on the right panel. 
The left panel of Fig.~\ref{fig:OIIIvsSFR} reports the comparison with the \rev{empirical relations derived} by \cite{DL14} \rev{and \cite{Harikane20}}, while the right panel shows the \rev{theoretical relations from} \cite{Olsen17} \rev{and \cite{Arata20}}. 
Because of the scarcity of observational constraints, there are only few data points reported in the left panel, all from observations at $z\gtrsim6$ (including an upper limit at $z\sim13$ \rev{by \cite{Popping22}}), hence \rev{they are not necessarily representative of the normal $z\sim6$ galaxy population that we probe with the \Ponos~simulation, since they could be biased towards the brightest [OIII] emitters.}
Our fiducial RT model lies slightly below the \cite{DL14} relation, and has a lower SFR than most of the observations of high redshift galaxies. The few observations of galaxies with similar SFR show higher [OIII] emission. \rev{Additional} observations of a broader population of high redshift galaxies are needed to further constrain the [OIII] - SFR relation. 

As shown in the right panel of Fig.~\ref{fig:OIIIvsSFR}, \rev{most other theoretical predictions lie below the observational data in the SFR-[OIII] relation, and indeed our RT model is in good agreement with them} \citep{Olsen17, Katz22,Pallottini19,Pallottini22}. \rev{Instead, our} SM and MPM models produce higher [OIII] luminosity values compared to other theoretical studies, \rev{which we ascribe to the hot and dense gas component created by the sub-grid blast wave feedback model of \Ponos~that, as already discussed in Section~\ref{sec:modelcomp}, is cooled down by the RT post-processing}.  

\subsection{\rev{CO(3-2) vs IR luminosity}}

\rev{To compare our CO(3-2) predictions with observational data, we computed the brightness-temperature dependent line luminosity ($L^{\prime}_{\rm CO}$, often used by observational works), using the following equation by \cite{Solomon97}:}
\begin{equation}
    L^{\prime}_{\rm CO} [\rm K~km~s^{-1}~pc^{2}] = (3.25 \cdot 10^{7}) \nu_{\rm obs}^{-2} D_{L}^{2} (1+z)^{-3} \int_{\Delta \nu} S_{\nu}dv
\end{equation} 
\rev{where $\nu_{\rm obs}$ is the observed frequency in units of GHz ($\nu_{\rm rest} = \nu_{\rm obs} \cdot (1+z)$), $\int_{\Delta \nu} S_{\nu}dv$ is  the velocity integrated flux in units of Jy km s$^{-1}$ (see equation~\ref{eq:Jy}),  and $D_{L}$ is the luminosity distance in units of Mpc (calculated with the Cosmology Calculator by \cite{CosmoCalc} for a flat universe).}
\rev{This results in $L^{\prime}_{\rm CO(3-2)} = 7.21 \cdot 10^{7}$ K~km~s$^{-1}$~pc$^{2}$, which is significantly lower than observations. Indeed, at $z>1$, the measured values are in the range $L^{\prime}_{\rm CO(3-2)} \sim 10^{9} - 10^{11}$ K~km~s$^{-1}$~pc$^{2}$ \citep{Yang17, Strandet17,Lenkic20,Decarli20,Boogaard23}. \cite{Knudsen17} reports a (3$\sigma$) upper limit of $\sim10^{9}$~K~km~s$^{-1}$~pc$^{2}$ for a dusty lensed $z=7.5$ galaxy with stellar mass of $\sim10^9~M_{\odot}$ and SFR$\sim12$~M$_{\odot}~yr^{-1}$.}
\rev{Such discrepancy with observational data is expected, as we known that our fiducial model underestimates CO emission.}

\rev{For a further comparison, we plot in Fig.~\ref{fig:COvsIR} the $L^{\prime}_{\rm CO}$-L$_{\rm IR}$ relation. For this, we have assumed the L$_{\rm IR}$ to relate to the SFR following a Chabrier initial mass function, as often done in observational works \citep{Chabrier03, Gowardhan19}:}
\begin{equation}
    \rm SFR~[M_{\odot} yr^{-1}] = 1.09 \cdot 10^{-10} \rm L_{\rm IR}~[L_{\odot}]
\end{equation} 
\rev{Due to the scarcity of $z>6$ observations, we compare our results with literature data between $z = 1 - 4$.}
\rev{As expected, the RT model lies below the observed values, as we do not reach the necessary densities in the RT model to fully represent molecular gas. All the MPM models lie within the scatter of the observations, with the MPM power law profile resembling the findings of \cite{Tacconi13} most closely, once again showing that for the modelling of molecular gas, sub-grid models are necessary.}
\rev{\cite{Vallini18} simulated the high-z CO emission for various transitions by modelling the emission from individual GMCs, and then applying this to a cosmological zoom-in simulation. Our RT model has lower emission in CO(3-2) in comparison. }

\section{Discussing our results in the context of theoretical studies}\label{sec:comp_sim} 

In this section we discuss how different modelling approaches and sub-grid \rev{recipes} can \rev{affect} the \rev{resulting} emission line luminosities. \rev{In literature there are currently two main approaches. The first one includes global radiative transfer and non-equilibrium chemistry of primordial gas \citep[e.g.][]{Pallottini19,Katz19, Pallottini22, Katz22} or (non)-equilibrium chemistry tracing the main coolants \citep{Arata20,Lupi2020}, which are most comparable to our fiducial RT model. The second approach includes more advanced sub-grid modelling of the dense gas, but usually with more approximative estimations of the FUV radiation field \citep[e.g.][]{Olsen15,Olsen17,Olsen18,Leung20} which are most comparable to our MPM model, next to some semi-analytical approaches \citep[e.g.][]{Popping19, Pizzati23}.}
\rev{We will focus on the [CII] and [OIII] emission lines, because, being the most prominent coolants of the ISM, and being observable at high-z using sub-mm ground-based telescopes, these lines have been modelled in a variety of theoretical studies. Important to note is that none of the following studies focus on the CGM emission in the same extent as we do, where we find 10\% of [CII] and 21\% of [OIII].}

\subsection{Global Radiative Transfer Studies}


\rev{Our fidicual RT model is most similar to the studies by \cite{Pallottini19,Pallottini22} using the SERRA cosmological zoom-in simulations. Their simulations have resolution of $\sim~25$~pc, somewhat lower but similar to \Ponos. 
The simulations are performed with \ramsesrt~ with on-the-fly radiative transfer and also used \krome~for primordial non-equilibrium chemistry. For the modelling of FIR emission lines, the authors use \Cloudy~grids for the density, metallicility and radiation field with a slab geometry similar to our grid. In addition, they employ a sub-grid log-normal density distribution profiles, to account for the clumpy structure of the ISM. Comparing to their model, \Ponos~does not have on-the-fly RT, which may result in inconsistencies in gas dynamics. Nevertheless, our post-processing with \kramsesrt do capture the ionisation sates and the FUV radiation accurately at a specific snapshot. Moreover, we include the radiation from the UVB for the first time to investigate the extended emission from the CGM. } 

Regarding [CII] emission, we have shown that over half of the emission is originate from the dense phase, and the emission peaks at intensity log(G$_{0}$)$\sim$~1.5. This is in partial agreement with \cite{Pallottini19}, where the authors found that predominant [CII] emission originates from their denser gas phase with $n\simeq160$~cm$^{-3}$, irradiated by an average radiation field of $\sim20$~$G_0$, with a subdominant contribution for the low density gas. However, unlike \cite{Pallottini19} and the previous work by \cite{Pallottini17}, our results indicate a significant fraction (44\%) of [CII] emission from the more diffuse gas (in neutral and ionised phase), and this diffuse phase [CII] dominates the CGM emission (see Fig.\ref{fig:Hist_mor}). From our definition of components, the CGM component accounts for about 10\% of the total [CII] emission, whereas \cite{Pallottini19} only find a very faint [CII] halo of $\Sigma_{[CII]} = 5 \cdot 10^{4}$~L$_{\odot}$kpc$^{-2}$. In \cite{Pallottini22}, where the authors found majority of [CII] emission originated from the central disc, they also do not find [CII] halos which are as extended as high redshift observations. This is most likely due the inclusion of the UVB in our modelling, which is demonstrated to be important in producing extended [CII] emission. Another possible reason is the sub-grid modelling of log-normal density distribution in \cite{Pallottini19,Pallottini22}. This can potentially result in more [CII] emission from the dense phase, because the high density gas beyond the resolution limit of the hydrodynamic simulation is sampled in such sub-grid models. Last but not least, we note that \Ponos~is one single system undergoing major merger, and mergers may enhance extended emission due to the large-scale flows they induce.

\rev{\cite{Pallottini19} found that in their models the [OIII] emission line peaks in dense but ionised regions, which is again in agreement with our findings, where most of the [OIII] originates in denser gas, but at higher temperatures compared to the other modelled emission lines, and in regions with higher G$_0$ values. The [OIII] emission modelled by \cite{Pallottini19} is mostly concentrated around the main galaxy, which is also where we find a significant fraction in our model, but we also find extended emission tracing the gas bridge between merger B and the main disc, as well as some outflowing gas. Similarly, \cite{Pallottini22} find a concentrated emission of [OIII], tracing the HII regions of the galaxies in their sample, which is similar to what we find in \Ponos.}

\rev{\cite{Katz19} investigated [CII] and [OIII] line emissions using the Aspen cosmological radiation hydrodynamics zoom-in simulation of a massive Lyman-break galaxy at z = 12 - 9. The simulation was performed with the \ramsesrt~code with on-the-fly RT and non-equilibrium chemistry, and it reaches a resolution of 13.6 pc. FIR line emissions were calculated using \Cloudy~(with grids for density, temperature, metallicity and radiation field, calculated with constant temperature) and machine learning. The authors found that the [CII] emission origins predominantly from the cold molecular disc, while [OIII] is concentrated around SF regions. Extended emissions are found to be connected to merger activities. While we also find that a significant amount of emission originates from tidal tails between the merging components, the extended emission in \Ponos~ also traces inflowing and outflowing gas. \cite{Katz19} found outflowing gas in their z = 10 galaxies traced by [OIII] but not by [CII], and there is a clear spatial offset between the two lines which was  confirmed by a later study with the same simulation, in \cite{Katz22}. The spatial offset is also observed in our study (see Fig.~\ref{fig:O3C2}) and [OIII] in \Ponos~is puffy in appearance. However, we do not find a significant difference in velocities between gas that is bright in [OIII] and gas that is bright in [CII]. In fact, inflowing gas dominates both [OIII] and [CII] systems in our simulation, except for the lowest density $n \lesssim 0.01 \rm cm^{-3}$ and high metallicity ($Z \gtrsim Z_{\odot} $) gas (Fig.~\ref{fig:CGM_hist}). As noted in a previous section, this inflowing gas is mainly found in the tidal bridge where ongoing star formation and SN events are located. 
Comparing the phase diagram for the RT model in Fig.~\ref{fig:phase} to the results of \cite{Katz19} at redshift z = 10, we find that the peaks of [CII] and [OIII] emissions in our simulation in the density-temperature space coincide those in \cite{Katz19}, although we also find emission from lower density gas. Their total [CII] and [OIII] luminosities are comparable to our derived values at z = 6. }

\rev{\cite{Lupi2020} compared various emission line luminosities ([CII]~158~$\mu$m, [OI]~63~$\mu$m, [OI]~146~$\mu$m, [OIII]~88~$\mu$m) computed based on established \Cloudy~models with the ones computed from on-the-fly non-equilibrium chemical network, using a GIZMO cosmological zoom-in simulation of a galaxy at $z=6$ (with a stellar mass of M$_{*} = 1.6 * 10^{10}$ ~M$_{\odot}$). The simulation has resolution m$_{\rm gas} = 2 \cdot 10^{4}$~M$_{\odot}$, and includes on-the-fly RT coupled with non-equilibrium chemistry using \krome, following 16 chemical species. In their models, the [CII] line luminosities derived with their "\krome~approach" differ from their \Cloudy~based models by about a factor of 3. The authors conclude that, because the effect of very different approaches on the [CII] luminosity is not very strong, [CII] is a good tool for constraining physics included in cosmological simulations. This is in agreement with our findings, which also suggest [CII] emission is rather insensitive to models. Nevertheless, the highest line luminosity found by \cite{Lupi2020} in their \krome~approach  (L$_{\rm [CII]} = 5.12 \cdot 10^{7}$~L$_{\odot}$) is lower than our results (L$_{\rm [CII]} = 1.95 \cdot 10^{9}$~L$_{\odot}$) by over an order of magnitude. This is also shown in Fig.~\ref{fig:CIIvsSFR} and in Fig.~\ref{fig:OIIIvsSFR}, where the [CII] and [OIII] luminosities from \cite{Lupi2020} for all their modelling approaches are lower compared to our models, even though their simulated galaxy is more massive and has a higher SFR (\textasciitilde 50 M$_{\odot}$~yr$^{-1}$) than \Ponos. This is likely to be caused by the difference in resolution, as the contrast in luminosities is stronger in [CII], which is multi-phase and sensitive to resolution, than in [OIII], which originates from diffuse, warm gas which converges at lower resolutions. }

\rev{\cite{Arata20} applied global radiative transfer in post-processing using the ART$^{2}$ code onto a suite of Gadget-3 cosmological zoom-in simulations of galaxies at z~=~6-15 where they traced 9 chemical species. They derived the emission of [OIII] and [CII] based on equilibrium heating and cooling calculations of O$^{++}$ and C$^{+}$, consistent with the equations used in \Cloudy. The authors found the results from their modelling approach consistent with \Cloudy~modelling. They found [CII] to be more extended than [OIII], and exhibit less variation with time, whereas [OIII] emission varies more significantly with time, and is brighter in eras of high star formation.  Their [CII] luminosity is consistent with the results of our study (see Fig.~\ref{fig:CIIvsSFR}), once again showing that [CII] is less model sensitive, while their [OIII] emission is higher, which could be due to the direct tracing of chemical species, and thus the more accurate treatment of abundance ratios within the gas. }

\subsection{Sub-grid modelling of dense gas}

\rev{By using a model based on cosmological simulations run with Gadget-2 (resolution of 60~pc), RT post-processing (using the UV RT code LICORICE) and sub-grid ISM modelling to account for a two-phase ISM, \cite{Vallini15} found that the PDR regions in their models of high redshift galaxies dominate the [CII] emission, while diffuse cold neutral gas only accounts for 10\%, and the authors attributed this behaviour to the high CMB temperature at these epochs, which suppress the flux contrast. The neutral diffuse gas in our RT model accounts for around \rev{18\%} of the [CII] emission. The two-phase ISM model in \cite{Vallini15} does not include a hot-ionised gas phase, and hence there is no contribution from ionised gas to [CII] luminosity. In our  fiducial RT model such ionised gas makes up \rev{25\%} of the emission, comparable to the contribution from the neutral diffuse phase.}

\cite{Olsen17, Olsen18} used the Mufasa cosmological simulations \citep{Dave16,Dave17} ($m_{\rm gas} = 1.9 \cdot 10^{5}~h^{-1}$~M$_{\odot}$) together with the \Sigame~approach (applying sub-grid GMCs to their simulation using the GMC mass function) to study [CII] emission from high redshift galaxies. They find the average contribution of GMCs to the [CII] emission to be around 50\% for galaxies at z = 6, with 44\% originating from the diffuse ionised gas, and only 6\% from the diffuse neutral gas. In our fiducial model, \rev{more than half} of the [CII] emission originates from the gas defined as a GMC, and \rev{25\%} from the diffuse ionised gas, with the remaining emission originating from the diffuse neutral gas. However, it is important to note that \cite{Olsen17, Olsen18} base their GMCs on the H$_2$ fraction, calculated according to \cite{Krumholz09}, and not directly from the gas density and temperature, as was done in our model. \rev{They also did not directly account for the propagation of radiation via RT, but use an algorithm to find the closest stars, to estimate the radiation field experienced by the gas. As stated in Section~\ref{sec:SM}, this method was unphysical to be applied in a simulation like \Ponos.} 
\rev{In comparison with our MPM model, they have much lower luminosities, in [CII]. The main reason for this is the resolution, as our assumptions, where we treat every particle as an individual GMC, and not as part of a GMC structure, lead to an increased PDR covering fraction. In addition, the non-inclusion of a local-radiation field could lead to an increased [CII] emission. Without global RT the approximation used to estimate the local radiation field by \cite{Olsen17,Olsen18} would overestimate the radiation experienced by the gas in \Ponos, and thus would be unphysical, as mentioned previously. Our results show that when reaching a certain mass and spatial resolution, global RT and a different sub-grid approach are necessary to reproduce the [CII] emission.}


\subsection{Semi-analytical modelling}

\rev{In the sphere of semi-analytical modelling, \cite{Popping19} discussed at length the} strong dependence on the implemented sub-grid cloud density profiles, \rev{and} suggested \rev{that} the Plummer profile leads to line luminosities that best reproduce current observational constraints. 
\rev{In our study, however, we do not find the Plummer profile to be the best at matching observations, which we ascribe to the high resolution of the simulation. Indeed, assuming every \rev{high density particle} to be a GMC, creates an artificially high PDR covering fraction, and thus strong [CII] emission, especially if the radiation field is approximated. Without RT to calculate the radiation field experienced in different parts of the galaxy, the emission will be overestimated for high resolution simulations, and thus lead to high PDR emission.}

\rev{\cite{Pizzati23} find in their semi-analytical modelling, which they applied to ALPINE data, that extended [CII] halos are a natural by-product of starburst-driven outflows. The outflows in their model are launched by SNe, expanding into the CGM with velocities of 200 - 500 km s$^{-1}$, where they first cool down to a few 100 K, before being heated up again by the UVB. This is in agreement with our results, where we find the UVB to be necessary for the extended [CII] emission. \cite{Pizzati23} also state that the ALMA sensitivity threshold for ALPINE observations explains the non-detections of the faint, extended emission.  }

\section{Discussion on future theoretical and observational prospects}\label{Dis}

\subsection{Challenges in modelling CGM emission}

Since the CGM is shaped by galaxy mergers, interactions, and cosmic accretion, as well as by internal feedback processes, only simulations of galaxies within a cosmological context can produce realistic predictions on the CGM properties, which is something idealised disc simulations cannot tackle. As of yet, no cosmological simulation has reached convergence when it comes to resolving the structure of cold gas in the CGM \citep{Scannapieco15, Schneider17, Mandelker18, Mccourt18, Hummels19, Suresh19, VanDeVoort19,Sparre19,Nelson21,Rey23}. A higher resolution leads to more substructures in the cold gas, hence high resolution simulations investigating the survival of cold streams and cold clumps including RT and chemistry (e.g. \citealt{Rey23}), magnetic fields (e.g. \citealt{Ponnada22,Heesen23}) and cosmic rays (e.g. \citealt{Butsky23}) are needed to get a better understanding of CGM properties.

In addition, better-resolved sub-structures will also affect the predicted emission of cold gas tracers, being highly sensitive to gas density, temperature, and ionization state. As an example of this, we find [CI](1-0) emission line to be weaker than [CI](2-1) line for the RT model, which is the opposite trend than in the MPM model: indeed [CI](1-0) traces denser gas than [CI](2-1), which is not resolved in \Ponos, but it is accounted for in analytical sub-grid density profiles. For the same reason, CO emission is  weak in the RT model, since the maximum resolved density (n$_{\rm max}\sim 2 \cdot 10^3$~cm$^{-3}$) is still not enough to allow for CO formation. According to \cite{Seifried20}, in order to have H$_2$ formation convergence and properly account for the H$_2$ non-equilibrium state, a resolution of at least 0.1~pc is needed, which is currently unattainable in cosmological simulations or even idealised disc simulations. Therefore, while RT post-processing seems to be an accurate way to estimate the radiation field in the simulation, multi-phase analytical sub-grid models allow to explore the contribution of dense and otherwise unresolved gas to the emission. In this respect, a promising approach could be to combine the two approaches, as we plan to investigate in a future study. 

Finally, we also point out that two other ingredients could improve the results: (1) accounting for variation of the stellar IMF through cosmic time and (2) including turbulent mixing effects on line emission. Neither the original \Ponos~simulation nor our post-processing take into account the time-dependency and metallicity-dependency of the stellar IMF. This is relevant since some observations have found a deficit in [CII] emission at high redshift (e.g. \citealt{Ota14, Maiolino15, Schaerer15, Pentericci16, Knudsen16,Bradac17}), which could be explained by a different C/O abundance ratio, caused by a top-heavy IMF in the early universe, enriching the ISM by low-metallicity core-collapse SNe \citep{Katz22}. 
While turbulent mixing is included in the SPH simulation, it is not accounted in any of our sub-grid modes (\Cloudy~post-processing \rev{includes some turbulence, but not accurate turbulent mixing on a global scale}). One consequence of this is that the [CI] emission only comes from PDR layers, and not from the molecular gas phase, as found in some observations \citep{Hartigan22, Papadopoulos04} \rev{and ascribed to} turbulent mixing and cosmic rays \citep{Papadopoulos18,Zhang18,Bisbas18}. This could explain why [CI] emission falls off for high values of the density (see Fig. \ref{fig:Hist}). We also do not resolve the very dense gas in \Ponos, which also lead to the very low CO emission. 

Despite these unavoidable limitations, the \Ponos~simulation, \rev{thanks to its high resolution in combination with RT modelling similar to state-of-the-art studies \citep{Pallottini19, Katz19,Lupi2020, Pallottini22, Katz22}, produces reasonable theoretical predictions of sub-mm and FIR line emission from the ISM and CGM of a high redshift galaxy, showing extended emission in filaments and tidal tales in [CII], consistent with observations. Furthermore, comparing multiple modelling recipes within the same simulation has provided insights on the effect of each model on the resulting emission, factoring out differences that would arise when comparing results from different simulation setups.} In the next section we exploit our results to discuss future avenues of observational astronomy.

\subsection{Future observational avenues and relevance to AtLAST}\label{Dis_mocks}

\rev{Detecting and imaging the cold ($T<10^4$~K)} and dense \rev{gas residing in extended CGM reservoirs} is \rev{extremely} challenging at high redshift and impossible for local galaxies, because of the aforementioned limitations of current interferometric facilities such as ALMA, which have a limited sensitivity to large angular scale structures. In this context, AtLAST will revolutionise this field, as it will be the first sub-mm telescope capable of detecting and imaging the cold diffuse and extended (by $\gtrsim50$~kpc) CGM at all redshifts and in multiple gas tracers.
In this section we discuss in more detail how a system such as \Ponos - i.e. a typical star forming galaxy at $z\sim6$ - would be observed by current sub-mm facilities. To simulate sub-mm observations, we need to derive observed fluxes (in units of Jy km s$^{-1}$) from the line luminosity values predicted by our models. We chose to focus on the fiducial RT model. We use the following relation (see e.g. \citealt{Solomon+05}):
\begin{equation}
    L_{\rm Line} [L_{\odot}] = (1.04 \cdot 10^{-3}) \frac{\nu_{\rm rest}}{(1 + z)} D_{L}^{2} \int_{\Delta \nu} S_{\nu}dv,
    \label{eq:Jy}
\end{equation} 
where $\nu_{\rm rest}$ is the rest-frame frequency in units of GHz, and $D_{L}$ is the luminosity distance in units of Mpc. The luminosity distance was calculated with the Cosmology Calculator by \citealt{CosmoCalc} for a flat universe. 

\Ponos~is a small system, and, although it shows FIR/sub-mm line emission extending on scales of \rev{$r\sim10$~kpc or more} due to merging companions, filaments, and other diffuse CGM components \rev{(see Fig.~\ref{fig:EMmaps_CGM}), these physical scales translate to rather compact angular sizes on the sky}.
At $z=6.5$, assuming the same WMAP cosmological parameters as the simulation (see Section~2), which deliver a scale of 5.627 kpc~arcsec$^{-1}$, the virial halo of \Ponos~($2\cdot R_{vir}=42$~kpc) would have a projected angular diameter of $7.46\arcsec$. 
\rev{The following considerations focus on the central 25~kpc region of \Ponos, corresponding to 4.4~arscec on the sky.}

\subsubsection{Spectroscopy: simulated emission line spectra}\label{sec:spectroscopy}

\begin{figure*}[tbh]
\centering
   \includegraphics[clip=true,trim=0.0cm 0cm 0.0cm 0.0cm,scale=.38]{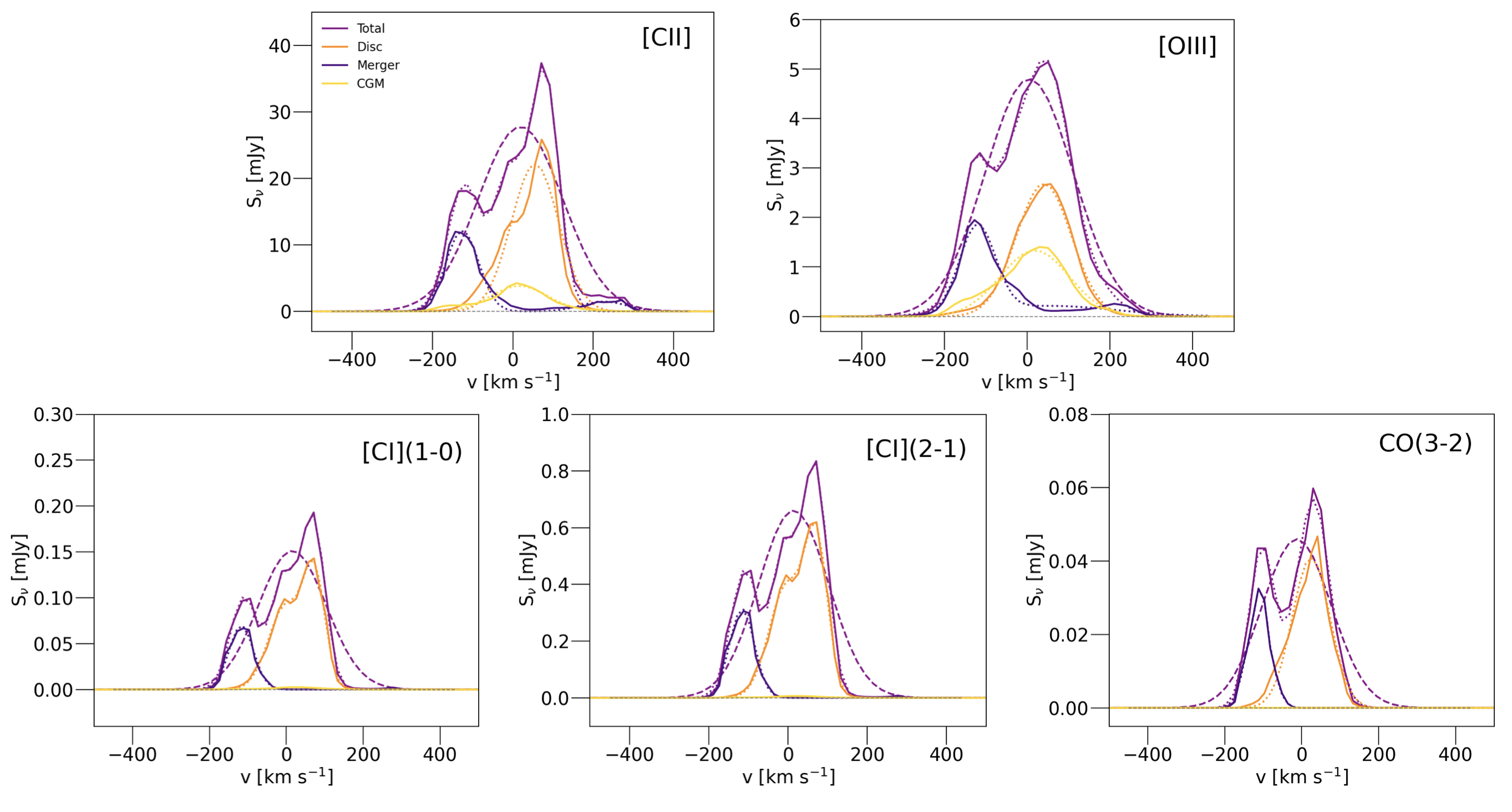}
      \caption{\rev{Simulated spectra of each of the modelled emission lines (fiducial RT model). The total emission is shown in purple, the emission from the central disc in orange, the ISM of the merger components in indigo, and the CGM in yellow. The dotted lines show the multi-component Gaussian line fits performed separately for the spectra of each component. The purple dashed lines show single-Gaussian fits of the total integrated spectra, which we executed with the goal of obtaining an FWHM and peak density value.}} 
      \label{fig:Spec}
\end{figure*}

\rev{Using the projected velocities from the simulation, we generated emission line spectra for each of the transitions modelled, which are reported in Fig.~\ref{fig:Spec}. Here, we show the total spectra (in purple) as well as the spectral decomposition of the different gas components: main disc (in orange), ISM of the merger companions (indigo), and CGM (in yellow).}
\rev{The spectra display clear spectral differences between the different components, in particular for the mergers, which can be easily resolved spectroscopically into two distinct peaks, as they orbit around the main disc of the central galaxy in \Ponos. The CGM shows a broader spectrum than the disc, with clear asymmetries.} 
\rev{The CO line profiles are narrower than the other lines, especially than [CII] and [OIII] which present broader spectra with high-velocity wings due to the mergers and to the CGM components.}

\rev{It is evident from Fig.~\ref{fig:Spec} that, considering the tracers explored in this work and our fiducial RT model, the CGM emission from a system like \Ponos~may be detected only in [CII] and [OIII] emission. To test this hypothesis, we used the ALMA Cycle~10 observing tool and the new AtLAST sensitivity calculator\footnote{\href{https://github.com/ukatc/AtLAST\_sensitivity\_calculator}{https://github.com/ukatc/AtLAST\_sensitivity\_calculator}} to estimate the observing time needed to detect the CGM emission from \Ponos~in these two lines. Through a Gaussian spectral fit (see yellow dotted curves in Fig.~\ref{fig:Spec}), we obtain peak CGM flux densities of 3.8~mJy and 1.34~mJy, with full-width at half maximum (FWHM) values of $\sim230$~km~s$^{-1}$ and $\sim280$~km~s$^{-1}$, respectively for [CII] and [OIII]. Because of its roughly uniform brightness that fills this entire area (see Fig.~\ref{fig:EMmaps_CGM}), to calculate the sensitivity per beam required to detect such a CGM component, we need to scale down the sensitivity required to detect the total CGM component by the number of ALMA (or AtLAST) resolution elements (beams) contained in the imaged region. \cite{Carniani20} dubbed this effects as `surface brightness dimming'.} 

\rev{As a result, we obtain that, to detect the CGM in [CII] at a S/N=10 measured in $\Delta v=50$~km~s$^{-1}$ channels, AtLAST would employ 9.7~hours ($5.1\arcsec$ beam), while ALMA would use 48 hours in the most compact configuration (C-1, $1.4\arcsec$ beam), and $>3550$ days in the C-5 configuration ($0.24\arcsec$ beam). Detecting the CGM of \Ponos~in [OIII]\footnote{\rev{In the case of the [OIII] line, we performed the sensitivity calculations at a central frequency of 460~GHz instead of 452.4~GHz (which would be the exact central frequency for a galaxy at $z=6.5$), to improve a bit the atmospheric transmission}.} would be a much more challenging task even for AtLAST, due to the faintness of this line (compared to [CII]) and to the low atmospheric transmission at its observing frequency. In the case of [OIII], we relax our sensitivity goal to a S/N=3 in $\Delta v=100$~km~s$^{-1}$ channels, and obtain an integration time of $\sim55$~hours with AtLAST ($2.9\arcsec$ beam), and 211 days with ALMA in compact configuration ($0.77\arcsec$ beam).}

\rev{This exercise showed that - despite the rather compact angular extent of \Ponos's CGM (a few arcsec), ALMA would struggle to observe it, even in [CII] which is the brightest available tracer. Instead, AtLAST can easily detect the CGM component in [CII] at high S/N in just a few hours. For the much more challenging  [OIII] line, tracing the ionised gas in the CGM, AtLAST can deliver a low S/N detection in about two days of integration, while this would be an impossible task for ALMA.}

\subsubsection{Imaging: mock maps}\label{sec:disc_imaging}

\begin{figure*}[tbh] 
\centering
   \includegraphics[scale=.39]{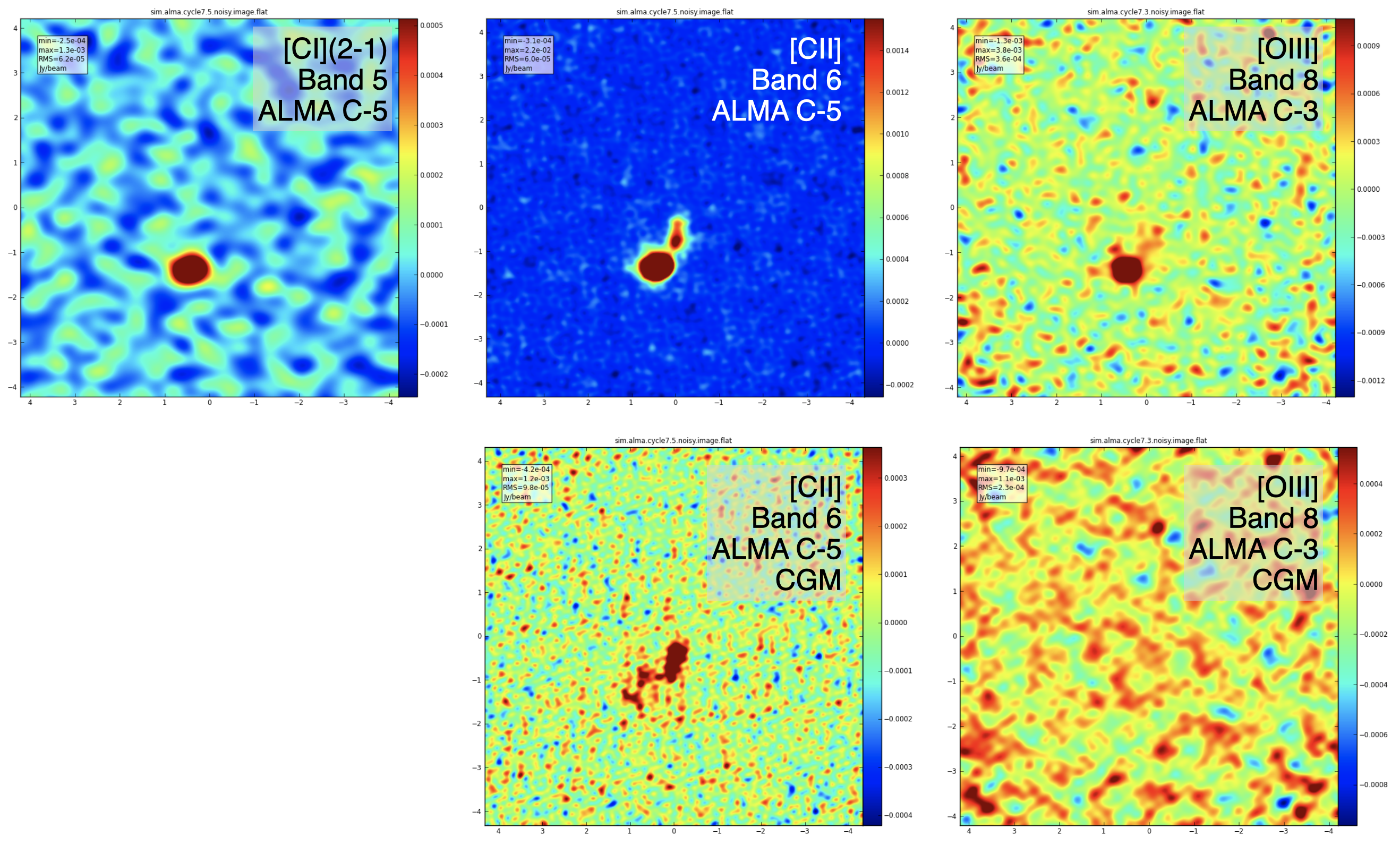}\\
      \caption{Simulated ALMA observations of \Ponos, performed using CASA. The field of view of the input image includes the main disc, and mergers A and B. \rev{Note that the output CASA mocks are flipped and rotated by 180~deg with respect to the maps shown in Fig.~\ref{fig:EMmaps} and Fig.~\ref{fig:EMmaps_CGM}: in these mocks, merger A is to the north and the accreting filament to the south, while the main galaxy disc is still to the left (east) and merger B to the right (west).} 
      {\it Upper left}: ALMA mock map of the total [CI](2-1) line emission, at an observed frequency of 107.91~GHz (Band 5). The simulation uses the ALMA C-5 array configuration ($0.30~\arcsec$ angular resolution, $3.62~\arcsec$ maximum recoverable scale (MRS)). 
      {\it Upper middle:} total [CII] line mock map, at an observed frequency of 253.40~GHz (Band 6). The simulation uses the ALMA C-5 array configuration ($0.24~\arcsec$ angular resolution, $2.91~\arcsec$ MRS). 
      {\it Upper right:} total [OIII] line mock map, at an observed frequency of 460~GHz (Band 8). The simulation uses the ALMA C-3 array configuration ($0.31~\arcsec$ angular resolution, $3.51~\arcsec$ MRS). 
      \rev{The {\it Bottom panels}: show the corresponding [CII] and [OIII] mocks obtained for the CGM component only (see Fig.~\ref{fig:EMmaps_CGM}).}
      \rev{In all CASA simulations, the total integration time was set equal to 10 hours (without overheads), and we simulated one velocity channel of 50~km~s$^{-1}$ centred at $v=0$~km~s$^{-1}$.}}
    \label{fig:CASA_mocks_z6.5}
\end{figure*}

\rev{To image the ISM components of} a system such as \Ponos~at $z=6.5$, \rev{most observers would use intermediate-resolution ALMA configurations ($\sim0.2-0.3\arcsec$ resolution).}
\rev{In Fig.~\ref{fig:CASA_mocks_z6.5} we} display ALMA mock observations of the emission lines observable from ground at this redshift, i.e. [CI](2-1), [CII], and [OIII]\footnote{[CI](1-0) at $z=6.5$ falls on an atmospheric absorption line between Band~2 and Band~3; CO is not included because of its intrinsic faintness in our RT model, as discussed earlier.} \rev{For [CII] and [OIII], we also report ALMA mock observations of only the CGM component of \Ponos~(see Fig.~\ref{fig:EMmaps_CGM}).} \rev{In all mocks, we simulated the central 50~km~s$^{-1}$, a spectral range where the CGM contribution relative to the ISM is maximised. In this velocity range, the ISM contribution arises from the main disc of the \Ponos~system. With a total integration time of 10~hours per line (not including overheads), ALMA can detect well the ISM of the main disc all three transitions, without however revealing any extended halo component. Only the [CII] map (top-central panel) shows an additional extended feature that is not ascribable to the disc, tracing the brightest CGM spot emitting in this transition. This component is however only marginally detected when removing the ISM contribution from the maps (bottom left panel)}. \rev{The CGM is not detected at all in [OIII], despite contributing to $\sim20$\% of the total flux in this tracer. These results are consistent with what discussed in Section~\ref{sec:spectroscopy}. This exercise showed that the detection of extended CGM components does not come for free with deep ALMA observations optimised for imaging the ISM of high-z galaxies, and dedicated efforts are needed.}
\rev{As mentioned earlier, \cite{Pizzati23} find in their semi-analytical model, that all high-z galaxies should have a [CII] halo, which would fall below the ALMA detection limit for lower masses. This can also be seen in our mock-observation, where the extended emission is too faint to be fully detected. }

\section{Conclusions}\label{sec:conclusions}

We presented a state-of-the-art theoretical effort at modelling the emission of several FIR/sub-mm cold and warm gas tracers ([CII], [CI](1-0), [CI](2-1), CO(3-2), [OIII]) in the ISM and CGM of galaxies. We have used the high-resolution (1.5~pc) cosmological zoom-in simulation \Ponos, which represents a typical star forming galaxy system at $z=6.5$, composed of a main disc with $M_*=2\times10^9$~M$_{\odot}$ and SFR=20~M$_{\odot}~yr^{-1}$, and several merging companions. We have explored different modelling approaches based on the photoionisation code \Cloudy, including recipes commonly adopted in the literature. Our fiducial model includes RT post-processing using \kramsesrt, allowing us to include heating by radiation and cooling via different species of H and He. Our main results can be summarised as follows:

\begin{itemize}
    \item The [CII] line is the least sensitive to the underlying modelling approach, as it is by nature multi-phase, and can be found co-spatially with ionised, neutral and molecular gas, due to the low excitation energy of carbon. As it does not trace the densest gas phases, the emission can be modelled without additional sub-grid models; even the simplest modelling approach (denoted simple model, SM) yields reasonable results that are consistent with more sophisticated approaches. \rev{In our modelling we found [CII] to be sensitive to the PDR covering fraction. } 
    \item Radiative transfer (RT) calculations are needed, as the gas emission is dependant on the underlying radiation field the gas experiences, and to account for the cooling and self-shielding that can occur; this can especially be seen for [OIII], where the RT process allows for heating and cooling, thus lowering the [OIII] emission compared to other models.
    \item The [CI](1-0) and CO(3-2) lines are very sensitive to post-processing recipes as \Ponos, like any other cosmological simulation, cannot resolve the scales relevant for $H_2$ formation ($<0.1$~pc). For the CO(3-2) line, only the multi-phase model (MPM) approach, which includes ad-hoc sub-resolution treatments of dense gas, produces reasonable results. Future efforts should  explore combining RT calculations with multi-phase analytical sub-grid models to explore the contribution of dense, otherwise unresolved, gas.
    \item  \rev{More than half} of [CII] in our fiducial model \rev{($\sim$55\%)} originates from gas at $T \leqslant 10^{4} $~K and n$_{H} \geqslant 10 $~cm$^{-3}$. Our fiducial RT model produces a lower density, $T \sim 10^{4} $~K tail of [CII] emission that is not seen in the other more simplistic models and that resides entirely in the CGM. \rev{This gas is ionised by the UV background.}
    \item Both [CII] and [OIII] have a non-negligible contribution from the CGM (\rev{10\%} and \rev{21\%} respectively in the fiducial RT model), but trace different components of the CGM. In particular, [CII] is brighter in the accreting filaments and in the merging companions and connected tidal features. The CGM contribution to [OIII] resides instead in puffy halo surrounding the main galaxy disc, ionised by newly formed stars and possibly linked to supernova feedback.
    \item \rev{In our study of the CGM emission, no gas tracer can be exclusively connected to in- or outflowing gas, showing that the interpretation of the origin of the extended emission is not straight forward. }
\end{itemize}

These results are highly relevant to the interpretation of current and future sub-millimeter observations, with ALMA and with the planned AtLAST. In particular, observations need to take into account that a significant ($\sim10\%$ and above) portion of the emission from typically employed sub-mm and FIR lines such as [CII] and [OIII] arises in the CGM component and not in galaxies' discs. This contribution increases if the angular resolution of the observations is insufficient for deblending the discs of merging satellites from the diffuse CGM emission. 
\rev{We have shown, that even for a relatively compact system such as \Ponos~at $z\sim6.5$ whose entire halo is contained within a $<10''$~angular region, imaging extended ISM and CGM components is extremely challenging with ALMA in [CII] and unfeasible in [OIII]. This is due to the diffuse and low surface brightness nature of these components.} For this type of experiments, it will be crucial to build AtLAST: a new large-aperture, large-FoV sub-mm single-dish telescope, in a site with excellent atmospheric transmission at sub-millimeter wavelengths as to enable multi-tracer observations of the ISM and CGM of galaxies. 


\begin{acknowledgements}
This project has received funding from the European Union’s Horizon 2020 research and innovation program under grant agreement No 951815 (AtLAST). D. Decataldo, S. Shen and B.Baumschlager acknowledge the support from the Research Council of Norway through NFR Young Research Talents Grant 276043. LDM is supported by the ERC-StG ``ClustersXCosmo'' grant agreement 716762 and acknowledges financial contribution from the agreement ASI-INAF n.2017-14-H.0. 
\revv{We thank the anonymous referee for providing insightful and constructive comments that significantly helped us improve the analysis and the overall presentation of the results.}
We thank Gerg{\"o} Popping for his helpful input during the early stages of this study. We also thank Carlos De Breuck for helping us navigate the observational results in literature. 
We acknowledge usage of the Python programming language \citep{van1995python,python3}, Matplotlib \citep{Hunter07}, NumPy \citep{Walt2011}, Pynbody \citep{Pynbody}, of the CASA simulator \citep{CASA} \revv{and of the AtLAST sensitivity calculator}.
\end{acknowledgements}

%
%
\bibliography{Ponos_text}

\begin{thebibliography}{191}
\expandafter\ifx\csname natexlab\endcsname\relax\def\natexlab#1{#1}\fi

\bibitem[{{Agertz} {et~al.}(2007){Agertz}, {Moore}, {Stadel}, {Potter},
  {Miniati}, {Read}, {Mayer}, {Gawryszczak}, {Kravtsov}, {Nordlund}, {Pearce},
  {Quilis}, {Rudd}, {Springel}, {Stone}, {Tasker}, {Teyssier}, {Wadsley}, \&
  {Walder}}]{Agertz07}
{Agertz}, O., {Moore}, B., {Stadel}, J., {et~al.} 2007, \mnras, 380, 963

\bibitem[{{Algera} {et~al.}(2023){Algera}, {Inami}, {Sommovigo}, {Fudamoto},
  {Schneider}, {Graziani}, {Dayal}, {Bouwens}, {Aravena}, {da Cunha},
  {Ferrara}, {Hygate}, {van Leeuwen}, {De Looze}, {Palla}, {Pallottini},
  {Smit}, {Stefanon}, {Topping}, \& {van der Werf}}]{Algera23}
{Algera}, H., {Inami}, H., {Sommovigo}, L., {et~al.} 2023, arXiv e-prints,
  arXiv:2301.09659

\bibitem[{{Arata} {et~al.}(2020){Arata}, {Yajima}, {Nagamine}, {Abe}, \&
  {Khochfar}}]{Arata20}
{Arata}, S., {Yajima}, H., {Nagamine}, K., {Abe}, M., \& {Khochfar}, S. 2020,
  \mnras, 498, 5541

\bibitem[{{Arzoumanian} {et~al.}(2011){Arzoumanian}, {Andr{\'e}}, {Didelon},
  {K{\"o}nyves}, {Schneider}, {Men'shchikov}, {Sousbie}, {Zavagno}, {Bontemps},
  {di Francesco}, {Griffin}, {Hennemann}, {Hill}, {Kirk}, {Martin}, {Minier},
  {Molinari}, {Motte}, {Peretto}, {Pezzuto}, {Spinoglio}, {Ward-Thompson},
  {White}, \& {Wilson}}]{Arzoumanian+11}
{Arzoumanian}, D., {Andr{\'e}}, P., {Didelon}, P., {et~al.} 2011, \aap, 529, L6

\bibitem[{{Asplund} {et~al.}(2009){Asplund}, {Grevesse}, {Sauval}, \&
  {Scott}}]{Asplund09}
{Asplund}, M., {Grevesse}, N., {Sauval}, A.~J., \& {Scott}, P. 2009, \araa, 47,
  481

\bibitem[{{Bergin} {et~al.}(1997){Bergin}, {Goldsmith}, {Snell}, \&
  {Langer}}]{Bergin97}
{Bergin}, E.~A., {Goldsmith}, P.~F., {Snell}, R.~L., \& {Langer}, W.~D. 1997,
  \apj, 482, 285

\bibitem[{{Bisbas} {et~al.}(2018){Bisbas}, {Tan}, {Csengeri}, {Wu}, {Lim},
  {Caselli}, {G{\"u}sten}, {Ricken}, \& {Riquelme}}]{Bisbas18}
{Bisbas}, T.~G., {Tan}, J.~C., {Csengeri}, T., {et~al.} 2018, \mnras, 478, L54

\bibitem[{{Bisbas} {et~al.}(2023){Bisbas}, {van Dishoeck}, {Hu}, \&
  {Schruba}}]{Bisbas23}
{Bisbas}, T.~G., {van Dishoeck}, E.~F., {Hu}, C.-Y., \& {Schruba}, A. 2023,
  \mnras, 519, 729

\bibitem[{{Bischetti} {et~al.}(2019){Bischetti}, {Maiolino}, {Carniani},
  {Fiore}, {Piconcelli}, \& {Fluetsch}}]{Bischetti+19}
{Bischetti}, M., {Maiolino}, R., {Carniani}, S., {et~al.} 2019, \aap, 630, A59

\bibitem[{{Boogaard} {et~al.}(2023){Boogaard}, {Decarli}, {Walter}, {Wei{\ss}},
  {Popping}, {Neri}, {Aravena}, {Riechers}, {Ellis}, {Carilli}, {Cox}, \&
  {Pety}}]{Boogaard23}
{Boogaard}, L.~A., {Decarli}, R., {Walter}, F., {et~al.} 2023, \apj, 945, 111

\bibitem[{{Bothwell} {et~al.}(2013){Bothwell}, {Smail}, {Chapman}, {Genzel},
  {Ivison}, {Tacconi}, {Alaghband-Zadeh}, {Bertoldi}, {Blain}, {Casey}, {Cox},
  {Greve}, {Lutz}, {Neri}, {Omont}, \& {Swinbank}}]{Bothwell13}
{Bothwell}, M.~S., {Smail}, I., {Chapman}, S.~C., {et~al.} 2013, \mnras, 429,
  3047

\bibitem[{{Bowen} {et~al.}(2016){Bowen}, {Chelouche}, {Jenkins}, {Tripp},
  {Pettini}, {York}, \& {Frye}}]{Bowen16}
{Bowen}, D.~V., {Chelouche}, D., {Jenkins}, E.~B., {et~al.} 2016, \apj, 826, 50

\bibitem[{{Brada{\v{c}}} {et~al.}(2017){Brada{\v{c}}}, {Garcia-Appadoo},
  {Huang}, {Vallini}, {Quinn Finney}, {Hoag}, {Lemaux}, {Borello Schmidt},
  {Treu}, {Carilli}, {Dijkstra}, {Ferrara}, {Fontana}, {Jones}, {Ryan}, {Wagg},
  \& {Gonzalez}}]{Bradac17}
{Brada{\v{c}}}, M., {Garcia-Appadoo}, D., {Huang}, K.-H., {et~al.} 2017, \apjl,
  836, L2

\bibitem[{{Brauher} {et~al.}(2008){Brauher}, {Dale}, \& {Helou}}]{Brauher08}
{Brauher}, J.~R., {Dale}, D.~A., \& {Helou}, G. 2008, \apjs, 178, 280

\bibitem[{{Butsky} {et~al.}(2023){Butsky}, {Nakum}, {Ponnada}, {Hummels}, {Ji},
  \& {Hopkins}}]{Butsky23}
{Butsky}, I.~S., {Nakum}, S., {Ponnada}, S.~B., {et~al.} 2023, \mnras
  [\eprint[arXiv]{2210.14232}]

\bibitem[{{Capak} {et~al.}(2015){Capak}, {Carilli}, {Jones}, {Casey},
  {Riechers}, {Sheth}, {Carollo}, {Ilbert}, {Karim}, {Lefevre}, {Lilly},
  {Scoville}, {Smolcic}, \& {Yan}}]{Capak15}
{Capak}, P.~L., {Carilli}, C., {Jones}, G., {et~al.} 2015, \nat, 522, 455

\bibitem[{{Carilli} \& {Walter}(2013)}]{Carilli13}
{Carilli}, C.~L. \& {Walter}, F. 2013, \araa, 51, 105

\bibitem[{{Carniani} {et~al.}(2020){Carniani}, {Ferrara}, {Maiolino},
  {Castellano}, {Gallerani}, {Fontana}, {Kohandel}, {Lupi}, {Pallottini},
  {Pentericci}, {Vallini}, \& {Vanzella}}]{Carniani20}
{Carniani}, S., {Ferrara}, A., {Maiolino}, R., {et~al.} 2020, \mnras, 499, 5136

\bibitem[{{Carniani} {et~al.}(2018){Carniani}, {Maiolino}, {Amorin},
  {Pentericci}, {Pallottini}, {Ferrara}, {Willott}, {Smit}, {Matthee},
  {Sobral}, {Santini}, {Castellano}, {De Barros}, {Fontana}, {Grazian}, \&
  {Guaita}}]{Carniani18}
{Carniani}, S., {Maiolino}, R., {Amorin}, R., {et~al.} 2018, \mnras, 478, 1170

\bibitem[{{CASA Team} {et~al.}(2022){CASA Team}, {Bean}, {Bhatnagar}, {Castro},
  {Donovan Meyer}, {Emonts}, {Garcia}, {Garwood}, {Golap}, {Gonzalez Villalba},
  {Harris}, {Hayashi}, {Hoskins}, {Hsieh}, {Jagannathan}, {Kawasaki},
  {Keimpema}, {Kettenis}, {Lopez}, {Marvil}, {Masters}, {McNichols},
  {Mehringer}, {Miel}, {Moellenbrock}, {Montesino}, {Nakazato}, {Ott}, {Petry},
  {Pokorny}, {Raba}, {Rau}, {Schiebel}, {Schweighart}, {Sekhar}, {Shimada},
  {Small}, {Steeb}, {Sugimoto}, {Suoranta}, {Tsutsumi}, {van Bemmel},
  {Verkouter}, {Wells}, {Xiong}, {Szomoru}, {Griffith}, {Glendenning}, \&
  {Kern}}]{CASA}
{CASA Team}, {Bean}, B., {Bhatnagar}, S., {et~al.} 2022, \pasp, 134, 114501

\bibitem[{{Chabrier}(2003)}]{Chabrier03}
{Chabrier}, G. 2003, \apjl, 586, L133

\bibitem[{Chavanis \& Sire(2006)}]{Logotrope}
Chavanis, P.-H. \& Sire, C. 2006, Physica A: Statistical Mechanics and its
  Applications, 375

\bibitem[{{Cicone} {et~al.}(2019){Cicone}, {De Breuck}, {Chen}, {van Kampen},
  {Narayanan}, {Mroczkowski}, {Andreani}, {Klaassen}, {Weiss}, {Kohno},
  {Kauffmann}, {Wagg}, {Riechers}, {Gullberg}, {Geach}, {Shen}, {Hill}, \&
  {Brownson}}]{Cicone19}
{Cicone}, C., {De Breuck}, C., {Chen}, C.-C., {et~al.} 2019, \baas, 51, 82

\bibitem[{{Cicone} {et~al.}(2021){Cicone}, {Mainieri}, {Circosta}, {Kakkad},
  {Vietri}, {Perna}, {Bischetti}, {Carniani}, {Cresci}, {Harrison}, {Mannucci},
  {Marconi}, {Piconcelli}, {Puglisi}, {Scholtz}, {Vignali}, {Zamorani},
  {Zappacosta}, \& {Arrigoni Battaia}}]{Cicone21}
{Cicone}, C., {Mainieri}, V., {Circosta}, C., {et~al.} 2021, \aap, 654, L8

\bibitem[{{Cicone} {et~al.}(2015){Cicone}, {Maiolino}, {Gallerani}, {Neri},
  {Ferrara}, {Sturm}, {Fiore}, {Piconcelli}, \& {Feruglio}}]{Cicone15}
{Cicone}, C., {Maiolino}, R., {Gallerani}, S., {et~al.} 2015, \aap, 574, A14

\bibitem[{{Circosta} {et~al.}(2021){Circosta}, {Mainieri}, {Lamperti},
  {Padovani}, {Bischetti}, {Harrison}, {Kakkad}, {Zanella}, {Vietri},
  {Lanzuisi}, {Salvato}, {Brusa}, {Carniani}, {Cicone}, {Cresci}, {Feruglio},
  {Husemann}, {Mannucci}, {Marconi}, {Perna}, {Piconcelli}, {Puglisi},
  {Saintonge}, {Schramm}, {Vignali}, \& {Zappacosta}}]{Circosta21}
{Circosta}, C., {Mainieri}, V., {Lamperti}, I., {et~al.} 2021, \aap, 646, A96

\bibitem[{{Cooper} {et~al.}(2019){Cooper}, {Simcoe}, {Cooksey}, {Bordoloi},
  {Miller}, {Furesz}, {Turner}, \& {Ba{\~n}ados}}]{Cooper19}
{Cooper}, T.~J., {Simcoe}, R.~A., {Cooksey}, K.~L., {et~al.} 2019, \apj, 882,
  77

\bibitem[{{Cormier} {et~al.}(2015){Cormier}, {Madden}, {Lebouteiller}, {Abel},
  {Hony}, {Galliano}, {R{\'e}my-Ruyer}, {Bigiel}, {Baes}, {Boselli},
  {Chevance}, {Cooray}, {De Looze}, {Doublier}, {Galametz}, {Hughes},
  {Karczewski}, {Lee}, {Lu}, \& {Spinoglio}}]{Cormier15}
{Cormier}, D., {Madden}, S.~C., {Lebouteiller}, V., {et~al.} 2015, \aap, 578,
  A53

\bibitem[{{da Cunha} {et~al.}(2013){da Cunha}, {Groves}, {Walter}, {Decarli},
  {Weiss}, {Bertoldi}, {Carilli}, {Daddi}, {Elbaz}, {Ivison}, {Maiolino},
  {Riechers}, {Rix}, {Sargent}, \& {Smail}}]{dacunha13}
{da Cunha}, E., {Groves}, B., {Walter}, F., {et~al.} 2013, \apj, 766, 13

\bibitem[{{Daddi} {et~al.}(2015){Daddi}, {Dannerbauer}, {Liu}, {Aravena},
  {Bournaud}, {Walter}, {Riechers}, {Magdis}, {Sargent}, {B{\'e}thermin},
  {Carilli}, {Cibinel}, {Dickinson}, {Elbaz}, {Gao}, {Gobat}, {Hodge}, \&
  {Krips}}]{Daddi15}
{Daddi}, E., {Dannerbauer}, H., {Liu}, D., {et~al.} 2015, \aap, 577, A46

\bibitem[{{Dav{\'e}} {et~al.}(2017){Dav{\'e}}, {Rafieferantsoa}, {Thompson}, \&
  {Hopkins}}]{Dave17}
{Dav{\'e}}, R., {Rafieferantsoa}, M.~H., {Thompson}, R.~J., \& {Hopkins}, P.~F.
  2017, \mnras, 467, 115

\bibitem[{{Dav{\'e}} {et~al.}(2016){Dav{\'e}}, {Thompson}, \&
  {Hopkins}}]{Dave16}
{Dav{\'e}}, R., {Thompson}, R., \& {Hopkins}, P.~F. 2016, \mnras, 462, 3265

\bibitem[{{De Breuck} {et~al.}(2022){De Breuck}, {Lundgren}, {Emonts}, {Kolwa},
  {Dannerbauer}, \& {Lehnert}}]{DeBreuck22}
{De Breuck}, C., {Lundgren}, A., {Emonts}, B., {et~al.} 2022, \aap, 658, L2

\bibitem[{{De Looze} {et~al.}(2014){De Looze}, {Cormier}, {Lebouteiller},
  {Madden}, {Baes}, {Bendo}, {Boquien}, {Boselli}, {Clements}, {Cortese},
  {Cooray}, {Galametz}, {Galliano}, {Graci{\'a}-Carpio}, {Isaak}, {Karczewski},
  {Parkin}, {Pellegrini}, {R{\'e}my-Ruyer}, {Spinoglio}, {Smith}, \&
  {Sturm}}]{DL14}
{De Looze}, I., {Cormier}, D., {Lebouteiller}, V., {et~al.} 2014, \aap, 568,
  A62

\bibitem[{{Decarli} {et~al.}(2020){Decarli}, {Aravena}, {Boogaard}, {Carilli},
  {Gonz{\'a}lez-L{\'o}pez}, {Walter}, {Cortes}, {Cox}, {da Cunha}, {Daddi},
  {D{\'\i}az-Santos}, {Hodge}, {Inami}, {Neeleman}, {Novak}, {Oesch},
  {Popping}, {Riechers}, {Smail}, {Uzgil}, {van der Werf}, {Wagg}, \&
  {Weiss}}]{Decarli20}
{Decarli}, R., {Aravena}, M., {Boogaard}, L., {et~al.} 2020, \apj, 902, 110

\bibitem[{{Decarli} {et~al.}(2017){Decarli}, {Walter}, {Venemans},
  {Ba{\~n}ados}, {Bertoldi}, {Carilli}, {Fan}, {Farina}, {Mazzucchelli},
  {Riechers}, {Rix}, {Strauss}, {Wang}, \& {Yang}}]{Decarli17}
{Decarli}, R., {Walter}, F., {Venemans}, B.~P., {et~al.} 2017, \nat, 545, 457

\bibitem[{{Decataldo} {et~al.}(2020){Decataldo}, {Lupi}, {Ferrara},
  {Pallottini}, \& {Fumagalli}}]{Decataldo20}
{Decataldo}, D., {Lupi}, A., {Ferrara}, A., {Pallottini}, A., \& {Fumagalli},
  M. 2020, \mnras, 497, 4718

\bibitem[{{Decataldo} {et~al.}(2019){Decataldo}, {Pallottini}, {Ferrara},
  {Vallini}, \& {Gallerani}}]{Decataldo19}
{Decataldo}, D., {Pallottini}, A., {Ferrara}, A., {Vallini}, L., \&
  {Gallerani}, S. 2019, \mnras, 487, 3377

\bibitem[{{Decataldo} {et~al.}(2023){Decataldo}, {Shen}, {Mayer},
  {Baumschlager}, \& {Madau}}]{Decataldo23}
{Decataldo}, D., {Shen}, S., {Mayer}, L., {Baumschlager}, B., \& {Madau}, P.
  2023, arXiv e-prints, arXiv:2306.03146

\bibitem[{{Dekel} {et~al.}(2009){Dekel}, {Birnboim}, {Engel}, {Freundlich},
  {Goerdt}, {Mumcuoglu}, {Neistein}, {Pichon}, {Teyssier}, \&
  {Zinger}}]{Dekel09}
{Dekel}, A., {Birnboim}, Y., {Engel}, G., {et~al.} 2009, \nat, 457, 451

\bibitem[{{D{\'\i}az-Santos} {et~al.}(2017){D{\'\i}az-Santos}, {Armus},
  {Charmandaris}, {Lu}, {Stierwalt}, {Stacey}, {Malhotra}, {van der Werf},
  {Howell}, {Privon}, {Mazzarella}, {Goldsmith}, {Murphy}, {Barcos-Mu{\~n}oz},
  {Linden}, {Inami}, {Larson}, {Evans}, {Appleton}, {Iwasawa}, {Lord},
  {Sanders}, \& {Surace}}]{Santos17}
{D{\'\i}az-Santos}, T., {Armus}, L., {Charmandaris}, V., {et~al.} 2017, \apj,
  846, 32

\bibitem[{{Dinerstein} {et~al.}(1985){Dinerstein}, {Lester}, \&
  {Werner}}]{Dinerstein85}
{Dinerstein}, H.~L., {Lester}, D.~F., \& {Werner}, M.~W. 1985, \apj, 291, 561

\bibitem[{{Dunne} {et~al.}(2022){Dunne}, {Maddox}, {Papadopoulos}, {Ivison}, \&
  {Gomez}}]{Dunne22}
{Dunne}, L., {Maddox}, S.~J., {Papadopoulos}, P.~P., {Ivison}, R.~J., \&
  {Gomez}, H.~L. 2022, arXiv e-prints, arXiv:2208.01622

\bibitem[{{Dutta} {et~al.}(2020){Dutta}, {Fumagalli}, {Fossati}, {Lofthouse},
  {Prochaska}, {Arrigoni Battaia}, {Bielby}, {Cantalupo}, {Cooke}, {Murphy}, \&
  {O'Meara}}]{Dutta20}
{Dutta}, R., {Fumagalli}, M., {Fossati}, M., {et~al.} 2020, \mnras, 499, 5022

\bibitem[{{Emonts} {et~al.}(2016){Emonts}, {Lehnert}, {Villar-Mart{\'\i}n},
  {Norris}, {Ekers}, {van Moorsel}, {Dannerbauer}, {Pentericci}, {Miley},
  {Allison}, {Sadler}, {Guillard}, {Carilli}, {Mao}, {R{\"o}ttgering}, {De
  Breuck}, {Seymour}, {Gullberg}, {Ceverino}, {Jagannathan}, {Vernet}, \&
  {Indermuehle}}]{Emonts16}
{Emonts}, B.~H.~C., {Lehnert}, M.~D., {Villar-Mart{\'\i}n}, M., {et~al.} 2016,
  Science, 354, 1128

\bibitem[{{Emonts} {et~al.}(2023){Emonts}, {Lehnert}, {Yoon}, {Mandelker},
  {Villar-Martin}, {Miley}, {De Breuck}, {Perez-Torres}, {Hatch}, \&
  {Guillard}}]{Emont23}
{Emonts}, B. H.~C., {Lehnert}, M.~D., {Yoon}, I., {et~al.} 2023, arXiv
  e-prints, arXiv:2303.17484

\bibitem[{{Ferkinhoff} {et~al.}(2010){Ferkinhoff}, {Hailey-Dunsheath},
  {Nikola}, {Parshley}, {Stacey}, {Benford}, \& {Staguhn}}]{Ferkinhoff10}
{Ferkinhoff}, C., {Hailey-Dunsheath}, S., {Nikola}, T., {et~al.} 2010, \apjl,
  714, L147

\bibitem[{{Ferland} {et~al.}(2017){Ferland}, {Chatzikos}, {Guzm{\'a}n},
  {Lykins}, {van Hoof}, {Williams}, {Abel}, {Badnell}, {Keenan}, {Porter}, \&
  {Stancil}}]{Cloudy}
{Ferland}, G.~J., {Chatzikos}, M., {Guzm{\'a}n}, F., {et~al.} 2017, \rmxaa, 53,
  385

\bibitem[{{Feruglio} {et~al.}(2010){Feruglio}, {Maiolino}, {Piconcelli},
  {Menci}, {Aussel}, {Lamastra}, \& {Fiore}}]{Feruglio10}
{Feruglio}, C., {Maiolino}, R., {Piconcelli}, E., {et~al.} 2010, \aap, 518,
  L155

\bibitem[{{Fiacconi} {et~al.}(2017){Fiacconi}, {Mayer}, {Madau}, {Lupi},
  {Dotti}, \& {Haardt}}]{Fiacconi17}
{Fiacconi}, D., {Mayer}, L., {Madau}, P., {et~al.} 2017, \mnras, 467, 4080

\bibitem[{{Field} {et~al.}(2011){Field}, {Blackman}, \& {Keto}}]{Field11}
{Field}, G.~B., {Blackman}, E.~G., \& {Keto}, E.~R. 2011, \mnras, 416, 710

\bibitem[{{Fischer} {et~al.}(2010){Fischer}, {Sturm}, {Gonz{\'a}lez-Alfonso},
  {Graci{\'a}-Carpio}, {Hailey-Dunsheath}, {Poglitsch}, {Contursi}, {Lutz},
  {Genzel}, {Sternberg}, {Verma}, \& {Tacconi}}]{Fischer10}
{Fischer}, J., {Sturm}, E., {Gonz{\'a}lez-Alfonso}, E., {et~al.} 2010, \aap,
  518, L41

\bibitem[{{Fudamoto} {et~al.}(2022){Fudamoto}, {Smit}, {Bowler}, {Oesch},
  {Bouwens}, {Stefanon}, {Inami}, {Endsley}, {Gonzalez}, {Schouws}, {Stark},
  {Algera}, {Aravena}, {Barrufet}, {da Cunha}, {Dayal}, {Ferrara}, {Graziani},
  {Hodge}, {Hygate}, {Inoue}, {Nanayakkara}, {Pallottini}, {Pizzati},
  {Schneider}, {Sommovigo}, {Sugahara}, {Topping}, {van der Werf}, {Bethermin},
  {Cassata}, {Dessauges-Zavadsky}, {Ibar}, {Faisst}, {Fujimoto}, {Ginolfi},
  {Hathi}, {Jones}, {Pozzi}, \& {Schaerer}}]{Fudamoto22}
{Fudamoto}, Y., {Smit}, R., {Bowler}, R.~A.~A., {et~al.} 2022, \apj, 934, 144

\bibitem[{{Fujimoto} {et~al.}(2021){Fujimoto}, {Oguri}, {Brammer}, {Yoshimura},
  {Laporte}, {Gonz{\'a}lez-L{\'o}pez}, {Caminha}, {Kohno}, {Zitrin}, {Richard},
  {Ouchi}, {Bauer}, {Smail}, {Hatsukade}, {Ono}, {Kokorev}, {Umehata},
  {Schaerer}, {Knudsen}, {Sun}, {Magdis}, {Valentino}, {Ao}, {Toft},
  {Dessauges-Zavadsky}, {Shimasaku}, {Caputi}, {Kusakabe}, {Morokuma-Matsui},
  {Shotaro}, {Egami}, {Lee}, {Rawle}, \& {Espada}}]{Fujimoto21}
{Fujimoto}, S., {Oguri}, M., {Brammer}, G., {et~al.} 2021, \apj, 911, 99

\bibitem[{{Fujimoto} {et~al.}(2019){Fujimoto}, {Ouchi}, {Ferrara},
  {Pallottini}, {Ivison}, {Behrens}, {Gallerani}, {Arata}, {Yajima}, \&
  {Nagamine}}]{Fujimoto19}
{Fujimoto}, S., {Ouchi}, M., {Ferrara}, A., {et~al.} 2019, \apj, 887, 107

\bibitem[{{Fujimoto} {et~al.}(2022){Fujimoto}, {Ouchi}, {Nakajima}, {Harikane},
  {Isobe}, {Brammer}, {Oguri}, {Gim{\'e}nez-Arteaga}, {Heintz}, {Kokorev},
  {Bauer}, {Ferrara}, {Kojima}, {Lagos}, {Laura}, {Schaerer}, {Shimasaku},
  {Hatsukade}, {Kohno}, {Sun}, {Valentino}, {Watson}, {Fudamoto}, {Inoue},
  {Gonz{\'a}lez-L{\'o}pez}, {Koekemoer}, {Knudsen}, {Lee}, {Magdis}, {Richard},
  {Strait}, {Sugahara}, {Tamura}, {Toft}, {Umehata}, \& {Walth}}]{Fujimoto22}
{Fujimoto}, S., {Ouchi}, M., {Nakajima}, K., {et~al.} 2022, arXiv e-prints,
  arXiv:2212.06863

\bibitem[{{Fujimoto} {et~al.}(2020){Fujimoto}, {Silverman}, {Bethermin},
  {Ginolfi}, {Jones}, {Le F{\`e}vre}, {Dessauges-Zavadsky}, {Rujopakarn},
  {Faisst}, {Fudamoto}, {Cassata}, {Morselli}, {Maiolino}, {Schaerer}, {Capak},
  {Yan}, {Vallini}, {Toft}, {Loiacono}, {Zamorani}, {Talia}, {Narayanan},
  {Hathi}, {Lemaux}, {Boquien}, {Amorin}, {Ibar}, {Koekemoer},
  {M{\'e}ndez-Hern{\'a}ndez}, {Bardelli}, {Vergani}, {Zucca}, {Romano}, \&
  {Cimatti}}]{Fujimoto20}
{Fujimoto}, S., {Silverman}, J.~D., {Bethermin}, M., {et~al.} 2020, \apj, 900,
  1

\bibitem[{{Gaia Collaboration} {et~al.}(2018){Gaia Collaboration}, {Brown},
  {Vallenari}, {Prusti}, {de Bruijne}, {Babusiaux}, {Bailer-Jones}, {Biermann},
  {Evans}, {Eyer}, {Jansen}, {Jordi}, {Klioner}, {Lammers}, {Lindegren},
  {Luri}, {Mignard}, {Panem}, {Pourbaix}, {Randich}, {Sartoretti}, {Siddiqui},
  {Soubiran}, {van Leeuwen}, {Walton}, {Arenou}, {Bastian}, {Cropper},
  {Drimmel}, {Katz}, {Lattanzi}, {Bakker}, {Cacciari}, {Casta{\~n}eda},
  {Chaoul}, {Cheek}, {De Angeli}, {Fabricius}, {Guerra}, {Holl}, {Masana},
  {Messineo}, {Mowlavi}, {Nienartowicz}, {Panuzzo}, {Portell}, {Riello},
  {Seabroke}, {Tanga}, {Th{\'e}venin}, {Gracia-Abril}, {Comoretto},
  {Garcia-Reinaldos}, {Teyssier}, {Altmann}, {Andrae}, {Audard},
  {Bellas-Velidis}, {Benson}, {Berthier}, {Blomme}, {Burgess}, {Busso},
  {Carry}, {Cellino}, {Clementini}, {Clotet}, {Creevey}, {Davidson}, {De
  Ridder}, {Delchambre}, {Dell'Oro}, {Ducourant},
  {Fern{\'a}ndez-Hern{\'a}ndez}, {Fouesneau}, {Fr{\'e}mat}, {Galluccio},
  {Garc{\'\i}a-Torres}, {Gonz{\'a}lez-N{\'u}{\~n}ez}, {Gonz{\'a}lez-Vidal},
  {Gosset}, {Guy}, {Halbwachs}, {Hambly}, {Harrison}, {Hern{\'a}ndez},
  {Hestroffer}, {Hodgkin}, {Hutton}, {Jasniewicz}, {Jean-Antoine-Piccolo},
  {Jordan}, {Korn}, {Krone-Martins}, {Lanzafame}, {Lebzelter}, {L{\"o}ffler},
  {Manteiga}, {Marrese}, {Mart{\'\i}n-Fleitas}, {Moitinho}, {Mora}, {Muinonen},
  {Osinde}, {Pancino}, {Pauwels}, {Petit}, {Recio-Blanco}, {Richards},
  {Rimoldini}, {Robin}, {Sarro}, {Siopis}, {Smith}, {Sozzetti}, {S{\"u}veges},
  {Torra}, {van Reeven}, {Abbas}, {Abreu Aramburu}, {Accart}, {Aerts},
  {Altavilla}, {{\'A}lvarez}, {Alvarez}, {Alves}, {Anderson}, {Andrei},
  {Anglada Varela}, {Antiche}, {Antoja}, {Arcay}, {Astraatmadja}, {Bach},
  {Baker}, {Balaguer-N{\'u}{\~n}ez}, {Balm}, {Barache}, {Barata}, {Barbato},
  {Barblan}, {Barklem}, {Barrado}, {Barros}, {Barstow}, {Bartholom{\'e}
  Mu{\~n}oz}, {Bassilana}, {Becciani}, {Bellazzini}, {Berihuete}, {Bertone},
  {Bianchi}, {Bienaym{\'e}}, {Blanco-Cuaresma}, {Boch}, {Boeche}, {Bombrun},
  {Borrachero}, {Bossini}, {Bouquillon}, {Bourda}, {Bragaglia}, {Bramante},
  {Breddels}, {Bressan}, {Brouillet}, {Br{\"u}semeister}, {Brugaletta},
  {Bucciarelli}, {Burlacu}, {Busonero}, {Butkevich}, {Buzzi}, {Caffau},
  {Cancelliere}, {Cannizzaro}, {Cantat-Gaudin}, {Carballo}, {Carlucci},
  {Carrasco}, {Casamiquela}, {Castellani}, {Castro-Ginard}, {Charlot},
  {Chemin}, {Chiavassa}, {Cocozza}, {Costigan}, {Cowell}, {Crifo}, {Crosta},
  {Crowley}, {Cuypers}, {Dafonte}, {Damerdji}, {Dapergolas}, {David}, {David},
  {de Laverny}, {De Luise}, {De March}, {de Martino}, {de Souza}, {de Torres},
  {Debosscher}, {del Pozo}, {Delbo}, {Delgado}, {Delgado}, {Di Matteo},
  {Diakite}, {Diener}, {Distefano}, {Dolding}, {Drazinos}, {Dur{\'a}n},
  {Edvardsson}, {Enke}, {Eriksson}, {Esquej}, {Eynard Bontemps}, {Fabre},
  {Fabrizio}, {Faigler}, {Falc{\~a}o}, {Farr{\`a}s Casas}, {Federici},
  {Fedorets}, {Fernique}, {Figueras}, {Filippi}, {Findeisen}, {Fonti},
  {Fraile}, {Fraser}, {Fr{\'e}zouls}, {Gai}, {Galleti}, {Garabato},
  {Garc{\'\i}a-Sedano}, {Garofalo}, {Garralda}, {Gavel}, {Gavras}, {Gerssen},
  {Geyer}, {Giacobbe}, {Gilmore}, {Girona}, {Giuffrida}, {Glass}, {Gomes},
  {Granvik}, {Gueguen}, {Guerrier}, {Guiraud}, {Guti{\'e}rrez-S{\'a}nchez},
  {Haigron}, {Hatzidimitriou}, {Hauser}, {Haywood}, {Heiter}, {Helmi}, {Heu},
  {Hilger}, {Hobbs}, {Hofmann}, {Holland}, {Huckle}, {Hypki}, {Icardi},
  {Jan{\ss}en}, {Jevardat de Fombelle}, {Jonker}, {Juh{\'a}sz}, {Julbe},
  {Karampelas}, {Kewley}, {Klar}, {Kochoska}, {Kohley}, {Kolenberg},
  {Kontizas}, {Kontizas}, {Koposov}, {Kordopatis}, {Kostrzewa-Rutkowska},
  {Koubsky}, {Lambert}, {Lanza}, {Lasne}, {Lavigne}, {Le Fustec}, {Le
  Poncin-Lafitte}, {Lebreton}, {Leccia}, {Leclerc}, {Lecoeur-Taibi},
  {Lenhardt}, {Leroux}, {Liao}, {Licata}, {Lindstr{\o}m}, {Lister}, {Livanou},
  {Lobel}, {L{\'o}pez}, {Managau}, {Mann}, {Mantelet}, {Marchal}, {Marchant},
  {Marconi}, {Marinoni}, {Marschalk{\'o}}, {Marshall}, {Martino}, {Marton},
  {Mary}, {Massari}, {Matijevi{\v{c}}}, {Mazeh}, {McMillan}, {Messina},
  {Michalik}, {Millar}, {Molina}, {Molinaro}, {Moln{\'a}r}, {Montegriffo},
  {Mor}, {Morbidelli}, {Morel}, {Morris}, {Mulone}, {Muraveva}, {Musella},
  {Nelemans}, {Nicastro}, {Noval}, {O'Mullane}, {Ord{\'e}novic},
  {Ord{\'o}{\~n}ez-Blanco}, {Osborne}, {Pagani}, {Pagano}, {Pailler},
  {Palacin}, {Palaversa}, {Panahi}, {Pawlak}, {Piersimoni}, {Pineau}, {Plachy},
  {Plum}, {Poggio}, {Poujoulet}, {Pr{\v{s}}a}, {Pulone}, {Racero}, {Ragaini},
  {Rambaux}, {Ramos-Lerate}, {Regibo}, {Reyl{\'e}}, {Riclet}, {Ripepi}, {Riva},
  {Rivard}, {Rixon}, {Roegiers}, {Roelens}, {Romero-G{\'o}mez}, {Rowell},
  {Royer}, {Ruiz-Dern}, {Sadowski}, {Sagrist{\`a} Sell{\'e}s}, {Sahlmann},
  {Salgado}, {Salguero}, {Sanna}, {Santana-Ros}, {Sarasso}, {Savietto},
  {Schultheis}, {Sciacca}, {Segol}, {Segovia}, {S{\'e}gransan}, {Shih},
  {Siltala}, {Silva}, {Smart}, {Smith}, {Solano}, {Solitro}, {Sordo}, {Soria
  Nieto}, {Souchay}, {Spagna}, {Spoto}, {Stampa}, {Steele},
  {Steidelm{\"u}ller}, {Stephenson}, {Stoev}, {Suess}, {Surdej}, {Szabados},
  {Szegedi-Elek}, {Tapiador}, {Taris}, {Tauran}, {Taylor}, {Teixeira},
  {Terrett}, {Teyssandier}, {Thuillot}, {Titarenko}, {Torra Clotet}, {Turon},
  {Ulla}, {Utrilla}, {Uzzi}, {Vaillant}, {Valentini}, {Valette}, {van Elteren},
  {Van Hemelryck}, {van Leeuwen}, {Vaschetto}, {Vecchiato}, {Veljanoski},
  {Viala}, {Vicente}, {Vogt}, {von Essen}, {Voss}, {Votruba}, {Voutsinas},
  {Walmsley}, {Weiler}, {Wertz}, {Wevers}, {Wyrzykowski}, {Yoldas},
  {{\v{Z}}erjal}, {Ziaeepour}, {Zorec}, {Zschocke}, {Zucker}, {Zurbach}, \&
  {Zwitter}}]{Gaia}
{Gaia Collaboration}, {Brown}, A.~G.~A., {Vallenari}, A., {et~al.} 2018, \aap,
  616, A1

\bibitem[{{Ginolfi} {et~al.}(2020{\natexlab{a}}){Ginolfi}, {Jones},
  {B{\'e}thermin}, {Faisst}, {Lemaux}, {Schaerer}, {Fudamoto}, {Oesch},
  {Dessauges-Zavadsky}, {Fujimoto}, {Carniani}, {Le F{\`e}vre}, {Cassata},
  {Silverman}, {Capak}, {Yan}, {Bardelli}, {Cucciati}, {Gal}, {Gruppioni},
  {Hathi}, {Lubin}, {Maiolino}, {Morselli}, {Pelliccia}, {Talia}, {Vergani}, \&
  {Zamorani}}]{Ginolfi20b}
{Ginolfi}, M., {Jones}, G.~C., {B{\'e}thermin}, M., {et~al.}
  2020{\natexlab{a}}, \aap, 643, A7

\bibitem[{{Ginolfi} {et~al.}(2020{\natexlab{b}}){Ginolfi}, {Jones},
  {B{\'e}thermin}, {Fudamoto}, {Loiacono}, {Fujimoto}, {Le F{\'e}vre},
  {Faisst}, {Schaerer}, {Cassata}, {Silverman}, {Yan}, {Capak}, {Bardelli},
  {Boquien}, {Carraro}, {Dessauges-Zavadsky}, {Giavalisco}, {Gruppioni},
  {Ibar}, {Khusanova}, {Lemaux}, {Maiolino}, {Narayanan}, {Oesch}, {Pozzi},
  {Rodighiero}, {Talia}, {Toft}, {Vallini}, {Vergani}, \&
  {Zamorani}}]{Ginolfi20a}
{Ginolfi}, M., {Jones}, G.~C., {B{\'e}thermin}, M., {et~al.}
  2020{\natexlab{b}}, \aap, 633, A90

\bibitem[{{Ginolfi} {et~al.}(2017){Ginolfi}, {Maiolino}, {Nagao}, {Carniani},
  {Belfiore}, {Cresci}, {Hatsukade}, {Mannucci}, {Marconi}, {Pallottini},
  {Schneider}, \& {Santini}}]{Ginolfi17}
{Ginolfi}, M., {Maiolino}, R., {Nagao}, T., {et~al.} 2017, \mnras, 468, 3468

\bibitem[{{Goldsmith} {et~al.}(2012){Goldsmith}, {Langer}, {Pineda}, \&
  {Velusamy}}]{Goldsmith12}
{Goldsmith}, P.~F., {Langer}, W.~D., {Pineda}, J.~L., \& {Velusamy}, T. 2012,
  \apjs, 203, 13

\bibitem[{{Gonz{\'a}lez-L{\'o}pez} {et~al.}(2014){Gonz{\'a}lez-L{\'o}pez},
  {Riechers}, {Decarli}, {Walter}, {Vallini}, {Neri}, {Bertoldi}, {Bolatto},
  {Carilli}, {Cox}, {da Cunha}, {Ferrara}, {Gallerani}, \& {Infante}}]{GL14}
{Gonz{\'a}lez-L{\'o}pez}, J., {Riechers}, D.~A., {Decarli}, R., {et~al.} 2014,
  \apj, 784, 99

\bibitem[{{Gowardhan} {et~al.}(2019){Gowardhan}, {Riechers}, {Pavesi}, {Daddi},
  {Dannerbauer}, \& {Neri}}]{Gowardhan19}
{Gowardhan}, A., {Riechers}, D., {Pavesi}, R., {et~al.} 2019, \apj, 875, 6

\bibitem[{{Grassi} {et~al.}(2014){Grassi}, {Bovino}, {Schleicher}, {Prieto},
  {Seifried}, {Simoncini}, \& {Gianturco}}]{KROME}
{Grassi}, T., {Bovino}, S., {Schleicher}, D.~R.~G., {et~al.} 2014, \mnras, 439,
  2386

\bibitem[{{Guedes} {et~al.}(2011){Guedes}, {Callegari}, {Madau}, \&
  {Mayer}}]{Guedes11}
{Guedes}, J., {Callegari}, S., {Madau}, P., \& {Mayer}, L. 2011, \apj, 742, 76

\bibitem[{{Haardt} \& {Madau}(2012)}]{HM12}
{Haardt}, F. \& {Madau}, P. 2012, \apj, 746, 125

\bibitem[{{Habing}(1968)}]{Habing68}
{Habing}, H.~J. 1968, \bain, 19, 421

\bibitem[{{Harikane} {et~al.}(2020){Harikane}, {Ouchi}, {Inoue}, {Matsuoka},
  {Tamura}, {Bakx}, {Fujimoto}, {Moriwaki}, {Ono}, {Nagao}, {Tadaki}, {Kojima},
  {Shibuya}, {Egami}, {Ferrara}, {Gallerani}, {Hashimoto}, {Kohno}, {Matsuda},
  {Matsuo}, {Pallottini}, {Sugahara}, \& {Vallini}}]{Harikane20}
{Harikane}, Y., {Ouchi}, M., {Inoue}, A.~K., {et~al.} 2020, \apj, 896, 93

\bibitem[{{Hartigan} {et~al.}(2022){Hartigan}, {Hummel}, {Isella}, \&
  {Downes}}]{Hartigan22}
{Hartigan}, P., {Hummel}, M., {Isella}, A., \& {Downes}, T. 2022, \aj, 164, 257

\bibitem[{Hashimoto {et~al.}(2019)Hashimoto, Inoue, Mawatari, Tamura, Matsuo,
  Furusawa, Harikane, Shibuya, Knudsen, Kohno, Ono, Zackrisson, Okamoto,
  Kashikawa, Oesch, Ouchi, Ota, Shimizu, Taniguchi, Umehata, \&
  Watson}]{Hashimoto19}
Hashimoto, T., Inoue, A.~K., Mawatari, K., {et~al.} 2019, Publications of the
  Astronomical Society of Japan, 71, 71

\bibitem[{{Hashimoto} {et~al.}(2018){Hashimoto}, {Laporte}, {Mawatari},
  {Ellis}, {Inoue}, {Zackrisson}, {Roberts-Borsani}, {Zheng}, {Tamura},
  {Bauer}, {Fletcher}, {Harikane}, {Hatsukade}, {Hayatsu}, {Matsuda}, {Matsuo},
  {Okamoto}, {Ouchi}, {Pell{\'o}}, {Rydberg}, {Shimizu}, {Taniguchi},
  {Umehata}, \& {Yoshida}}]{Hashimoto18}
{Hashimoto}, T., {Laporte}, N., {Mawatari}, K., {et~al.} 2018, \nat, 557, 392

\bibitem[{{Heesen} {et~al.}(2023){Heesen}, {O'Sullivan}, {Br{\"u}ggen}, {Basu},
  {Beck}, {Seta}, {Carretti}, {Krause}, {Haverkorn}, {Hutschenreuter},
  {Bracco}, {Stein}, {Bomans}, {Dettmar}, {Chy{\.z}y}, {Heald}, {Paladino}, \&
  {Horellou}}]{Heesen23}
{Heesen}, V., {O'Sullivan}, S.~P., {Br{\"u}ggen}, M., {et~al.} 2023, \aap, 670,
  L23

\bibitem[{{Herrera-Camus} {et~al.}(2015){Herrera-Camus}, {Bolatto}, {Wolfire},
  {Smith}, {Croxall}, {Kennicutt}, {Calzetti}, {Helou}, {Walter}, {Leroy},
  {Draine}, {Brandl}, {Armus}, {Sandstrom}, {Dale}, {Aniano}, {Meidt},
  {Boquien}, {Hunt}, {Galametz}, {Tabatabaei}, {Murphy}, {Appleton}, {Roussel},
  {Engelbracht}, \& {Beirao}}]{Herrera15}
{Herrera-Camus}, R., {Bolatto}, A.~D., {Wolfire}, M.~G., {et~al.} 2015, \apj,
  800, 1

\bibitem[{{Herrera-Camus, R.} {et~al.}(2021){Herrera-Camus, R.}, {F\"orster
  Schreiber, N.}, {Genzel, R.}, {Tacconi, L.}, {Bolatto, A.}, {Davies, R. L.},
  {Fisher, D.}, {Lutz, D.}, {Naab, T.}, {Shimizu, T.}, {Tadaki, K.}, \&
  {\"Ubler, H.}}]{Herrera21}
{Herrera-Camus, R.}, {F\"orster Schreiber, N.}, {Genzel, R.}, {et~al.} 2021,
  A\&A, 649, A31

\bibitem[{{Hinshaw} {et~al.}(2013){Hinshaw}, {Larson}, {Komatsu}, {Spergel},
  {Bennett}, {Dunkley}, {Nolta}, {Halpern}, {Hill}, {Odegard}, {Page}, {Smith},
  {Weiland}, {Gold}, {Jarosik}, {Kogut}, {Limon}, {Meyer}, {Tucker}, {Wollack},
  \& {Wright}}]{Hinshaw13}
{Hinshaw}, G., {Larson}, D., {Komatsu}, E., {et~al.} 2013, \apjs, 208, 19

\bibitem[{{Hollenbach} \& {Tielens}(1999)}]{HT99}
{Hollenbach}, D.~J. \& {Tielens}, A.~G.~G.~M. 1999, Reviews of Modern Physics,
  71, 173

\bibitem[{{Hummels} {et~al.}(2019){Hummels}, {Smith}, {Hopkins}, {O'Shea},
  {Silvia}, {Werk}, {Lehner}, {Wise}, {Collins}, \& {Butsky}}]{Hummels19}
{Hummels}, C.~B., {Smith}, B.~D., {Hopkins}, P.~F., {et~al.} 2019, \apj, 882,
  156

\bibitem[{Hunter(2007)}]{Hunter07}
Hunter, J.~D. 2007, Computing in Science \& Engineering, 9, 90

\bibitem[{{Inoue} {et~al.}(2016){Inoue}, {Tamura}, {Matsuo}, {Mawatari},
  {Shimizu}, {Shibuya}, {Ota}, {Yoshida}, {Zackrisson}, {Kashikawa}, {Kohno},
  {Umehata}, {Hatsukade}, {Iye}, {Matsuda}, {Okamoto}, \&
  {Yamaguchi}}]{Inoue16}
{Inoue}, A.~K., {Tamura}, Y., {Matsuo}, H., {et~al.} 2016, Science, 352, 1559

\bibitem[{{Izumi} {et~al.}(2021){Izumi}, {Matsuoka}, {Fujimoto}, {Onoue},
  {Strauss}, {Umehata}, {Imanishi}, {Kohno}, {Kawaguchi}, {Kawamuro}, {Baba},
  {Nagao}, {Toba}, {Inayoshi}, {Silverman}, {Inoue}, {Ikarashi}, {Iwasawa},
  {Kashikawa}, {Hashimoto}, {Nakanishi}, {Ueda}, {Schramm}, {Lee}, \&
  {Suh}}]{Izumi21b}
{Izumi}, T., {Matsuoka}, Y., {Fujimoto}, S., {et~al.} 2021, \apj, 914, 36

\bibitem[{{Jones} {et~al.}(2023){Jones}, {Maiolino}, {Carniani}, {Circosta},
  {Fudamoto}, \& {Scholtz}}]{Jones23}
{Jones}, G.~C., {Maiolino}, R., {Carniani}, S., {et~al.} 2023, arXiv e-prints,
  arXiv:2303.17488

\bibitem[{{Kanekar} {et~al.}(2013){Kanekar}, {Wagg}, {Chary}, \&
  {Carilli}}]{Kanekar13}
{Kanekar}, N., {Wagg}, J., {Chary}, R.~R., \& {Carilli}, C.~L. 2013, \apjl,
  771, L20

\bibitem[{{Katz} {et~al.}(2019){Katz}, {Galligan}, {Kimm}, {Rosdahl},
  {Haehnelt}, {Blaizot}, {Devriendt}, {Slyz}, {Laporte}, \& {Ellis}}]{Katz19}
{Katz}, H., {Galligan}, T.~P., {Kimm}, T., {et~al.} 2019, \mnras, 487, 5902

\bibitem[{{Katz} {et~al.}(2022){Katz}, {Rosdahl}, {Kimm}, {Garel}, {Blaizot},
  {Haehnelt}, {Michel-Dansac}, {Martin-Alvarez}, {Devriendt}, {Slyz},
  {Teyssier}, {Ocvirk}, {Laporte}, \& {Ellis}}]{Katz22}
{Katz}, H., {Rosdahl}, J., {Kimm}, T., {et~al.} 2022, \mnras, 510, 5603

\bibitem[{{Khusanova} {et~al.}(2022){Khusanova}, {Ba{\~n}ados}, {Mazzucchelli},
  {Rojas-Ruiz}, {Momjian}, {Walter}, {Decarli}, {Venemans}, {Farina}, {Meyer},
  {Wang}, \& {Yang}}]{Khusanova22}
{Khusanova}, Y., {Ba{\~n}ados}, E., {Mazzucchelli}, C., {et~al.} 2022, arXiv
  e-prints, arXiv:2204.08973

\bibitem[{{Killi} {et~al.}(2023){Killi}, {Watson}, {Fujimoto}, {Akins},
  {Knudsen}, {Richard}, {Harikane}, {Rigopoulou}, {Rizzo}, {Ginolfi},
  {Popping}, \& {Kokorev}}]{Killi23}
{Killi}, M., {Watson}, D., {Fujimoto}, S., {et~al.} 2023, \mnras, 521, 2526

\bibitem[{{Kim} {et~al.}(2014){Kim}, {Abel}, {Agertz}, {Bryan}, {Ceverino},
  {Christensen}, {Conroy}, {Dekel}, {Gnedin}, {Goldbaum}, {Guedes}, {Hahn},
  {Hobbs}, {Hopkins}, {Hummels}, {Iannuzzi}, {Keres}, {Klypin}, {Kravtsov},
  {Krumholz}, {Kuhlen}, {Leitner}, {Madau}, {Mayer}, {Moody}, {Nagamine},
  {Norman}, {Onorbe}, {O'Shea}, {Pillepich}, {Primack}, {Quinn}, {Read},
  {Robertson}, {Rocha}, {Rudd}, {Shen}, {Smith}, {Szalay}, {Teyssier},
  {Thompson}, {Todoroki}, {Turk}, {Wadsley}, {Wise}, {Zolotov}, \& {AGORA
  Collaboration29}}]{Kim14}
{Kim}, J.-h., {Abel}, T., {Agertz}, O., {et~al.} 2014, \apjs, 210, 14

\bibitem[{{Klaassen} {et~al.}(2020){Klaassen}, {Mroczkowski}, {Cicone},
  {Hatziminaoglou}, {Sartori}, {De Breuck}, {Bryan}, {Dicker}, {Duran},
  {Groppi}, {Kaercher}, {Kawabe}, {Kohno}, \& {Geach}}]{AtLAST2020}
{Klaassen}, P.~D., {Mroczkowski}, T.~K., {Cicone}, C., {et~al.} 2020, in
  Society of Photo-Optical Instrumentation Engineers (SPIE) Conference Series,
  Vol. 11445, Society of Photo-Optical Instrumentation Engineers (SPIE)
  Conference Series, 114452F

\bibitem[{{Knudsen} {et~al.}(2016){Knudsen}, {Richard}, {Kneib}, {Jauzac},
  {Cl{\'e}ment}, {Drouart}, {Egami}, \& {Lindroos}}]{Knudsen16}
{Knudsen}, K.~K., {Richard}, J., {Kneib}, J.-P., {et~al.} 2016, \mnras, 462, L6

\bibitem[{{Knudsen} {et~al.}(2017){Knudsen}, {Watson}, {Frayer}, {Christensen},
  {Gallazzi}, {Micha{\l}owski}, {Richard}, \& {Zavala}}]{Knudsen17}
{Knudsen}, K.~K., {Watson}, D., {Frayer}, D., {et~al.} 2017, \mnras, 466, 138

\bibitem[{{Komatsu} {et~al.}(2011){Komatsu}, {Smith}, {Dunkley}, {Bennett},
  {Gold}, {Hinshaw}, {Jarosik}, {Larson}, {Nolta}, {Page}, {Spergel},
  {Halpern}, {Hill}, {Kogut}, {Limon}, {Meyer}, {Odegard}, {Tucker}, {Weiland},
  {Wollack}, \& {Wright}}]{Komatsu11}
{Komatsu}, E., {Smith}, K.~M., {Dunkley}, J., {et~al.} 2011, \apjs, 192, 18

\bibitem[{{Kroupa}(2001)}]{Kroupa01}
{Kroupa}, P. 2001, \mnras, 322, 231

\bibitem[{{Krumholz} {et~al.}(2009){Krumholz}, {McKee}, \&
  {Tumlinson}}]{Krumholz09}
{Krumholz}, M.~R., {McKee}, C.~F., \& {Tumlinson}, J. 2009, \apj, 699, 850

\bibitem[{{Kumari} {et~al.}(2023){Kumari}, {Smit}, {Leitherer}, {Witstok},
  {Irwin}, {Sirianni}, \& {Aloisi}}]{Kumari23}
{Kumari}, N., {Smit}, R., {Leitherer}, C., {et~al.} 2023, arXiv e-prints,
  arXiv:2307.00059

\bibitem[{{Lagache} {et~al.}(2018){Lagache}, {Cousin}, \&
  {Chatzikos}}]{Lagache18}
{Lagache}, G., {Cousin}, M., \& {Chatzikos}, M. 2018, \aap, 609, A130

\bibitem[{{Lehner} {et~al.}(2015){Lehner}, {Howk}, \& {Wakker}}]{Lehner15}
{Lehner}, N., {Howk}, J.~C., \& {Wakker}, B.~P. 2015, \apj, 804, 79

\bibitem[{{Leike} {et~al.}(2020){Leike}, {Glatzle}, \& {En{\ss}lin}}]{Leike20}
{Leike}, R.~H., {Glatzle}, M., \& {En{\ss}lin}, T.~A. 2020, \aap, 639, A138

\bibitem[{{Leitherer} {et~al.}(1999){Leitherer}, {Schaerer}, {Goldader},
  {Delgado}, {Robert}, {Kune}, {de Mello}, {Devost}, \& {Heckman}}]{SB99}
{Leitherer}, C., {Schaerer}, D., {Goldader}, J.~D., {et~al.} 1999, \apjs, 123,
  3

\bibitem[{{Lenki{\'c}} {et~al.}(2020){Lenki{\'c}}, {Bolatto}, {F{\"o}rster
  Schreiber}, {Tacconi}, {Neri}, {Combes}, {Walter}, {Garc{\'\i}a-Burillo},
  {Genzel}, {Lutz}, \& {Cooper}}]{Lenkic20}
{Lenki{\'c}}, L., {Bolatto}, A.~D., {F{\"o}rster Schreiber}, N.~M., {et~al.}
  2020, \aj, 159, 190

\bibitem[{{Leroy} {et~al.}(2022){Leroy}, {Rosolowsky}, {Usero}, {Sandstrom},
  {Schinnerer}, {Schruba}, {Bolatto}, {Sun}, {Barnes}, {Belfiore}, {Bigiel},
  {den Brok}, {Cao}, {Chiang}, {Chevance}, {Dale}, {Eibensteiner}, {Faesi},
  {Glover}, {Hughes}, {Jim{\'e}nez Donaire}, {Klessen}, {Koch}, {Kruijssen},
  {Liu}, {Meidt}, {Pan}, {Pety}, {Puschnig}, {Querejeta}, {Saito}, {Sardone},
  {Watkins}, {Weiss}, \& {Williams}}]{Leroy22}
{Leroy}, A.~K., {Rosolowsky}, E., {Usero}, A., {et~al.} 2022, \apj, 927, 149

\bibitem[{{Leung} {et~al.}(2020){Leung}, {Olsen}, {Somerville}, {Dav{\'e}},
  {Greve}, {Hayward}, {Narayanan}, \& {Popping}}]{Leung20}
{Leung}, T.~K.~D., {Olsen}, K.~P., {Somerville}, R.~S., {et~al.} 2020, \apj,
  905, 102

\bibitem[{Li {et~al.}(2021)Li, Emonts, Cai, Prochaska, Yoon, Lehnert, Zhang,
  Wu, Li, Li, Lacy, \& Villar-Mart{\'{\i}}n}]{Li21}
Li, J., Emonts, B. H.~C., Cai, Z., {et~al.} 2021, The Astrophysical Journal
  Letters, 922, L29

\bibitem[{{Lupi} {et~al.}(2020){Lupi}, {Pallottini}, {Ferrara}, {Bovino},
  {Carniani}, \& {Vallini}}]{Lupi2020}
{Lupi}, A., {Pallottini}, A., {Ferrara}, A., {et~al.} 2020, \mnras, 496, 5160

\bibitem[{{Maiolino} {et~al.}(2015){Maiolino}, {Carniani}, {Fontana},
  {Vallini}, {Pentericci}, {Ferrara}, {Vanzella}, {Grazian}, {Gallerani},
  {Castellano}, {Cristiani}, {Brammer}, {Santini}, {Wagg}, \&
  {Williams}}]{Maiolino15}
{Maiolino}, R., {Carniani}, S., {Fontana}, A., {et~al.} 2015, \mnras, 452, 54

\bibitem[{{Maiolino} {et~al.}(2012){Maiolino}, {Gallerani}, {Neri}, {Cicone},
  {Ferrara}, {Genzel}, {Lutz}, {Sturm}, {Tacconi}, {Walter}, {Feruglio},
  {Fiore}, \& {Piconcelli}}]{Maiolino12}
{Maiolino}, R., {Gallerani}, S., {Neri}, R., {et~al.} 2012, \mnras, 425, L66

\bibitem[{{Mandelker} {et~al.}(2018){Mandelker}, {van Dokkum}, {Brodie}, {van
  den Bosch}, \& {Ceverino}}]{Mandelker18}
{Mandelker}, N., {van Dokkum}, P.~G., {Brodie}, J.~P., {van den Bosch}, F.~C.,
  \& {Ceverino}, D. 2018, \apj, 861, 148

\bibitem[{{McCourt} {et~al.}(2018){McCourt}, {Oh}, {O'Leary}, \&
  {Madigan}}]{Mccourt18}
{McCourt}, M., {Oh}, S.~P., {O'Leary}, R., \& {Madigan}, A.-M. 2018, \mnras,
  473, 5407

\bibitem[{{Meyer} {et~al.}(2022){Meyer}, {Walter}, {Cicone}, {Cox}, {Decarli},
  {Neri}, {Novak}, {Pensabene}, {Riechers}, \& {Weiss}}]{Meyer22}
{Meyer}, R.~A., {Walter}, F., {Cicone}, C., {et~al.} 2022, \apj, 927, 152

\bibitem[{{Montoya Arroyave} {et~al.}(2023){Montoya Arroyave}, {Cicone},
  {Makroleivaditi}, {Weiss}, {Lundgren}, {Severgnini}, {De Breuck},
  {Baumschlager}, {Schimek}, {Shen}, \& {Aravena}}]{Montoya-Arroyave23}
{Montoya Arroyave}, I., {Cicone}, C., {Makroleivaditi}, E., {et~al.} 2023,
  arXiv e-prints, arXiv:2302.06629

\bibitem[{{Nagao} {et~al.}(2013){Nagao}, {Maiolino}, {De Breuck}, {Caselli},
  {Hatsukade}, \& {Saigo}}]{Nagao13}
{Nagao}, T., {Maiolino}, R., {De Breuck}, C., {et~al.} 2013, in Astronomical
  Society of the Pacific Conference Series, Vol. 476, New Trends in Radio
  Astronomy in the ALMA Era: The 30th Anniversary of Nobeyama Radio
  Observatory, ed. R.~{Kawabe}, N.~{Kuno}, \& S.~{Yamamoto}, 29

\bibitem[{{Nagao} {et~al.}(2011){Nagao}, {Maiolino}, {Marconi}, \&
  {Matsuhara}}]{Nagao11}
{Nagao}, T., {Maiolino}, R., {Marconi}, A., \& {Matsuhara}, H. 2011, \aap, 526,
  A149

\bibitem[{{Nelson} {et~al.}(2021){Nelson}, {Byrohl}, {Peroux}, {Rubin}, \&
  {Burchett}}]{Nelson21}
{Nelson}, D., {Byrohl}, C., {Peroux}, C., {Rubin}, K. H.~R., \& {Burchett},
  J.~N. 2021, \mnras, 507, 4445

\bibitem[{{Olsen} {et~al.}(2017){Olsen}, {Greve}, {Narayanan}, {Thompson},
  {Dav{\'e}}, {Niebla Rios}, \& {Stawinski}}]{Olsen17}
{Olsen}, K., {Greve}, T.~R., {Narayanan}, D., {et~al.} 2017, \apj, 846, 105

\bibitem[{{Olsen} {et~al.}(2018){Olsen}, {Pallottini}, {Wofford}, {Chatzikos},
  {Revalski}, {Guzm{\'a}n}, {Popping}, {V{\'a}zquez-Semadeni}, {Magdis},
  {Richardson}, {Hirschmann}, \& {Gray}}]{Olsen18}
{Olsen}, K., {Pallottini}, A., {Wofford}, A., {et~al.} 2018, Galaxies, 6, 100

\bibitem[{{Olsen} {et~al.}(2015){Olsen}, {Greve}, {Narayanan}, {Thompson},
  {Toft}, \& {Brinch}}]{Olsen15}
{Olsen}, K.~P., {Greve}, T.~R., {Narayanan}, D., {et~al.} 2015, \apj, 814, 76

\bibitem[{{Ota} {et~al.}(2014){Ota}, {Walter}, {Ohta}, {Hatsukade}, {Carilli},
  {da Cunha}, {Gonz{\'a}lez-L{\'o}pez}, {Decarli}, {Hodge}, {Nagai}, {Egami},
  {Jiang}, {Iye}, {Kashikawa}, {Riechers}, {Bertoldi}, {Cox}, {Neri}, \&
  {Weiss}}]{Ota14}
{Ota}, K., {Walter}, F., {Ohta}, K., {et~al.} 2014, \apj, 792, 34

\bibitem[{{Ouchi} {et~al.}(2013){Ouchi}, {Ellis}, {Ono}, {Nakanishi}, {Kohno},
  {Momose}, {Kurono}, {Ashby}, {Shimasaku}, {Willner}, {Fazio}, {Tamura}, \&
  {Iono}}]{Ouchi13}
{Ouchi}, M., {Ellis}, R., {Ono}, Y., {et~al.} 2013, \apj, 778, 102

\bibitem[{{Pallottini} {et~al.}(2017){Pallottini}, {Ferrara}, {Bovino},
  {Vallini}, {Gallerani}, {Maiolino}, \& {Salvadori}}]{Pallottini17}
{Pallottini}, A., {Ferrara}, A., {Bovino}, S., {et~al.} 2017, \mnras, 471, 4128

\bibitem[{{Pallottini} {et~al.}(2019){Pallottini}, {Ferrara}, {Decataldo},
  {Gallerani}, {Vallini}, {Carniani}, {Behrens}, {Kohandel}, \&
  {Salvadori}}]{Pallottini19}
{Pallottini}, A., {Ferrara}, A., {Decataldo}, D., {et~al.} 2019, \mnras, 487,
  1689

\bibitem[{{Pallottini} {et~al.}(2022){Pallottini}, {Ferrara}, {Gallerani},
  {Behrens}, {Kohandel}, {Carniani}, {Vallini}, {Salvadori}, {Gelli},
  {Sommovigo}, {D'Odorico}, {Di Mascia}, \& {Pizzati}}]{Pallottini22}
{Pallottini}, A., {Ferrara}, A., {Gallerani}, S., {et~al.} 2022, \mnras, 513,
  5621

\bibitem[{{Papadopoulos}(2004)}]{Papadopoulos04}
{Papadopoulos}, P. 2004, in Astronomical Society of the Pacific Conference
  Series, Vol. 320, The Neutral ISM in Starburst Galaxies, ed. S.~{Aalto},
  S.~{Huttemeister}, \& A.~{Pedlar}, 101

\bibitem[{{Papadopoulos} {et~al.}(2022){Papadopoulos}, {Dunne}, \&
  {Maddox}}]{Papadopoulos22}
{Papadopoulos}, P., {Dunne}, L., \& {Maddox}, S. 2022, \mnras, 510, 725

\bibitem[{{Papadopoulos} {et~al.}(2018){Papadopoulos}, {Bisbas}, \&
  {Zhang}}]{Papadopoulos18}
{Papadopoulos}, P.~P., {Bisbas}, T.~G., \& {Zhang}, Z.-Y. 2018, \mnras, 478,
  1716

\bibitem[{{Pentericci} {et~al.}(2016){Pentericci}, {Carniani}, {Castellano},
  {Fontana}, {Maiolino}, {Guaita}, {Vanzella}, {Grazian}, {Santini}, {Yan},
  {Cristiani}, {Conselice}, {Giavalisco}, {Hathi}, \&
  {Koekemoer}}]{Pentericci16}
{Pentericci}, L., {Carniani}, S., {Castellano}, M., {et~al.} 2016, \apjl, 829,
  L11

\bibitem[{Pereira-Santaella {et~al.}(2017)Pereira-Santaella, Rigopoulou,
  Farrah, Lebouteiller, \& Li}]{Pereira17}
Pereira-Santaella, M., Rigopoulou, D., Farrah, D., Lebouteiller, V., \& Li, J.
  2017, Monthly Notices of the Royal Astronomical Society, 470, 1218

\bibitem[{{Pizzati} {et~al.}(2020){Pizzati}, {Ferrara}, {Pallottini},
  {Gallerani}, {Vallini}, {Decataldo}, \& {Fujimoto}}]{Pizzati20}
{Pizzati}, E., {Ferrara}, A., {Pallottini}, A., {et~al.} 2020, \mnras, 495, 160

\bibitem[{{Pizzati} {et~al.}(2023){Pizzati}, {Ferrara}, {Pallottini},
  {Sommovigo}, {Kohandel}, \& {Carniani}}]{Pizzati23}
{Pizzati}, E., {Ferrara}, A., {Pallottini}, A., {et~al.} 2023, \mnras, 519,
  4608

\bibitem[{{Plummer}(1911)}]{Plummer11}
{Plummer}, H.~C. 1911, \mnras, 71, 460

\bibitem[{{Ponnada} {et~al.}(2022){Ponnada}, {Panopoulou}, {Butsky}, {Hopkins},
  {Loebman}, {Hummels}, {Ji}, {Wetzel}, {Faucher-Gigu{\`e}re}, \&
  {Hayward}}]{Ponnada22}
{Ponnada}, S.~B., {Panopoulou}, G.~V., {Butsky}, I.~S., {et~al.} 2022, \mnras,
  516, 4417

\bibitem[{{Pontzen} {et~al.}(2013){Pontzen}, {Ro{\v{s}}kar}, {Stinson}, \&
  {Woods}}]{Pynbody}
{Pontzen}, A., {Ro{\v{s}}kar}, R., {Stinson}, G., \& {Woods}, R. 2013,
  {pynbody: N-Body/SPH analysis for python}

\bibitem[{{Popping}(2023)}]{Popping22}
{Popping}, G. 2023, \aap, 669, L8

\bibitem[{{Popping} {et~al.}(2019){Popping}, {Narayanan}, {Somerville},
  {Faisst}, \& {Krumholz}}]{Popping19}
{Popping}, G., {Narayanan}, D., {Somerville}, R.~S., {Faisst}, A.~L., \&
  {Krumholz}, M.~R. 2019, \mnras, 482, 4906

\bibitem[{{Prochaska} {et~al.}(2014){Prochaska}, {Lau}, \&
  {Hennawi}}]{Prochaska14}
{Prochaska}, J.~X., {Lau}, M.~W., \& {Hennawi}, J.~F. 2014, \apj, 796, 140

\bibitem[{{Raiteri} {et~al.}(1996){Raiteri}, {Villata}, \& {Navarro}}]{RVN96}
{Raiteri}, C.~M., {Villata}, M., \& {Navarro}, J.~F. 1996, \aap, 315, 105

\bibitem[{{Ramasawmy} {et~al.}(2022){Ramasawmy}, {Klaassen}, {Cicone},
  {Mroczkowski}, {Chen}, {Cornish}, {da Cunha}, {Hatziminaoglou}, {Johnstone},
  {Liu}, {Perrott}, {Schimek}, {Stanke}, \& {Wedemeyer}}]{AtLAST22}
{Ramasawmy}, J., {Klaassen}, P.~D., {Cicone}, C., {et~al.} 2022, in Society of
  Photo-Optical Instrumentation Engineers (SPIE) Conference Series, Vol. 12190,
  Millimeter, Submillimeter, and Far-Infrared Detectors and Instrumentation for
  Astronomy XI, ed. J.~{Zmuidzinas} \& J.-R. {Gao}, 1219007

\bibitem[{{Ren} {et~al.}(2023){Ren}, {Fudamoto}, {Inoue}, {Sugahara},
  {Tokuoka}, {Tamura}, {Matsuo}, {Kohno}, {Umehata}, {Hashimoto}, {Bouwens},
  {Smit}, {Kashikawa}, {Okamoto}, {Shibuya}, \& {Shimizu}}]{Ren23}
{Ren}, Y.~W., {Fudamoto}, Y., {Inoue}, A.~K., {et~al.} 2023, \apj, 945, 69

\bibitem[{{Rey} {et~al.}(2023){Rey}, {Katz}, {Cameron}, {Devriendt}, \&
  {Slyz}}]{Rey23}
{Rey}, M.~P., {Katz}, H.~B., {Cameron}, A.~J., {Devriendt}, J., \& {Slyz}, A.
  2023, arXiv e-prints, arXiv:2302.08521

\bibitem[{{Rigopoulou} {et~al.}(2018){Rigopoulou}, {Pereira-Santaella},
  {Magdis}, {Cooray}, {Farrah}, {Marques-Chaves}, {Perez-Fournon}, \&
  {Riechers}}]{Rigo18}
{Rigopoulou}, D., {Pereira-Santaella}, M., {Magdis}, G.~E., {et~al.} 2018,
  \mnras, 473, 20

\bibitem[{{Rosdahl} {et~al.}(2013){Rosdahl}, {Blaizot}, {Aubert}, {Stranex}, \&
  {Teyssier}}]{RamsesRT}
{Rosdahl}, J., {Blaizot}, J., {Aubert}, D., {Stranex}, T., \& {Teyssier}, R.
  2013, \mnras, 436, 2188

\bibitem[{{Rubin} {et~al.}(1994){Rubin}, {Simpson}, {Lord}, {Colgan},
  {Erickson}, \& {Haas}}]{Rubin94}
{Rubin}, R.~H., {Simpson}, J.~P., {Lord}, S.~D., {et~al.} 1994, \apj, 420, 772

\bibitem[{{Scannapieco} \& {Br{\"u}ggen}(2015)}]{Scannapieco15}
{Scannapieco}, E. \& {Br{\"u}ggen}, M. 2015, \apj, 805, 158

\bibitem[{{Schaerer} {et~al.}(2015){Schaerer}, {Boone}, {Zamojski}, {Staguhn},
  {Dessauges-Zavadsky}, {Finkelstein}, \& {Combes}}]{Schaerer15}
{Schaerer}, D., {Boone}, F., {Zamojski}, M., {et~al.} 2015, \aap, 574, A19

\bibitem[{{Schaerer} {et~al.}(2020){Schaerer}, {Ginolfi}, {B{\'e}thermin},
  {Fudamoto}, {Oesch}, {Le F{\`e}vre}, {Faisst}, {Capak}, {Cassata},
  {Silverman}, {Yan}, {Jones}, {Amorin}, {Bardelli}, {Boquien}, {Cimatti},
  {Dessauges-Zavadsky}, {Giavalisco}, {Hathi}, {Fujimoto}, {Ibar}, {Koekemoer},
  {Lagache}, {Lemaux}, {Loiacono}, {Maiolino}, {Narayanan}, {Morselli},
  {M{\'e}ndez-Hern{\`a}ndez}, {Pozzi}, {Riechers}, {Talia}, {Toft}, {Vallini},
  {Vergani}, {Zamorani}, \& {Zucca}}]{Schaerer20}
{Schaerer}, D., {Ginolfi}, M., {B{\'e}thermin}, M., {et~al.} 2020, \aap, 643,
  A3

\bibitem[{{Schneider} \& {Robertson}(2017)}]{Schneider17}
{Schneider}, E.~E. \& {Robertson}, B.~E. 2017, \apj, 834, 144

\bibitem[{{Scholtz} {et~al.}(2023){Scholtz}, {Maiolino}, {Jones}, \&
  {Carniani}}]{Scholtz23}
{Scholtz}, J., {Maiolino}, R., {Jones}, G.~C., \& {Carniani}, S. 2023, \mnras,
  519, 5246

\bibitem[{{Seifried} {et~al.}(2020){Seifried}, {Haid}, {Walch}, {Borchert}, \&
  {Bisbas}}]{Seifried20}
{Seifried}, D., {Haid}, S., {Walch}, S., {Borchert}, E.~M.~A., \& {Bisbas},
  T.~G. 2020, \mnras, 492, 1465

\bibitem[{{Seon} {et~al.}(2011){Seon}, {Witt}, {Kim}, {Shinn}, {Edelstein},
  {Min}, \& {Han}}]{Seon11}
{Seon}, K.-I., {Witt}, A., {Kim}, I.-J., {et~al.} 2011, \apj, 743, 188

\bibitem[{{Shen} {et~al.}(2014){Shen}, {Madau}, {Conroy}, {Governato}, \&
  {Mayer}}]{Shen14}
{Shen}, S., {Madau}, P., {Conroy}, C., {Governato}, F., \& {Mayer}, L. 2014,
  \apj, 792, 99

\bibitem[{{Shen} {et~al.}(2010){Shen}, {Wadsley}, \& {Stinson}}]{Shen10}
{Shen}, S., {Wadsley}, J., \& {Stinson}, G. 2010, \mnras, 407, 1581

\bibitem[{{Solomon} {et~al.}(1997){Solomon}, {Downes}, {Radford}, \&
  {Barrett}}]{Solomon97}
{Solomon}, P.~M., {Downes}, D., {Radford}, S.~J.~E., \& {Barrett}, J.~W. 1997,
  \apj, 478, 144

\bibitem[{{Solomon} \& {Vanden Bout}(2005)}]{Solomon+05}
{Solomon}, P.~M. \& {Vanden Bout}, P.~A. 2005, \araa, 43, 677

\bibitem[{{Sparre} {et~al.}(2019){Sparre}, {Pfrommer}, \&
  {Vogelsberger}}]{Sparre19}
{Sparre}, M., {Pfrommer}, C., \& {Vogelsberger}, M. 2019, \mnras, 482, 5401

\bibitem[{Stacey {et~al.}(1991)Stacey, Townes, Poglitsch, Madden, Jackson,
  Herrmann, Genzel, \& Geis}]{Stacey91}
Stacey, G., Townes, C., Poglitsch, A., {et~al.} 1991, The Astrophysical
  Journal, 382, L37

\bibitem[{{Stinson} {et~al.}(2006){Stinson}, {Seth}, {Katz}, {Wadsley},
  {Governato}, \& {Quinn}}]{Stinson06}
{Stinson}, G., {Seth}, A., {Katz}, N., {et~al.} 2006, \mnras, 373, 1074

\bibitem[{{Strandet} {et~al.}(2017){Strandet}, {Weiss}, {De Breuck}, {Marrone},
  {Vieira}, {Aravena}, {Ashby}, {B{\'e}thermin}, {Bothwell}, {Bradford},
  {Carlstrom}, {Chapman}, {Cunningham}, {Chen}, {Fassnacht}, {Gonzalez},
  {Greve}, {Gullberg}, {Hayward}, {Hezaveh}, {Litke}, {Ma}, {Malkan}, {Menten},
  {Miller}, {Murphy}, {Narayanan}, {Phadke}, {Rotermund}, {Spilker}, \&
  {Sreevani}}]{Strandet17}
{Strandet}, M.~L., {Weiss}, A., {De Breuck}, C., {et~al.} 2017, \apjl, 842, L15

\bibitem[{{Sturm} {et~al.}(2011{\natexlab{a}}){Sturm}, {Gonz{\'a}lez-Alfonso},
  {Veilleux}, {Fischer}, {Graci{\'a}-Carpio}, {Hailey-Dunsheath}, {Contursi},
  {Poglitsch}, {Sternberg}, {Davies}, {Genzel}, {Lutz}, {Tacconi}, {Verma},
  {Maiolino}, \& {de Jong}}]{Sturm11}
{Sturm}, E., {Gonz{\'a}lez-Alfonso}, E., {Veilleux}, S., {et~al.}
  2011{\natexlab{a}}, \apjl, 733, L16

\bibitem[{{Sturm} {et~al.}(2011{\natexlab{b}}){Sturm}, {Gonz{\'a}lez-Alfonso},
  {Veilleux}, {Fischer}, {Graci{\'a}-Carpio}, {Hailey-Dunsheath}, {Contursi},
  {Poglitsch}, {Sternberg}, {Davies}, {Genzel}, {Lutz}, {Tacconi}, {Verma},
  {Maiolino}, \& {de Jong}}]{Cicone14}
{Sturm}, E., {Gonz{\'a}lez-Alfonso}, E., {Veilleux}, S., {et~al.}
  2011{\natexlab{b}}, \apjl, 733, L16

\bibitem[{{Suresh} {et~al.}(2019){Suresh}, {Nelson}, {Genel}, {Rubin}, \&
  {Hernquist}}]{Suresh19}
{Suresh}, J., {Nelson}, D., {Genel}, S., {Rubin}, K. H.~R., \& {Hernquist}, L.
  2019, \mnras, 483, 4040

\bibitem[{{Tacconi} {et~al.}(2013){Tacconi}, {Neri}, {Genzel}, {Combes},
  {Bolatto}, {Cooper}, {Wuyts}, {Bournaud}, {Burkert}, {Comerford}, {Cox},
  {Davis}, {F{\"o}rster Schreiber}, {Garc{\'\i}a-Burillo}, {Gracia-Carpio},
  {Lutz}, {Naab}, {Newman}, {Omont}, {Saintonge}, {Shapiro Griffin}, {Shapley},
  {Sternberg}, \& {Weiner}}]{Tacconi13}
{Tacconi}, L.~J., {Neri}, R., {Genzel}, R., {et~al.} 2013, \apj, 768, 74

\bibitem[{{Teyssier}(2002)}]{Ramses}
{Teyssier}, R. 2002, \aap, 385, 337

\bibitem[{{Thi} {et~al.}(2009){Thi}, {van Dishoeck}, {Bell}, {Viti}, \&
  {Black}}]{Thi09}
{Thi}, W.~F., {van Dishoeck}, E.~F., {Bell}, T., {Viti}, S., \& {Black}, J.
  2009, \mnras, 400, 622

\bibitem[{{Tielens} \& {Hollenbach}(1985)}]{Tielens85}
{Tielens}, A.~G.~G.~M. \& {Hollenbach}, D. 1985, \apj, 291, 722

\bibitem[{{Tumlinson} {et~al.}(2017){Tumlinson}, {Peeples}, \&
  {Werk}}]{Tumlinson17}
{Tumlinson}, J., {Peeples}, M.~S., \& {Werk}, J.~K. 2017, \araa, 55, 389

\bibitem[{{Valentino} {et~al.}(2020){Valentino}, {Magdis}, {Daddi}, {Liu},
  {Aravena}, {Bournaud}, {Cortzen}, {Gao}, {Jin}, {Juneau}, {Kartaltepe},
  {Kokorev}, {Lee}, {Madden}, {Narayanan}, {Popping}, \&
  {Puglisi}}]{Valentino20}
{Valentino}, F., {Magdis}, G.~E., {Daddi}, E., {et~al.} 2020, \apj, 890, 24

\bibitem[{{Vallini} {et~al.}(2021){Vallini}, {Ferrara}, {Pallottini},
  {Carniani}, \& {Gallerani}}]{Vallini21}
{Vallini}, L., {Ferrara}, A., {Pallottini}, A., {Carniani}, S., \& {Gallerani},
  S. 2021, \mnras, 505, 5543

\bibitem[{{Vallini} {et~al.}(2017){Vallini}, {Ferrara}, {Pallottini}, \&
  {Gallerani}}]{Vallini17}
{Vallini}, L., {Ferrara}, A., {Pallottini}, A., \& {Gallerani}, S. 2017,
  \mnras, 467, 1300

\bibitem[{{Vallini} {et~al.}(2015){Vallini}, {Gallerani}, {Ferrara},
  {Pallottini}, \& {Yue}}]{Vallini15}
{Vallini}, L., {Gallerani}, S., {Ferrara}, A., {Pallottini}, A., \& {Yue}, B.
  2015, \apj, 813, 36

\bibitem[{{Vallini} {et~al.}(2018){Vallini}, {Pallottini}, {Ferrara},
  {Gallerani}, {Sobacchi}, \& {Behrens}}]{Vallini18}
{Vallini}, L., {Pallottini}, A., {Ferrara}, A., {et~al.} 2018, \mnras, 473, 271

\bibitem[{{van de Voort} {et~al.}(2019){van de Voort}, {Springel}, {Mandelker},
  {van den Bosch}, \& {Pakmor}}]{VanDeVoort19}
{van de Voort}, F., {Springel}, V., {Mandelker}, N., {van den Bosch}, F.~C., \&
  {Pakmor}, R. 2019, \mnras, 482, L85

\bibitem[{{van der Walt} {et~al.}(2011){van der Walt}, {Colbert}, \&
  {Varoquaux}}]{Walt2011}
{van der Walt}, S., {Colbert}, S.~C., \& {Varoquaux}, G. 2011, Computing in
  Science and Engineering, 13, 22

\bibitem[{Van~Rossum \& Drake(2009)}]{python3}
Van~Rossum, G. \& Drake, F.~L. 2009, Python 3 Reference Manual (Scotts Valley,
  CA: CreateSpace)

\bibitem[{Van~Rossum \& Drake~Jr(1995)}]{van1995python}
Van~Rossum, G. \& Drake~Jr, F.~L. 1995, Python reference manual (Centrum voor
  Wiskunde en Informatica Amsterdam)

\bibitem[{{Vanderhoof} {et~al.}(2022){Vanderhoof}, {Faisst}, {Shen}, {Lemaux},
  {B{\'e}thermin}, {Capak}, {Cassata}, {Le F{\`e}vre}, {Schaerer}, {Silverman},
  {Yan}, {Boquien}, {Gal}, {Kartaltepe}, {Lubin}, {Dessauges-Zavadsky},
  {Fudamoto}, {Ginolfi}, {Hathi}, {Jones}, {Koekemoer}, {Narayanan}, {Romano},
  {Talia}, {Vergani}, \& {Zamorani}}]{Vanderhoof22}
{Vanderhoof}, B.~N., {Faisst}, A.~L., {Shen}, L., {et~al.} 2022, \mnras, 511,
  1303

\bibitem[{{Veilleux} {et~al.}(2020){Veilleux}, {Maiolino}, {Bolatto}, \&
  {Aalto}}]{Veilleux20}
{Veilleux}, S., {Maiolino}, R., {Bolatto}, A.~D., \& {Aalto}, S. 2020, \aapr,
  28, 2

\bibitem[{{Wadsley} {et~al.}(2017){Wadsley}, {Keller}, \& {Quinn}}]{Wadsley17}
{Wadsley}, J.~W., {Keller}, B.~W., \& {Quinn}, T.~R. 2017, \mnras, 471, 2357

\bibitem[{{Wadsley} {et~al.}(2004){Wadsley}, {Stadel}, \& {Quinn}}]{Wadsley04}
{Wadsley}, J.~W., {Stadel}, J., \& {Quinn}, T. 2004, \na, 9, 137

\bibitem[{{Walker} {et~al.}(1990){Walker}, {Adams}, \& {Lada}}]{Walker90}
{Walker}, C.~K., {Adams}, F.~C., \& {Lada}, C.~J. 1990, \apj, 349, 515

\bibitem[{{Webber} \& {Soutoul}(1998)}]{Webber98}
{Webber}, W.~R. \& {Soutoul}, A. 1998, \apj, 506, 335

\bibitem[{{Werk} {et~al.}(2014){Werk}, {Prochaska}, {Tumlinson}, {Peeples},
  {Tripp}, {Fox}, {Lehner}, {Thom}, {O'Meara}, {Ford}, {Bordoloi}, {Katz},
  {Tejos}, {Oppenheimer}, {Dav{\'e}}, \& {Weinberg}}]{Werk14}
{Werk}, J.~K., {Prochaska}, J.~X., {Tumlinson}, J., {et~al.} 2014, \apj, 792, 8

\bibitem[{{Whitworth} \& {Ward-Thompson}(2001)}]{Whitworth01}
{Whitworth}, A.~P. \& {Ward-Thompson}, D. 2001, \apj, 547, 317

\bibitem[{{Willott} {et~al.}(2015){Willott}, {Carilli}, {Wagg}, \&
  {Wang}}]{Willott15}
{Willott}, C.~J., {Carilli}, C.~L., {Wagg}, J., \& {Wang}, R. 2015, \apj, 807,
  180

\bibitem[{{Wilson} {et~al.}(2008){Wilson}, {Petitpas}, {Iono}, {Baker}, {Peck},
  {Krips}, {Warren}, {Golding}, {Atkinson}, {Armus}, {Cox}, {Ho}, {Juvela},
  {Matsushita}, {Mihos}, {Pihlstrom}, \& {Yun}}]{Wilson08}
{Wilson}, C.~D., {Petitpas}, G.~R., {Iono}, D., {et~al.} 2008, \apjs, 178, 189

\bibitem[{{Witstok} {et~al.}(2022){Witstok}, {Smit}, {Maiolino}, {Kumari},
  {Aravena}, {Boogaard}, {Bouwens}, {Carniani}, {Hodge}, {Jones}, {Stefanon},
  {van der Werf}, \& {Schouws}}]{Witstok22}
{Witstok}, J., {Smit}, R., {Maiolino}, R., {et~al.} 2022, \mnras, 515, 1751

\bibitem[{Wolfire {et~al.}(2022)Wolfire, Vallini, \& Chevance}]{Wolfire22}
Wolfire, M.~G., Vallini, L., \& Chevance, M. 2022, Photodissociation and X-Ray
  Dominated Regions

\bibitem[{{Wright}(2006)}]{CosmoCalc}
{Wright}, E.~L. 2006, \pasp, 118, 1711

\bibitem[{{Yang} {et~al.}(2017){Yang}, {Omont}, {Beelen}, {Gao}, {van der
  Werf}, {Gavazzi}, {Zhang}, {Ivison}, {Lehnert}, {Liu}, {Oteo},
  {Gonz{\'a}lez-Alfonso}, {Dannerbauer}, {Cox}, {Krips}, {Neri}, {Riechers},
  {Baker}, {Micha{\l}owski}, {Cooray}, \& {Smail}}]{Yang17}
{Yang}, C., {Omont}, A., {Beelen}, A., {et~al.} 2017, \aap, 608, A144

\bibitem[{{Zanella} {et~al.}(2023){Zanella}, {Valentino}, {Gallazzi}, {Belli},
  {Magdis}, \& {Bolamperti}}]{Zanella23}
{Zanella}, A., {Valentino}, F., {Gallazzi}, A., {et~al.} 2023, \mnras, 524, 923

\bibitem[{{Zhang} {et~al.}(2018){Zhang}, {Xu}, {Vasyunin}, {Semenov}, {Wang},
  {Dib}, {Liu}, {Liu}, {Zhang}, {Liu}, {Wang}, {Li}, {Wu}, {Yuan}, {Li}, \&
  {Gao}}]{Zhang18}
{Zhang}, G.-Y., {Xu}, J.-L., {Vasyunin}, A.~I., {et~al.} 2018, \aap, 620, A163

\bibitem[{{Zhang} {et~al.}(2016){Zhang}, {Papadopoulos}, {Ivison}, {Galametz},
  {Smith}, \& {Xilouris}}]{Zhang16}
{Zhang}, Z.-Y., {Papadopoulos}, P.~P., {Ivison}, R.~J., {et~al.} 2016, Royal
  Society Open Science, 3, 160025

\bibitem[{{Zucker} {et~al.}(2021){Zucker}, {Goodman}, {Alves}, {Bialy}, {Koch},
  {Speagle}, {Foley}, {Finkbeiner}, {Leike}, {En{\ss}lin}, {Peek}, \&
  {Edenhofer}}]{Zucker21}
{Zucker}, C., {Goodman}, A., {Alves}, J., {et~al.} 2021, \apj, 919, 35

\end{thebibliography}
\bibliographystyle{aa}

\clearpage
\begin{appendix} 

\section{Multiphase model maps}\label{app:MPM}

\begin{figure*}[tbh]
\centering
  \includegraphics[width = 0.83\columnwidth]{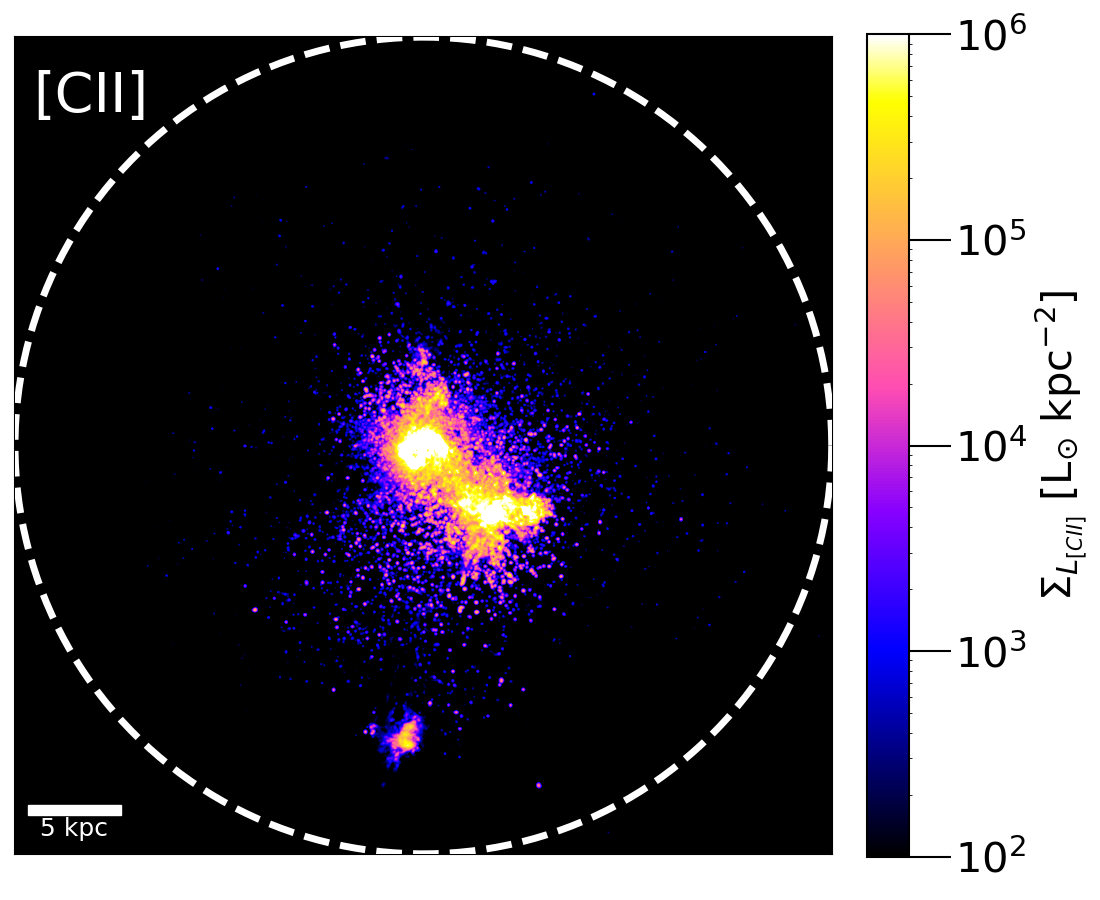}
  \includegraphics[width = 0.83\columnwidth]{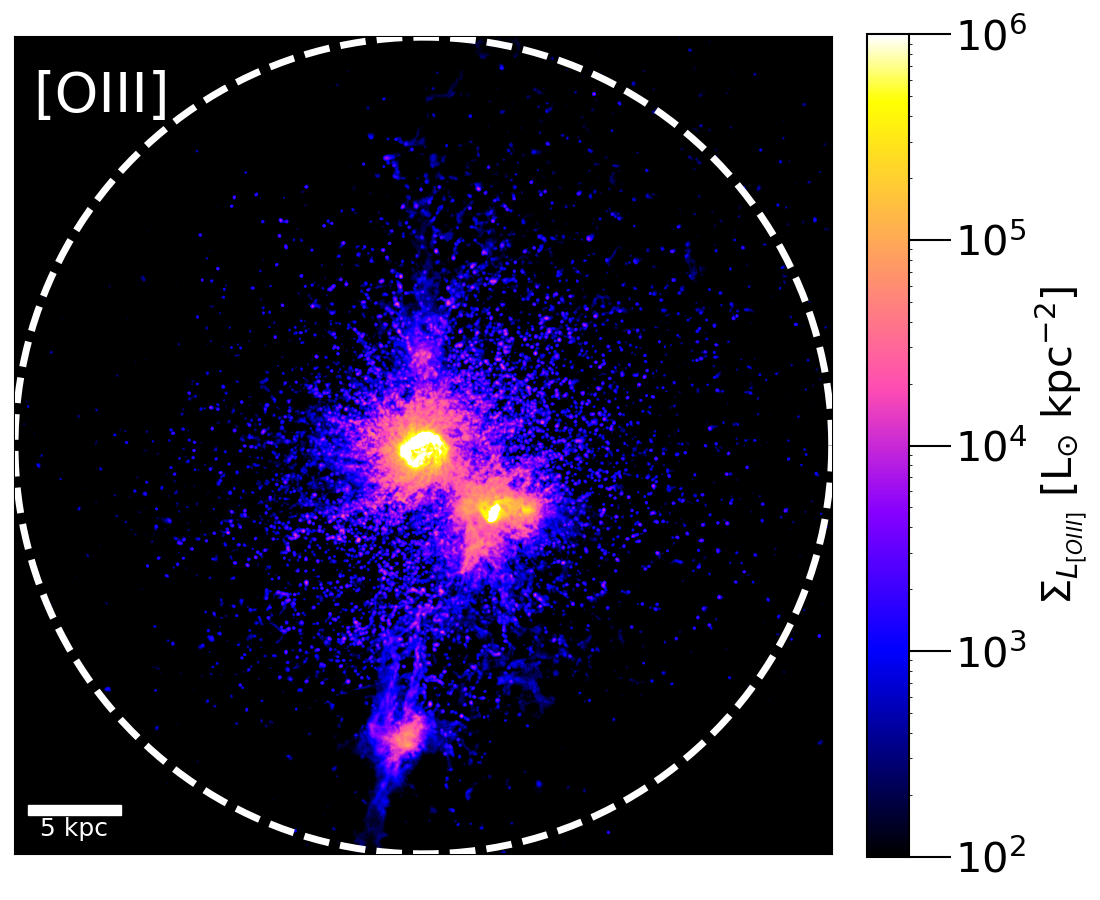}
  \includegraphics[width = 0.83\columnwidth]{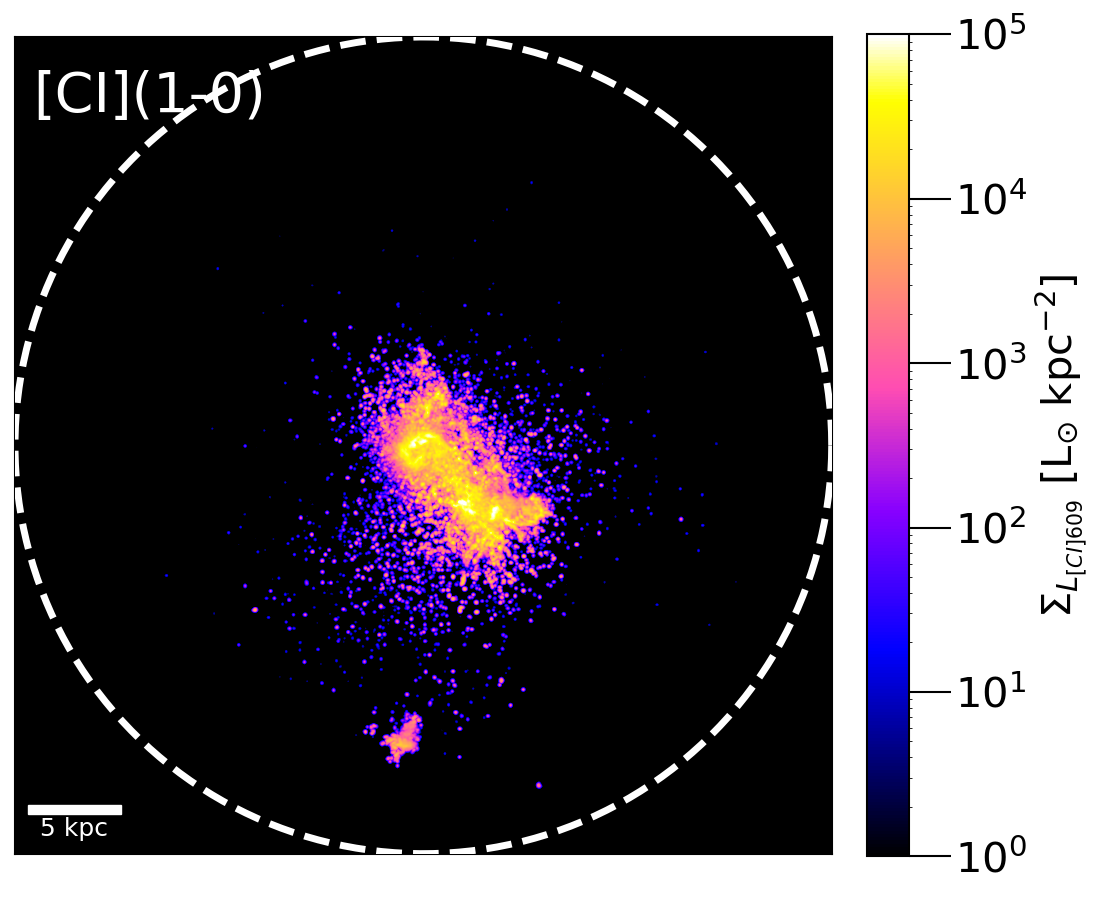}
  \includegraphics[width = 0.83\columnwidth]{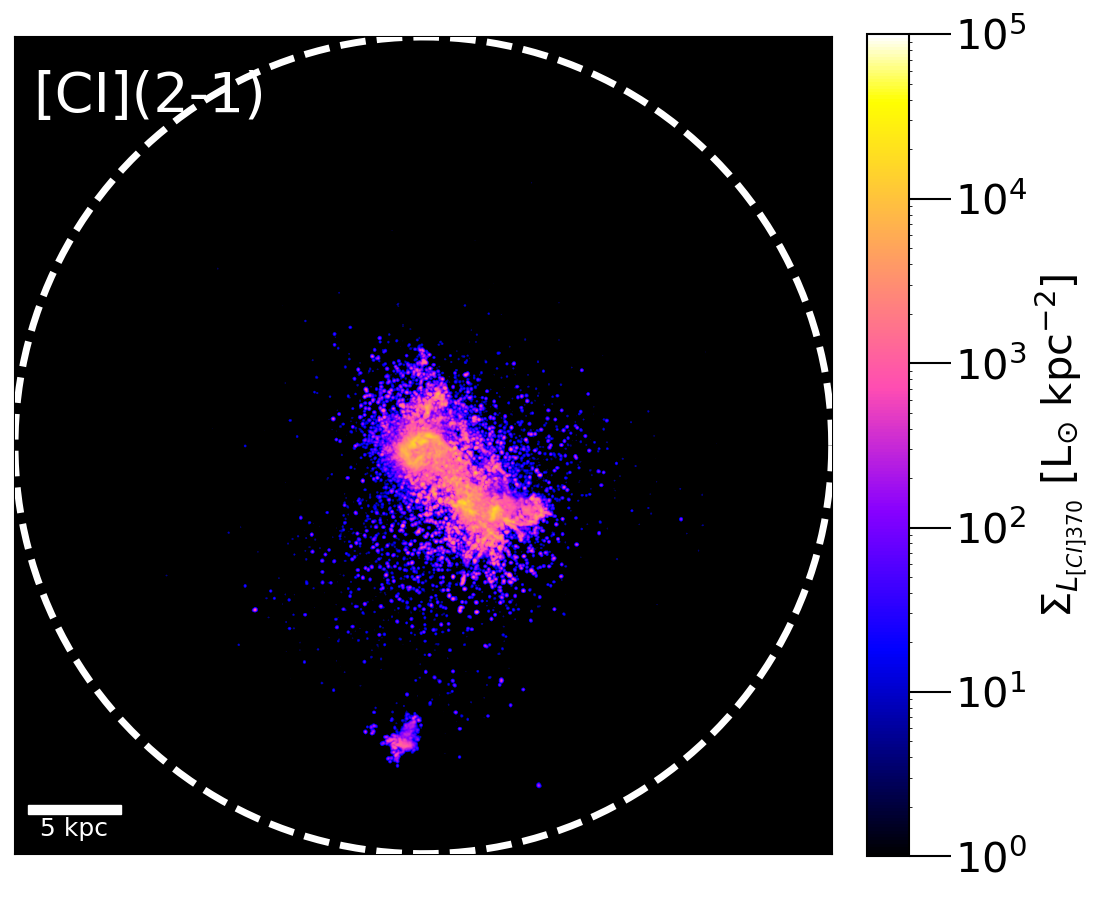}

  \includegraphics[width = 0.83\columnwidth]{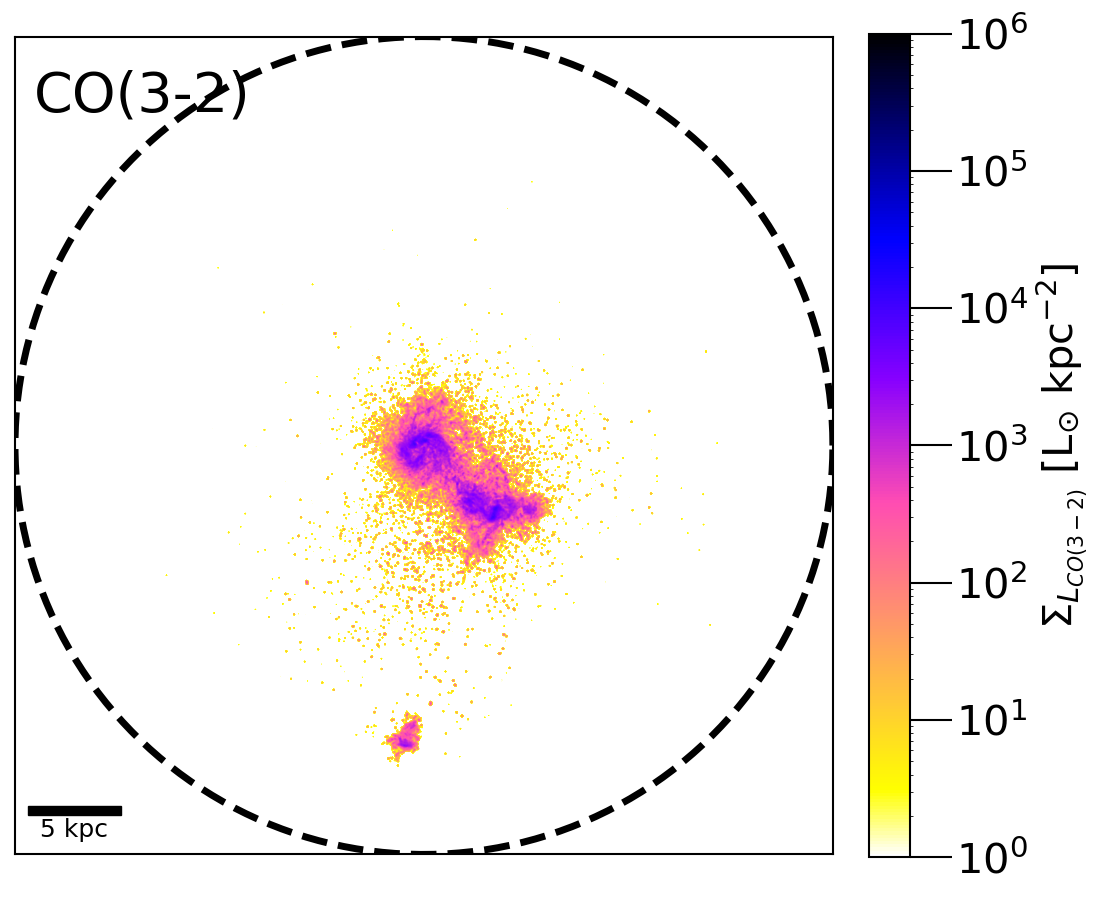}\\
      \caption{Line emission maps obtained with the MPM 2CG model. The circles in each plot mark R$_{\rm vir}$.}
         \label{fig:EMmapsMPM}
\end{figure*}

In Fig.~\ref{fig:EMmapsMPM} we show the different emission line maps obtained with the 2CG MPM, to compare it with the corresponding maps obtained with our fiducial RT model, that can be seen in Fig.~\ref{fig:EMmaps}. All maps of the MPMs look very similar in morphology, so we selected to only show one example. In the MPM we see a far more extended emission in [OIII] compared to the RT model, due to the contribution of cold and diffuse, as well as hot and dense gas, which heats up and cools down in the RT post-processing respectively, as commented in Section~\ref{sec:modelcomp} and as shown in the phase diagram in Fig.~\ref{fig:phase}. 

The [CII] and [CI] line emission in the MPM model do not trace the filaments as well as the RT model, and their resulting emission are very clumpy in appearance and less concentrated to the central disc of the galaxy than in the RT model, which is due to the assumption of every dense gas particle being treated as a GMC. The same behaviour can be seen in the CO(3-2) maps, which is shown in the bottom panel of Fig.~\ref{fig:EMmapsMPM}, where the left plot shows the map in the same scaling as the RT CO map, while the right panel shows a rescaled map to be able to see the differences in emission within the model. The reasons for this differences are explained in more detail in the main text of this paper. 

As in our modelling the dense gas is defined according to the star formation criteria of \Ponos, the particles where a sub-grid model was applied are the same particles that are able to form stars. Thus, the particles traced by CO in the MPM emission maps are also the places where we found recent star formation, including inside the CGM. 

\section{Emission line table}\label{app:table}

Table~\ref{table:linelums} lists the line luminosities in units of L$_{\odot}$  for all models discussed in this paper, together with the corresponding fractions for gas phase and morphology. We use the same definitions for the dense and diffuse gas in all models for consistency, which is described in detail in Section~\ref{compsep}. 

\clearpage
\onecolumn
\begin{landscape}
\begin{longtable}{lcccccccccc} 
\caption{Line luminosities for all models}          
\label{table:linelums}   \\   
\hline\hline  
& $L_{[\rm CII]}$ [L$_{\odot}$] & \%  & $L_{[\rm CI]609\mu m}$ [$L_{\odot}$] & \% & $L_{\rm [CI]370\mu m}$/ [$L_{\odot}$] & \% & $L_{\rm CO(3-2)}$ $[L_{\odot}]$ & \% & $L_{\rm [OIII]88\mu m}$ [$L_{\odot}$] & \% \\    
\hline
 \endfirsthead
 \multicolumn{11}{c}{Following from previous page} \\ 
 \hline 
& $L_{[\rm CII]}$ [L$_{\odot}$] & \%  & $L_{[\rm CI]609\mu m}$ [$L_{\odot}$] & \% & $L_{\rm [CI]370\mu m}$/ [$L_{\odot}$] & \% & $L_{\rm CO(3-2)}$ $[L_{\odot}]$ & \% & $L_{\rm [OIII]88\mu m}$ [$L_{\odot}$] & \% \\ \hline    
 \endhead
 \hline 
 \endfoot
 \hline
   \multicolumn{11}{c}{\textbf{Simple Model (SM)}} \\ \hline
   Total Luminosity & $2.20 * 10^{9}$ & & $3.63 * 10^{5}$ & & $2.25 * 10^{6}$ & & $4.53 * 10^{-4}$ & & $4.54 * 10^{9}$ & \\ 
   Dense Phase  & $1.81 * 10^{9}$ &  $82.14\%$  & $3.56 * 10^{5}$ &  $98.10\%$ & $2.19 * 10^{6}$ &  $97.42\%$ & $3.86 * 10^{-4}$ &  $85.21\%$ & $7.90 * 10^{8}$ &  $17.41\%$\\ 
   Diffuse Phase  & $3.93 * 10^{8}$ &  $17.86\%$  & $6.89 * 10^{3}$ &  $1.90\%$   & $5.81 * 10^{4}$ &  $2.58\%$  & $6.69 * 10^{-5}$ &  $14.79\%$ & $3.75 * 10^{9}$ &  $82.59\%$ \\ 
   - Diffuse Ionised & $1.73 * 10^{8}$ &  $7.85\%$  & $3.30 * 10^{3}$ &  $0.91\%$ & $2.89 * 10^{4}$ &  $1.28\%$  & $4.19 * 10^{-6}$ &  $0.93\%$ & $3.11 * 10^{9}$ &  $68.46\%$ \\ 
   - Diffuse Neutral & $2.20 * 10^{8}$ &  $10.01\%$ & $3.58 * 10^{3}$ &  $0.99\%$ & $2.92 * 10^{4}$ &  $1.30\%$ & $6.28 * 10^{-5}$ &  $13.86\%$ & $6.41 * 10^{8}$ &  $14.13\%$        \\ 
   Disc & $1.53 * 10^{9}$ &  $69.66\%$ & $2.89 * 10^{5}$ &  $79.73\%$ & $1.76 * 10^{6}$ &  $78.36\%$ & $3.61 * 10^{-4}$ &  $79.70\%$ & $2.95 * 10^{9}$ &  $64.92\%$        \\ 
   Merger & $5.85 * 10^{8}$ &  $26.59\%$ & $7.27 * 10^{4}$ &  $20.03\%$ & $4.83 * 10^{5}$ &  $21.45\%$ & $9.20 * 10^{-5}$ &  $20.31\%$ & $1.20 * 10^{9}$ &  $26.32\%$        \\ 
   CGM & $8.51 * 10^{7}$ &  $3.87\%$ & $1.29 * 10^{3}$ &  $0.36\%$ & $7.35 * 10^{3}$ &  $0.33\%$ & $1.08 * 10^{-8}$ &  $<0.01\%$ & $3.95 * 10^{8}$ &  $8.71\%$        \\ \hline 
   \multicolumn{11}{c}{\textbf{Multiphase model (MPM) - Plummer density profile }} \\ \hline
   Total Luminosity & $6.80 * 10^{9}$ & &$2.23 * 10^{8}$ && $2.73 * 10^{7}$ &  & $1.99 * 10^{8}$ &  & $3.75* 10^{9}$ & \\ 
   Dense Phase  & $6.44 * 10^{9}$ &  $94.25\%$  & $2.23 * 10^{8}$ &  $>99.99\%$  & $2.73 * 10^{7}$ &  $99.79\%$ & $1.99 * 10^{8}$ &  $>99.99\%$ & $9.32 * 10^{3}$ &  $<0.01\%$ \\ 
   Diffuse Phase  & $3.93 * 10^{8}$ &  $5.75\%$  & $6.89 * 10^{3}$ &  $<0.01\%$ & $5.81 * 10^{4}$ &  $0.21\%$  & $6.69* 10^{-5}$ &  $<0.01\%$  & $3.75 * 10^{9}$ &  $>99.99\%$      \\ 
   - Diffuse Ionised & $1.73 * 10^{8}$ &  $2.53\%$  & $3.30 * 10^{3}$ &  $<0.01\%$ & $2.89 * 10^{4}$ &  $0.11\%$ & $4.19* 10^{-6}$ &  $<0.01\%$ & $3.11 * 10^{9}$ &  $82.90\%$\\ 
   - Diffuse Neutral &  $2.20 * 10^{8}$ &  $3.22\%$ & $3.58 * 10^{3}$ &  $<0.01\%$  & $2.92 * 10^{4}$ &  $0.11\%$ & $6.28 * 10^{-5}$ &  $<0.01\%$   & $6.42 * 10^{8}$ &  $17.09\%$     \\ 
   Disc & $3.63 * 10^{9}$ &  $53.13\%$ & $8.54 * 10^{7}$ &  $38.36\%$ & $1.15 * 10^{7}$ &  $42.03\%$ & $1.04 * 10^{8}$ &  $52.14\%$ & $2.55 * 10^{9}$ &  $68.00\%$        \\ 
   Merger & $2.59 * 10^{9}$ &  $37.91\%$ & $8.48 * 10^{7}$ &  $38.03\%$ & $9.81 * 10^{6}$ &  $35.93\%$ & $7.75 * 10^{7}$ &  $38.94\%$ & $8.87 * 10^{8}$ &  $23.65\%$        \\ 
   CGM & $6.11 * 10^{8}$ &  $8.94\%$ & $5.28 * 10^{7}$ &  $23.71\%$ & $5.99 * 10^{6}$ &  $21.89\%$ & $1.75 * 10^{7}$ &  $8.83\%$ & $3.13 * 10^{8}$ &  $8.35\%$        \\ \hline
\multicolumn{11}{c}{\textbf{Multiphase model (MPM) - Logotropic density profile}} \\ \hline
   Total Luminosity & $1.23 * 10^{9}$ & & $1.09 * 10^{8}$ & & $1.64 * 10^{7}$ & & $7.60 * 10^{7}$ & & $3.75 * 10^{9}$  & \\ 
   Dense Phase & $8.36 * 10^{8}$ &  $68.04\%$ & $1.09 * 10^{8}$ &  $99.99\%$ & $1.63 * 10^{7}$ &  $99.65\%$ & $7.60 * 10^{7}$ &  $>99.99\%$ & $2.00 * 10^{-1}$ &  $<0.01\%$ \\ 
   Diffuse Phase  & $3.93* 10^{8}$ &  $31.96\%$ & $6.89 * 10^{3}$ &  $0.01\%$ & $5.81 * 10^{4}$ &  $0.35\%$ & $6.69 * 10^{-5}$ &  $<0.01\%$ & $3.75 * 10^{9}$ &  $>99.99\%$ \\ 
   - Diffuse Ionised & $1.73 * 10^{8}$ &  $14.05\%$ & $3.30 * 10^{3}$ &  $<0.01\%$ & $2.89 * 10^{4}$ &  $0.18\%$ & $4.19 * 10^{-6}$ &  $<0.01\%$ & $3.11 * 10^{9}$ &  $82.91\%$ \\ 
   - Diffuse Neutral & $2.20 * 10^{8}$ &  $17.91\%$ & $3.58 * 10^{3}$ &  $<0.01\%$  & $2.92 * 10^{4}$ &  $0.18\%$ & $6.28 * 10^{-5}$ &  $<0.01\%$   & $6.42 * 10^{8}$ &  $17.09\%$       \\ 
   Disc & $6.93 * 10^{8}$ &  $56.39\%$ & $6.04 * 10^{7}$ &  $55.21\%$ & $1.01 * 10^{7}$ &  $61.66\%$ & $4.69 * 10^{7}$ &  $61.64\%$ & $2.55 * 10^{9}$ &  $68.00\%$        \\ 
   Merger & $4.22 * 10^{8}$ &  $34.31\%$ & $3.68 * 10^{7}$ &  $33.76\%$ & $4.91 * 10^{6}$ &  $29.94\%$ & $2.50 * 10^{7}$ &  $32.95\%$ & $8.87 * 10^{8}$ &  $23.65\%$        \\ 
   CGM & $1.15 * 10^{8}$ &  $9.33\%$ & $1.18 * 10^{7}$ &  $10.76\%$ & $1.39 * 10^{6}$ &  $8.47\%$ & $4.06 * 10^{6}$ &  $5.34\%$ & $3.13 * 10^{8}$ &  $8.35\%$        \\ \hline
   \multicolumn{11}{c}{\textbf{Multiphase model (MPM) - Power Law density profile}} \\ \hline
   Total Luminosity & $3.21 * 10^{9}$ & & $3.04 * 10^{8}$ & & $3.84 * 10^{7}$ & & $4.93 * 10^{8}$ & & $3.75 * 10^{9}$ &\\ 
   Dense Phase  & $2.81 * 10^{9}$ &  $87.75\%$ & $3.04 * 10^{8}$ &  $>99.99\%$ & $3.84* 10^{7}$ &  $99.85\%$ & $4.93 * 10^{8}$ &  $>99.99\%$ & $3.89 * 10^{-1}$ &  $<0.01\%$            \\ 
   Diffuse Phase  & $3.93* 10^{8}$ &  $12.25\%$ & $6.89 * 10^{3}$ &  $<0.01\%$ & $5.81 * 10^{4}$ &  $0.33\%$ & $6.69 * 10^{-5}$ &  $<0.01\%$ & $3.75 * 10^{9}$ &  $>99.99\%$     \\ 
   - Diffuse Ionised & $1.73 * 10^{8}$ &  $5.38\%$ & $3.30 * 10^{3}$ &  $<0.01\%$ & $2.89 * 10^{4}$ &  $0.08\%$ & $4.19 * 10^{-6}$ &  $<0.01\%$ & $3.11 * 10^{9}$ &  $82.91\%$ \\ 
   - Diffuse Neutral & $2.20 * 10^{8}$ &  $6.86\%$ & $3.58 * 10^{3}$ &  $<0.01\%$  & $2.92 * 10^{4}$ &  $0.08\%$ & $6.28 * 10^{-5}$ &  $<0.01\%$   & $6.42 * 10^{8}$ &  $17.09\%$       \\ 
   Disc & $2.23 * 10^{9}$ &  $69.43\%$ & $1.35 * 10^{8}$ &  $44.41\%$ & $1.85 * 10^{7}$ &  $48.14\%$ & $2.16 * 10^{8}$ &  $43.78\%$ & $2.55 * 10^{9}$ &  $68.00\%$        \\ 
   Merger & $8.11 * 10^{8}$ &  $25.26\%$ & $1.10 * 10^{8}$ &  $36.22\%$ & $1.31 * 10^{7}$ &  $34.19\%$ & $2.15 * 10^{8}$ &  $43.69\%$ & $8.87 * 10^{8}$ &  $23.65\%$        \\ 
   CGM & $1.69 * 10^{8}$ &  $5.26\%$ & $5.89 * 10^{7}$ &  $19.38\%$ & $6.77 * 10^{6}$ &  $17.63\%$ & $6.16 * 10^{7}$ &  $12.49\%$ & $3.13 * 10^{8}$ &  $8.35\%$        \\ \hline 
   \multicolumn{11}{c}{\textbf{Multiphase model (MPM) - Two component Gaussian (2CG) density profile }} \\ \hline
   Total Luminosity & $8.41 * 10^{9}$ & & $2.08 * 10^{8}$ & & $2.55 * 10^{7}$ & & $1.93 * 10^{8}$ & & $3.75 * 10^{9}$ & \\ 
   Dense Phase & $28.02 * 10^{9}$ &  $95.33\%$ & $2.08 * 10^{8}$ &  $>99.99\%$ & $2.54 * 10^{7}$ &  $99.77\%$  & $1.93 * 10^{8}$ &  $>99.99\%$ & $1.22 * 10^{5}$ &  $<0.01\%$ \\ 
   Diffuse Phase & $3.93 * 10^{8}$ &  $4.67\%$ & $6.89 * 10^{3}$ &  $<0.01\%$ & $5.81 * 10^{4}$ &  $0.23\%$  & $6.69 * 10^{-5}$ &  $<0.01\%$ & $3.75 * 10^{9}$ &  $>99.99\%$ \\ 
   - Diffuse Ionised & $1.73 * 10^{8}$ &  $2.05\%$ & $3.30 * 10^{3}$ &  $<0.01\%$ & $2.89 * 10^{4}$ &  $0.11\%$ & $4.19 * 10^{-6}$ &  $<0.01\%$ & $3.11 * 10^{9}$ &  $82.90\%$ \\ 
   - Diffuse Neutral & $2.20 * 10^{8}$ &  $2.62\%$ & $3.58 * 10^{3}$ &  $<0.01\%$  & $2.92 * 10^{4}$ &  $0.11\%$ & $6.28 * 10^{-5}$ &  $<0.01\%$   & $6.42 * 10^{8}$ &  $17.09\%$   \\ 
   Disc & $4.39 * 10^{9}$ &  $52.24\%$ & $7.68 * 10^{7}$ &  $36.86\%$ & $1.03 * 10^{7}$ &  $40.44\%$ & $9.24 * 10^{7}$ &  $47.94\%$ & $2.55 * 10^{9}$ &  $68.00\%$        \\ 
   Merger & $3.26 * 10^{9}$ &  $38.72\%$ & $7.64 * 10^{7}$ &  $36.70\%$ & $8.97 * 10^{6}$ &  $35.18\%$ & $8.05 * 10^{7}$ &  $41.71\%$ & $8.87 * 10^{8}$ &  $23.65\%$        \\ 
   CGM & $7.64 * 10^{8}$ &  $9.08\%$ & $5.48 * 10^{7}$ &  $26.29\%$ & $6.23 * 10^{6}$ &  $24.45\%$ & $2.01 * 10^{7}$ &  $10.40\%$ & $3.13 * 10^{8}$ &  $8.35\%$        \\ \hline 
   \multicolumn{11}{c}{\textbf{\rev{Global Radiative Transfer - RamsesRT}}} \\ \hline
   Total Luminosity     & $3.91 * 10^{8}$ &             & $1.76 * 10^{6}$ &            & $7.70 * 10^{6}$ &             & $5.25 * 10^{5}$  &            & $7.09 * 10^{7}$ &          \\ 
   Dense Phase          & $1.72 * 10^{8}$ &  $55.99\%$  & $1.71 * 10^{6}$ &  $96.98\%$ & $7.54 * 10^{6}$ &  $97.90\%$  & $5.25 * 10^{5}$  &  $99.97\%$ & $2.27 * 10^{5}$ &  $0.32\%$ \\ 
   Diffuse Phase        & $1.72 * 10^{8}$ &  $44.01\%$  & $5.32 * 10^{4}$ &  $3.02\%$  & $1.62 * 10^{5}$ &  $2.10\%$   & $1.41 * 10^{2}$  &  $0.03\%$  & $7.06 * 10^{7}$ &  $99.68\%$ \\ 
   - Diffuse Ionised    & $9.87 * 10^{7}$ &  $25.26\%$  & $2.92 * 10^{4}$ &  $1.65\%$  & $8.48 * 10^{4}$ &  $1.10\%$   & $1.11 * 10^{1}$  &  $<0.01\%$ & $6.79 * 10^{7}$ &  $95.79\%$ \\ 
   - Diffuse Neutral    & $7.33 * 10^{7}$ &  $18.75\%$  & $2.40 * 10^{4}$ &  $1.36\%$  & $7.67 * 10^{4}$ &  $1.00\%$   & $1.29 * 10^{2}$  &  $0.02\%$  & $2.76 * 10^{6}$ &  $3.89\%$ \\ 
   Disc                 & $2.41 * 10^{8}$ &  $61.65\%$  & $1.33 * 10^{6}$ &  $75.36\%$ & $5.84 * 10^{6}$ &  $75.81\%$  & $3.66 * 10^{5}$  &  $69.79\%$ & $3.25 * 10^{7}$ &  $45.85\%$ \\ 
   Merger               & $1.08 * 10^{8}$ &  $27.64\%$  & $4.07 * 10^{5}$ &  $23.08\%$ & $1.80 * 10^{6}$ &  $23.32\%$  & $1.58 * 10^{5}$  &  $30.20\%$ & $1.95 * 10^{7}$ &  $27.46\%$ \\ 
   CGM                  & $3.87 * 10^{7}$ &  $9.90\%$   & $2.33 * 10^{4}$ &  $1.32\%$  & $5.83 * 10^{4}$ &  $0.76\%$   & $4.16 * 10^{1}$  &  $0.01\%$  & $1.52 * 10^{7}$ &  $21.44\%$ \\ 
\hline                                   
\end{longtable}
\end{landscape}
\end{appendix}

\end{document}